\documentclass[12pt]{article}
\pdfoutput=1
\usepackage{latexsym}
\usepackage{amssymb, amsfonts, amsmath}
\usepackage{indentfirst}
\usepackage{bbm}
\usepackage{amssymb}
\usepackage{verbatim}
\usepackage{amsmath, amsthm, amssymb}
\usepackage{mathrsfs}
\usepackage{amsfonts}
\usepackage{dsfont}
\usepackage{csquotes}
\usepackage{hyperref}
\usepackage{url}
\usepackage{xcolor}
\usepackage{xpatch}
\usepackage[english]{babel}

\usepackage{graphicx} 
\usepackage[export]{adjustbox}
\usepackage{float}
\graphicspath{ {images/} }

\usepackage[style=phys, articletitle=true, biblabel=brackets, backend=bibtex,
chaptertitle=false, pageranges=false, eprint=true,%
natbib=true, maxbibnames=10, giveninits=true, sorting=none]{biblatex} 

\DeclareFieldFormat[article]{journaltitle}{\mkbibitalic{#1\isdot}} 
\makeatletter
\newrobustcmd{\mkbibfixedbrackets}[1]{%
	\begingroup
	\blx@blxinit
	\blx@setsfcodes
	\bibleftbracket#1\bibrightbracket
	\endgroup}

\renewbibmacro*{doi+eprint+url}{%
	\iftoggle{bbx:doi}
	{\printfield{doi}}
	{}%
	\ifboolexpr{test {\iffieldequalstr{eprinttype}{arxiv}} or test {\iffieldequalstr{eprinttype}{arXiv}}}
	{\setunit{\addspace}\newblock}
	{\newunit\newblock}%
	\iftoggle{bbx:eprint}
	{\usebibmacro{eprint}}
	{}%
	\newunit\newblock
	\iftoggle{bbx:url}
	{\usebibmacro{url+urldate}}
	{}}

\xpatchbibdriver{online}
{\newunit\newblock
	\iftoggle{bbx:eprint}
	{\usebibmacro{eprint}}
	{}}
{\ifboolexpr{test {\iffieldequalstr{eprinttype}{arxiv}} or test {\iffieldequalstr{eprinttype}{arXiv}}}
	{\setunit{\addspace}\newblock}
	{\newunit\newblock}%
	\iftoggle{bbx:eprint}
	{\usebibmacro{eprint}}
	{}}
{}{}

\DeclareFieldFormat{eprint:arxiv}{%
	\mkbibbrackets{%
		\ifhyperref
		{\href{http://arxiv.org/\abx@arxivpath/#1}{%
				arXiv\addcolon
				\nolinkurl{#1}%
				\iffieldundef{eprintclass}
				{}
				{\addspace\UrlFont{\mkbibfixedbrackets{\thefield{eprintclass}}}}}}
		{arXiv\addcolon
			\nolinkurl{#1}%
			\iffieldundef{eprintclass}
			{}
			{\addspace\UrlFont{\mkbibfixedbrackets{\thefield{eprintclass}}}}}}}
\makeatother

\parskip=\medskipamount

\DeclareMathAlphabet{\mathbbmsl}{U}{bbm}{m}{sl}

\newcommand{\cA}{{\cal A}}
\newcommand{\cB}{{\cal B}}

\newcommand{\cD}{{\cal D}}

\newcommand{\cF}{{\cal F}}
\newcommand{\cG}{{\cal G}}
\newcommand{\cH}{{\cal H}}
\newcommand{\cI}{{\cal I}}

\newcommand{\cN}{{\cal N}}
\newcommand{\cO}{{\cal O}}

\newcommand{\cQ}{{\cal Q}}

\newcommand{\cZ}{{\cal Z}}

\def\a{\alpha}

\def\b{\beta}

\def\d{\delta}
\def\e{\epsilon}
\def\f{\phi}
\def\g{\gamma}
\def\G{\Gamma}

\def\l{\lambda}

\def\q{\theta}
\def\r{\rho}
\def\s{\sigma}

\def\x{\xi}

\def\D{\Delta}
\def\F{\Phi}
\def\J{\Psi}

\def\P{\Pi}
\def\Q{\Theta}

\newcommand{\ve}{\varepsilon}                            

\newcommand{\pa}{\partial}                           

\newcommand{\be}{\begin{equation}}
\newcommand{\ee}{\end{equation}}
\newcommand{\bea}{\begin{eqnarray}}
\newcommand{\eea}{\end{eqnarray}}

\newcommand{\ba}{\begin{array}}
\newcommand{\ea}{\end{array}}

\def\double #1{#1{\hbox{\kern-2pt $#1$}}}

\newcommand{\bsubeq}{\begin{subequations}}
\newcommand{\esubeq}{\end{subequations}}

\numberwithin{equation}{section}

\begin{document}

\begin{titlepage}
\begin{flushright}
Feb, 2023
\end{flushright}
\vspace{2mm}

\begin{center}
\Large \bf Three-point functions of conserved supercurrents \\ in 3D $\cN=1$ SCFT: general formalism \\ for arbitrary superspins
\end{center}

\begin{center}
{\bf
Evgeny I. Buchbinder and Benjamin J. Stone}

{\footnotesize{
{\it Department of Physics M013, The University of Western Australia\\
35 Stirling Highway, Crawley W.A. 6009, Australia}} ~\\
}
\end{center}
\begin{center}
\texttt{Email: evgeny.buchbinder@uwa.edu.au, \\ benjamin.stone@research.uwa.edu.au}
\end{center}

\vspace{4mm}
\begin{abstract}
\baselineskip=14pt
\noindent 

We analyse the general structure of the three-point functions of conserved higher-spin supercurrents in 3D, $\cN=1$ superconformal field theory. 
It is shown that supersymmetry imposes additional restrictions on correlation functions of conserved higher-spin currents. We develop a manifestly supersymmetric formalism 
to compute the three-point function $\langle \mathbf{J}^{}_{s_{1}} \mathbf{J}'_{s_{2}} \mathbf{J}''_{s_{3}} \rangle$, where $\mathbf{J}^{}_{s_{1}}$, $\mathbf{J}'_{s_{2}}$ and $\mathbf{J}''_{s_{3}}$ 
are conserved higher-spin supercurrents with superspins $s_{1}$, $s_{2}$ and $s_{3}$ respectively (integer or half-integer).
Using a computational approach limited only by computer power, we analytically impose the constraints arising from the superfield conservation equations and symmetries under permutations of superspace points. Explicit solutions for three-point functions are presented and we provide a complete classification of the results for $s_{i} \leq 20 $; the pattern is very clear, and we propose that our classification holds for arbitrary superspins. We demonstrate that Grassmann-even three-point functions are fixed up to one parity-even structure and one parity-odd structure, 
while Grassmann-odd three-point functions are fixed up to a single parity-even structure. The existence of the parity-odd structure in the Grassmann-even correlation functions is subject to a set of triangle inequalities in the superspins. For completeness, we also analyse the structure of three-point functions involving conserved higher-spin supercurrents and scalar superfields.



\end{abstract}
\end{titlepage}

\newpage
\renewcommand{\thefootnote}{\arabic{footnote}}
\setcounter{footnote}{0}

\tableofcontents
\vspace{1cm}
\bigskip\hrule


\section{Introduction}\label{section1}

A well known implication of conformal 
symmetry~\cite{Polyakov:1970xd, Schreier:1971um, Migdal:1971xh, Migdal:1971fof,Ferrara:1972cq,Ferrara:1973yt, Koller:1974ut, Mack:1976pa, Fradkin:1978pp, Stanev:1988ft,Osborn:1993cr} 
is that the general form of two- and three-point correlation 
functions of primary operators is fixed up to finitely many parameters. However, constructing explicit solutions for three-point functions of conserved current operators such as the 
energy-momentum tensor, vector currents, and more generally, higher-spin currents, remains an open problem. An interesting feature of three-dimensional conformal field theories is the existence of 
parity-odd structures in the three-point functions of conserved currents. These structures were overlooked in the seminal work by 
Osborn \& Petkou \cite{Osborn:1993cr} (see also~\cite{Erdmenger:1996yc}), which 
introduced the group-theoretic formalism to study the three-point functions of the energy-momentum tensor and vector currents. The parity-odd structures were discovered later using a 
polarisation spinor approach in \cite{Giombi:2011rz}, where results for three-point functions of conserved (bosonic) higher-spin currents were obtained. 
Soon after, it was proven by Maldacena and Zhiboedov in~\cite{Maldacena:2011jn} that correlation functions involving the energy-momentum tensor and higher-spin currents are equal to those 
of free field theories.\footnote{An assumption of the Maldacena-Zhiboedov theorem is that the conformal theory under consideration possesses a unique spin-2 conserved current -- the 
energy-momentum tensor. This assumption, however, does not hold in the presence of fermionic higher-spin currents. Hence, it also does not hold in superconformal theories 
possessing conserved higher-spin supercurrents.}
This can be viewed as an extension of the Coleman-Mandula theorem \cite{Coleman:1967ad} to conformal field theories; 
it was originally proven in three dimensions and was generalised to four- and higher-dimensional cases in~\cite{Zhiboedov:2012bm, Stanev:2012nq, Stanev:2013qra,Boulanger:2013zza} (see also \cite{Alba:2013yda, Alba:2015upa}). 
In three dimensional theories the general structure of the three-point function $\langle J^{}_{s_{1}} J'_{s_{2}} J''_{s_{3}} \rangle$, 
where $J^{}_{s}$ denotes a conserved current of arbitrary spin-$s$, is fixed up to the following form~\cite{Giombi:2011rz, Maldacena:2011jn}:\footnote{Recall: in a $d$-dimensional CFT, 
a conserved current of spin-$s$ is a totally 
symmetric and traceless tensor $J_{m_{1} ... m_{s}}$ of scale dimension $\D_{J} = s+d-2$, satisfying the conservation equation $\partial^{m_{1}} J_{m_{1} ... m_{s}} = 0$.}
\begin{equation}
	\langle J^{}_{s_{1}} J'_{s_{2}} J''_{s_{3}} \rangle = a_{1} \langle J^{}_{s_{1}} J'_{s_{2}} J''_{s_{3}} \rangle_{E_{1}} + a_{2} \langle J^{}_{s_{1}} J'_{s_{2}} J''_{s_{3}} \rangle_{E_{2}} + b \langle J^{}_{s_{1}} J'_{s_{2}} J''_{s_{3}} \rangle_{O} \, ,
\end{equation}
where $\langle J^{}_{s_{1}} J'_{s_{2}} J''_{s_{3}} \rangle_{E_{1}}$, $\langle J^{}_{s_{1}} J'_{s_{2}} J''_{s_{3}} \rangle_{E_{2}}$ are parity-even solutions corresponding to free field theories, and $\langle J^{}_{s_{1}} J'_{s_{2}} J''_{s_{3}} \rangle_{O}$ is a parity-violating, or parity-odd solution which is not generated by a free CFT. The existence of the parity-odd solution is subject to the following triangle inequalities on the spins:
\begin{align}
	s_{1} \leq s_{2} + s_{3} \, , && s_{2} \leq s_{1} + s_{3} \, , && s_{3} \leq s_{1} + s_{2} \, .
\end{align}
If any of the above inequalities are not satisfied, then the odd solution is incompatible with current conservation. Parity-odd solutions are unique to three dimensions, and have been shown to arise in Chern-Simons theories interacting with parity-violating matter \cite{Aharony:2011jz, Giombi:2011kc, Maldacena:2012sf, Jain:2012qi, GurAri:2012is, Aharony:2012nh, Giombi:2016zwa, Chowdhury:2017vel, Sezgin:2017jgm, Skvortsov:2018uru, Inbasekar:2019wdw}. Existence and uniqueness of the odd solution has been proven in \cite{Giombi:2016zwa}, while methods to obtain explicit solutions for arbitrary spin are contained in \cite{Zhiboedov:2012bm,Stanev:2012nq,Stanev:2013eha,Buchbinder:2022mys}. 


A natural follow-up question arises: in conformal field theories, what are the implications of supersymmetry on the general structure of three-point correlation functions? The study of correlation functions in superconformal theories has been carried out in diverse dimensions using the group-theoretic approach developed in the following 
publications \cite{Osborn:1998qu, Park:1998nra, Park:1999pd, Park:1999cw, Kuzenko:1999pi, Nizami:2013tpa, Buchbinder:2015qsa, Buchbinder:2015wia, 
Kuzenko:2016cmf, Buchbinder:2021gwu, Buchbinder:2021izb, Buchbinder:2021kjk, Buchbinder:2021qlb, Jain:2022izp,Buchbinder:2021qlb,Buchbinder:2022cqp,Buchbinder:2022kmj,Buchbinder:2022mys}. It has been shown that
superconformal symmetry imposes additional restrictions on the three-point functions of conserved currents compared to non-supersymmetric theories. For example, it was pointed out
 in~\cite{Buchbinder:2021gwu} that there is an apparent tension between supersymmetry and the existence of parity-violating structures. 
 In contrast with the non-supersymmetric case, parity-odd structures are not found in the three-point functions of the energy-momentum tensor and conserved 
 vector currents \cite{Buchbinder:2015qsa,Buchbinder:2015wia,Kuzenko:2016cmf,Buchbinder:2021gwu}.
 For three-point functions of higher-spin currents the results are more unclear, however, it was shown in \cite{Buchbinder:2021qlb} that parity-odd structures can appear in the 
 three-point functions of currents belonging to a superspin-$2$ current multiplet. Such a multiplet contains independent conserved currents of spin-$2$ and 
 spin-$\tfrac{5}{2}$ (the spin-$2$ current is not equal to but possesses the same properties as the energy-momentum tensor). In general, for three-point functions 
 involving conserved higher-spin currents, the conditions under which parity-violating structures can arise in supersymmetric theories are not well understood.

The intent of this paper is to address these concerns and provide a complete classification of conserved three-point functions in 3D $\cN=1$ superconformal field theory. To do this we develop a general formalism to study the three-point function
\begin{equation} \label{3D N=1 three-point function}
	\langle \mathbf{J}^{}_{s_{1}}(z_{1}) \, \mathbf{J}'_{s_{2}}(z_{2}) \, \mathbf{J}''_{s_{3}}(z_{3}) \rangle \, ,
\end{equation}
where $z_{1}, z_{2}, z_{3}$ are points in 3D $\cN=1$ Minkowski superspace, and the superfield $\mathbf{J}_{s}(z)$ is a conserved higher-spin supercurrent of superspin-$s$ (integer or half-integer). These currents are primary superfields transforming in an irreducible representation of the 3D $\cN=1$ superconformal algebra, $\mathfrak{so}(3,2|1) \cong \mathfrak{osp}(1|2;\mathbb{R})$. They are described by totally symmetric spin-tensors of rank $2s$, $\mathbf{J}_{\a_{1} ... \a_{2s}}(z) = \mathbf{J}_{(\a_{1} ... \a_{2s})}(z)$, and satisfy the following superfield conservation equation:
\begin{equation} \label{Conserved supercurrent}
	D^{\a_{1}} \mathbf{J}_{\a_{1} \a_{2} ... \a_{2s}}(z) = 0\, ,
\end{equation}
where $D^{\a}$ is the conventional covariant spinor derivative in $\cN=1$ superspace. As a result of the superfield conservation equation \eqref{Conserved supercurrent}, conserved supercurrents have scale dimension $\D_{\mathbf{J}} = s + 1$ (saturating the unitary bound), and at the component level contain independent conserved currents of spin-$s$ and $s+\tfrac{1}{2}$ respectively. The most important examples of conserved supercurrents in superconformal field theory are the supercurrent and flavour current multiplets, corresponding to the cases $s=\tfrac{3}{2}$ and $s = \tfrac{1}{2}$ respectively (for a review of the properties of supercurrent and flavour current multiplets in 3D theories, see \cite{Buchbinder:2015qsa,Korovin:2016tsq} and the references there-in). The supercurrent multiplet contains the energy-momentum tensor and the supersymmetry current.\footnote{In $\cN$-extended superconformal theories, the supercurrent multiplet also contains the $R$-symmetry currents.} Likewise, the flavour current multiplet contains a conserved vector current. Three-point correlation functions of these currents contain important physical information about a given superconformal field theory and are highly constrained by superconformal symmetry. 

The general structure of three-point functions of conserved (higher-spin) currents in 3D $\cN=1$ superconformal field theory was proposed in~\cite{Nizami:2013tpa} to be fixed up to the 
following form:
%
\begin{equation}
	\langle \mathbf{J}^{}_{s_{1}} \mathbf{J}'_{s_{2}} \mathbf{J}''_{s_{3}} \rangle = a \, \langle \mathbf{J}^{}_{s_{1}} \mathbf{J}'_{s_{2}} \mathbf{J}''_{s_{3}} \rangle_{E} + b \, \langle \mathbf{J}^{}_{s_{1}} \mathbf{J}'_{s_{2}} \mathbf{J}''_{s_{3}} \rangle_{O} \, ,
\label{zh1}	
\end{equation}
where $\langle \mathbf{J}^{}_{s_{1}} \mathbf{J}'_{s_{2}} \mathbf{J}''_{s_{3}} \rangle_{E}$ is a parity-even solution, and $\langle \mathbf{J}^{}_{s_{1}} \mathbf{J}'_{s_{2}} \mathbf{J}''_{s_{3}} \rangle_{O}$ 
is a parity-odd solution. However, as was pointed out above there is a tension between supersymmetry and existence of parity-odd structures, which means that the coefficient $b$ in~\eqref{zh1} 
vanishes in many correlators. In this paper we provide a complete classification for when the parity-odd structures are allowed and when they are not. In particular, we show that 
the odd solution does not appear in correlation functions that are overall Grassmann-odd (or fermionic). 
In the Grassmann-even (bosonic) three-point functions the existence of the parity-odd solution is subject to the following superspin triangle inequalities:
\begin{align}
	s_{1} \leq s_{2} + s_{3} \, , && s_{2} \leq s_{1} + s_{3} \, , && s_{3} \leq s_{1} + s_{2} \, .
\label{zh2}	
\end{align}
When the triangle inequalities are simultaneously satisfied there is one even solution and one odd solution, however, if any of the above inequalities
are not satisfied then the odd solution is incompatible with the superfield conservation equations. Our classification is in perfect agreement with our previous results 
in~\cite{Buchbinder:2015qsa, Buchbinder:2021gwu} for the three-point functions of the energy-momentum tensor and conserved vector currents. They belong to the supermultiplets 
of superspins $s=\tfrac{3}{2}$ and $s = \tfrac{1}{2}$ respectively and, hence, their three-point functions in superspace are Grassmann-odd. 
Based on our classification, it is implied that they do not possess parity-odd
contributions, which is in agreement with the earlier results. Our classification is also in agreement with our previous result in~\cite{Buchbinder:2021qlb} for the three-point function 
of the conserved supercurrent of superspin-2. This three-point function is Grassmann-even in superspace and since the triangle inequalities~\eqref{zh2} are satisfied a parity-odd contribution is allowed.

Our method assumes only the constraints imposed by superconformal symmetry and superfield conservation equations; within the framework of our formalism we reproduce all known 
results concerning the structure of three-point functions of conserved supercurrents in 3D $\cN=1$ SCFT. We present new results for three-point functions involving higher-spin supercurrents, 
obtaining explicit and completely analytic results. We also analyse three-point functions involving scalar superfields, thus covering essentially all possible three-point function in 3D $\cN=1$ 
superconformal field theory. Our method is based on a computational approach (by means of analytic/symbolic computer algebra in \textit{Mathematica}) which constructs all possible structures for 
the correlation function for a given set of superspins $s_1, s_2$ and $s_3$, consistent with its superconformal properties. Next, we extract the linearly independent structures by systematic 
application of linear dependence relations and then impose the superfield conservation equations and symmetries under permutations of superspace points. As a result we obtain the three-point 
function in a very explicit form which can be presented for relatively high superspins.
The method can be applied for arbitrary superspins and is limited only by computer power. Due to these limitations we were able to 
carry out computations up to $s_{i} = 20$ (a ``soft" limit, after which the calculations take many hours), however, with a sufficiently powerful computer one could extend this bound even further. 
The computational approach we have developed (based on the same method as in~\cite{Buchbinder:2022mys}) is completely algorithmic; one simply chooses the superspins of the fields and the 
solution for the three-point function consistent with conservation and point-switch symmetries is generated. 

The analysis is computationally intensive for higher-spins; to streamline the calculations we develop a hybrid, index-free formalism which combines the group-theoretic superspace formalism 
introduced by Osborn~\cite{Osborn:1998qu}
and Park~ \cite{Park:1999cw,Park:1999pd}, and a method based on contraction of tensor indices with auxiliary spinors. 
This method is widely used throughout the literature to construct correlation functions of higher-spin currents (see e.g.~\cite{Giombi:2011rz, Costa:2011mg, 
Stanev:2012nq, Zhiboedov:2012bm, Nizami:2013tpa, Elkhidir:2014woa}), however, this particular approach describes the correlation function completely in terms of a polynomial, $\cH(\boldsymbol{X},\Q; u,v,w)$, which is a function of 
two superconformally covariant three-point building blocks, $\boldsymbol{X}$ and $\Q$, and the auxiliary spinor variables $u$, $v$, and $w$. As a result one does not have 
to work with the superspace points explicitly when imposing the superfield conservation equations.


The results of this paper are organised as follows. In section \ref{section2} we review the essentials of the group theoretic formalism used to construct correlation functions of 
primary superfields in 3D $\cN=1$ SCFT. In section \ref{section3} we outline a method to impose all constraints arising from superfield conservation equations and point-switch symmetries on three-point functions of conserved higher-spin supercurrents. 
In particular, we introduce an index-free, auxiliary spinor formalism which allows us to construct a generating function for the three-point functions and we outline the important aspects of our computational approach. Section \ref{section4} is then devoted to the analysis of three-point functions involving conserved supercurrents. As a test of our approach, we present an explicit analysis for three-point correlation functions involving 
combinations of supercurrent and flavour current multiplets, reproducing the known results \cite{Buchbinder:2015qsa, Buchbinder:2021gwu}. 
The results are then expanded to include conserved higher-spin supercurrents, for which we provide many examples and confirm the results of \cite{Buchbinder:2021qlb}. Here we also 
resolve a contradiction in the literature concerning the structure of the three-point function $\langle \mathbf{J}^{}_{1/2} \mathbf{J}'_{1/2} \mathbf{J}''_{2} \rangle$; it was found 
in~\cite{Nizami:2013tpa} that this three-point function contains a parity-odd solution, however, it was shown later in~\cite{Buchbinder:2021qlb} that parity-odd structures are inconsistent 
with conservation equations. In this paper we re-examine this three-point function and provide a straightforward explanation, based on the triangle 
inequalities~\eqref{zh2}, for why this structure cannot appear. In section \ref{section5}, 
for completeness, we perform the analysis of correlation functions involving 
combinations of scalar superfields and conserved higher-spin supercurrents. Finally, in section \ref{section6} we comment on the general results in the context of superconformal field theories. 
The appendices are devoted to mathematical conventions and various useful identities.

\section{Superconformal symmetry in three-dimensions}\label{section2}

In this section we will review the pertinent aspects of the group-theoretic formalism used to compute three-point correlation functions of primary superfields in 3D $\cN=1$ superconformal field theories. For a more detailed review of the formalism the reader may consult \cite{Park:1999cw,Buchbinder:2015qsa}. 

\subsection{Superconformal transformations and primary superfields}


Let us begin by reviewing infinitesimal superconformal transformations and the transformation laws of primary superfields. This section closely follows the notation of \cite{Kuzenko:2006mv,Kuzenko:2010rp,Kuzenko:2010bd}. Now consider 3D, $\cN=1$ Minkowski superspace $\mathbb{M}^{3 | 2}$, parameterised by coordinates $z^{A} = (x^{a} , \q^{\a})$, where $a = 0,1,2$, $\a = 1,2$ are Lorentz and spinor indices respectively. 
We consider infinitesimal superconformal transformations
\begin{equation}
	\d z^{A} = \x z^{A}  \hspace{3mm} \Longleftrightarrow \hspace{3mm} \d x^{a} = \x^{a}(z) + \text{i} (\g^{a})_{\a \b} \, \x^{\a}(z) \, \q^{\b} \, , 
	\hspace{8mm} \d \q^{\a} = \x^{\a}(z) , \, 
	\label{new1}	
\end{equation}
which are associated with the real first-order differential operator
\begin{equation}
	\x = \x^{A}(z) \, \partial_{A} = \x^{a}(z) \, \partial_{a} + \x^{\a}(z) D_{\a} \, . \label{Superconformal Killing vector field}
\end{equation}
This operator satisfies the master equation $[\x , D_{\a} ] \propto D_{\b}$, from which we obtain
\begin{equation}
	\x^{\a} = \frac{\text{i}}{6} D_{\b} \x^{\a \b} \, .
\end{equation}
As a consequence, the conformal Killing equation is satisfied
\begin{equation}
	\partial_{a} \x_{b} + \partial_{b} \x_{a} = \frac{2}{3} \eta_{a b} \partial_{c} \x^{c} \, .
	\label{new2}	
\end{equation}
The solutions to the master equation are called the superconformal Killing vector fields of Minkowski superspace \cite{Buchbinder:1998qv,Kuzenko:2010rp}. They span a Lie algebra isomorphic to the superconformal algebra $\mathfrak{osp}(1 | 2 ; \mathbb{R})$. 
The components of the operator $\x$ were calculated explicitly in \cite{Park:1999cw,Buchbinder:2015qsa}, and are found to be
\begin{subequations}
	\begin{align}
		\begin{split}
			\x^{\a \b} &= a^{\a \b} - \l^{\a}{}_{\g} x^{\g \b} - x^{\a \g} \l_{\g}{}^{\b} + \s x^{\a \b} + 4 \text{i} \e^{(\a} \q^{\b)} \\
			& \hspace{20mm} + x^{\a \g} x^{\b \d} b_{\g \d} + \text{i} b_{\d}^{(\a } x^{\b) \d} \q^{2} - 4 \text{i} \eta_{\g} x^{\g(\a} \q^{\b)} \, , \label{Superconformal killing vector - component 1}
		\end{split}
	\end{align}
	\vspace{-5mm}
	\begin{align}	
		\x^{\a} &= \e^{\a} - \l^{\a}{}_{\b} \q^{\b} + \frac{1}{2} \s \q^{\a} + b_{\b \g} \boldsymbol{x}^{\b \a} \q^{\g} + \eta_{\b} ( 2 \text{i} \q^{\b} \q^{\a} - \boldsymbol{x}^{\b \a} ) \, , \label{Superconformal killing vector - component 2}
	\end{align}
	\begin{equation}
		a_{\a \b} = a_{\b \a} \, , \hspace{5mm} \l_{\a \b} = \l_{\b \a} \, , \hspace{2mm} \l^{\a}{}_{\a} = 0 \, , \hspace{5mm} b_{\a \b} = b_{\b \a} \, .
	\end{equation}
\end{subequations}
The bosonic parameters $a_{\a \b}$, $\l_{\a \b}$, $\s$, $b_{\a \b}$ correspond to infinitesimal translations, 
Lorentz transformations, scale transformations and special conformal transformations respectively, while the
fermionic parameters $\e^{\a}$ and $\eta^{\a}$ correspond to $Q$-supersymmetry and $S$-su\-per\-sym\-met\-ry transformations. 
Furthermore, the identity $D_{[\a} \x_{\b]} \propto \ve_{\a \b} $ implies that 
\begin{equation}
	[\x , D_{\a} ] = - ( D_{\a} \x^{\b}) D_{\b} = \l_{\a}{}^{\b}(z) D_{\b} - \frac{1}{2} \s(z) D_{\a} \, ,
\end{equation}
\begin{equation}
	\l_{\a \b}(z) = - D_{(\a} \x_{\b)} \, , \hspace{5mm} \s(z) = D_{\a} \x^{\a} \, . 
	\label{new4}
\end{equation}
The local parameters $\l^{\a \b}(z)$, $\s(z)$ are interpreted as being associated with combined special-conformal/Lorentz and scale transformations respectively, and appear in the transformation laws for primary tensor superfields. For later use let's also introduce the $z$-dependent $S$-supersymmetry parameter
\begin{equation}
	\eta_{\a}(z) = -\frac{\text{i}}{2} D_{\a} \s(z) \,.
	\label{new5}
\end{equation}
Explicit calculations of the local parameters give \cite{Park:1999cw,Buchbinder:2015qsa}
\begin{subequations}
	\begin{align}
		\l^{\a \b}(z) &= \l^{\a \b} - x^{\g (\a} b^{\b)}_{\g} + 2 \text{i} \eta^{(\a} \q^{\b)} - \frac{\text{i}}{2} b^{\a \b} \q^{2} \, , \label{Local parameter 1} \\ 
		\s(z) &= \s + b_{\a \b} x^{\a \b} + 2 \text{i} \q^{\a} \eta_{\a} \, , \label{Local parameter 3} \\[2mm]
		\eta_{\a}(z) &= \eta_{\a} - b_{\a \b} \q^{\b} \, . \label{Local parameter 4}
	\end{align}
\end{subequations}
Now consider a tensor superfield $\F_{\cA}(z)$ transforming in an irreducible representation of the Lorentz group with respect to the index $\cA$. Such a superfield is called primary with dimension $\D$ if it possesses the following superconformal transformation properties
\begin{equation}
	\d \F_{\cA} = - \x \F_{\cA} - \D \s(z) \F_{\cA} + \l^{\a \b}(z) (M_{\a \b})_{\cA}{}^{\cB} \F_{\cB} \,,
	\label{new6}
\end{equation}
where $\x$ is the superconformal Killing vector, $\s(z)$, $\l^{\a \b}(z)$ are $z$-dependent parameters associated with $\x$, and the matrix $M_{\a \b}$ is a Lorentz generator.


\subsubsection{Conserved supercurrents}\label{subsection2.3}

In this paper we are primarily interested in the structure of three-point correlation functions involving conserved higher-spin supercurrents. In 3D, $\cN=1$ theories, a conserved higher-spin supercurrent of superspin-$s$ (integer or half-integer), is defined as a totally symmetric spin-tensor of rank $2s$, $\mathbf{J}_{\a_{1} \dots \a_{2s} }(z) = \mathbf{J}_{(\a_{1} \dots \a_{2s}) }(z) = \mathbf{J}_{\a(2s) }(z)$, satisfying a conservation equation of the form:
\begin{equation} \label{Conserved current}
	D^{\a_{1}} \mathbf{J}_{\a_{1} \a_{2} \dots \a_{2s}}(z) = 0 \, ,
\end{equation}
where $D^{\a}$ is the conventional covariant spinor derivative \eqref{Covariant spinor derivatives}. Conserved currents are primary superfields as they possesses the following infinitesimal superconformal transformation properties \cite{Buchbinder:1998qv,Park:1999cw,Buchbinder:2015qsa}:
\begin{equation}
	\delta \mathbf{J}_{\a_{1} \dots \a_{2s}}(z) = - \xi \mathbf{J}_{\a_{1} \dots \a_{2s}}(z) - \Delta_{\mathbf{J}} \, \s(z) \, \mathbf{J}_{\a_{1} \dots \a_{2s}}(z) + 2s \, \l_{( \a_{1} }{}^{\delta}(z) \, \mathbf{J}_{\a_{2} \dots \a_{2s}) \delta}(z) \, .
\end{equation}
The dimension $\Delta_{\mathbf{J}}$ is constrained by the conservation condition \eqref{Conserved current} to $\D_{\mathbf{J}} = s+1$. Higher-spin supercurrents possess the following component structure:
\begin{equation}
\mathbf{J}_{\a(2s)}(z) = J^{(0)}_{\a(2s)}(x) + J^{(1)}_{\a(2s+1)}(x) \, \q^{\a_{2s+1}} + \tilde{J}^{(1)}_{(\a_{1} ... \a_{2s-1}}(x) \, \q^{}_{\a_{2s})}  + J^{(2)}_{\a(2s)}(x) \, \q^{2} \, .
\end{equation}
After imposing \eqref{Conserved current}, a short calculation gives $\tilde{J}^{(1)} = 0$, while $J^{(2)}$ is a function of $J^{(0)}_{\a(2s)}$.
On the other hand, the components $J^{(0)}$, $J^{(1)}$ satisfy the following conservation equations:
\begin{equation}
	\pa^{\a_{1} \a_{2}} J^{(0)}_{\a_{1} \a_{2} \a(2s-2)}(x) = 0 \, , \hspace{10mm} \pa^{\a_{1} \a_{2}} J^{(1)}_{\a_{1} \a_{2} \a(2s -1)}(x) = 0 \, .
\end{equation}
Hence, at the component level, a higher-spin supercurrent of superspin-$s$ contains conserved conformal currents of spin-$s$ and spin-$(s+\tfrac{1}{2})$ respectively.

\subsection{Two-point building blocks}


Given two superspace points $z_{1}$ and $z_{2}$, we define the two-point functions
\begin{equation}
	\boldsymbol{x}_{12}^{\alpha \beta} = (x_{1} - x_{2})^{\alpha \beta} + 2 \text{i} \theta^{(\alpha}_{1} \theta^{\beta)}_{2} - \text{i} \theta^{\a}_{12} \theta^{\b}_{12} \, ,  \hspace{10mm} \theta^{\alpha}_{12} = \theta_{1}^{\alpha} - \theta_{2}^{\alpha} \, , \label{Two-point building blocks 1}
\end{equation}
which transform under the superconformal group as follows
\begin{subequations}
	\begin{align}
		\tilde{\d} \boldsymbol{x}_{12}^{\a \b} &= - \bigg( \l^{\a}{}_{\g}(z_{1}) - \frac{1}{2} \d^{\a}{}_{\g} \, \s(z_{1}) \bigg) \boldsymbol{x}_{12}^{\g \b} - \boldsymbol{x}_{12}^{\a \g} \bigg( \l_{\g}{}^{\b}(z_{2}) - \frac{1}{2} \d_{\g}{}^{\b} \s(z_{2}) \bigg) \, , \label{Two-point building blocks 1 - transformation law 1} \\[2mm]
		\tilde{\d} \q_{12 }^{\a} &= - \bigg( \l^{\a}{}_{\b}(z_{1}) - \frac{1}{2} \d^{\a}{}_{\b} \, \s(z_{1}) \bigg) \q_{12}^{\b} - \boldsymbol{x}_{12}^{\a \b} \, \eta_{\b}(z_{2}) \,. \label{Two-point building blocks 1 - transformation law 2}
	\end{align}
\end{subequations}
Here the total variation $\tilde{\d}$ is defined by its action on an $n$-point function $\F(z_{1},...,z_{n})$ as
\begin{equation}
	\tilde{\d} \F(z_{1},...,z_{n}) = \sum_{i=1}^{n} \x_{z_{i}} \F(z_{1},...,z_{n}) \, . \label{Total variation} 
\end{equation}
Only \eqref{Two-point building blocks 1 - transformation law 1} transforms covariantly under superconformal transformations, as \eqref{Two-point building blocks 1 - transformation law 2} 
contains an inhomogeneous piece in its transformation law. Therefore it will not appear as a building block in two- or three-point correlation functions. 
Due to the useful property, $\boldsymbol{x}_{21}^{\a \b} = - \boldsymbol{x}_{12}^{\b \a}$, the two-point function \eqref{Two-point building blocks 1} can be split into symmetric and antisymmetric parts as follows:
\begin{equation}
	\boldsymbol{x}_{12}^{\a \b} = x_{12}^{\a \b} + \frac{\text{i}}{2} \ve^{\alpha \beta} \theta^{2}_{12} \, , \hspace{10mm} \q_{12}^{2} = \q_{12}^{\a} \q^{}_{12 \, \a} \, . \label{Two-point building blocks 1 - properties 1}
\end{equation}
The symmetric component
\begin{equation}
	x_{12}^{\a \b} = (x_{1} - x_{2})^{\alpha \beta} + 2 \text{i} \theta^{(\alpha}_{1} \theta^{\beta)}_{2} \, , \label{Two-point building blocks 1 - properties 2}
\end{equation}
is recognised as the bosonic part of the standard two-point superspace interval. The two-point functions possess the property:
\begin{align}  \label{Two-point building blocks - properties 1} 
\boldsymbol{x}_{12}^{\a \s} \boldsymbol{x}^{}_{21 \, \s \b} = \boldsymbol{x}_{12}^{2} \d_{\b}^{\a} \, , \hspace{5mm} \boldsymbol{x}_{12}^{2} = - \frac{1}{2} \boldsymbol{x}_{12}^{\a \b}  \boldsymbol{x}^{}_{12  \, \a \b} \, .
\end{align}
Hence, we find
\begin{equation} \label{Two-point building blocks 4}
	(\boldsymbol{x}_{12}^{-1})^{\a \b} = - \frac{\boldsymbol{x}_{12}^{ \b \a}}{\boldsymbol{x}_{12}^{2}} \, .
\end{equation}
It is now useful to introduce the normalised two-point functions, denoted by $\hat{\boldsymbol{x}}_{12}$,
\begin{align} \label{Two-point building blocks 3}
	\hat{\boldsymbol{x}}_{12 \, \a \b} = \frac{\boldsymbol{x}_{12 \, \a \b}}{( \boldsymbol{x}_{12}^{2})^{1/2}} \, , \hspace{10mm} \hat{\boldsymbol{x}}_{12}^{\a \s} \hat{\boldsymbol{x}}^{}_{21 \, \s \b} = \d_{\b}^{\a} \, . 
\end{align}
Under superconformal transformations, $\boldsymbol{x}_{12}^{2}$ transforms with local scale parameters, while \eqref{Two-point building blocks 3} transforms with local Lorentz parameters
\begin{subequations}
	\begin{align}
		\tilde{\d} \boldsymbol{x}_{12}^{2} &= ( \s(z_{1}) + \s(z_{2}) ) \, \boldsymbol{x}_{12}^{2} \, , \label{Two-point building blocks 2 - transformation law 1} \\
		\tilde{\d} \hat{\boldsymbol{x}}_{12}^{\a \b} &= - \l^{\a}{}_{\g}(z_{1}) \, \hat{\boldsymbol{x}}_{12}^{\g \b} - \hat{\boldsymbol{x}}_{12}^{\a \g} \, \l_{\g}{}^{\b}(z_{2}) \, . \label{Two-point building blocks 3 - transformation law 1}
	\end{align}
\end{subequations}
There are also the following differential identities for the action of covariant spinor derivatives on the two-point functions:
\begin{equation}
	D_{(1) \g} \boldsymbol{x}_{12}^{\a \b} = - 2 \text{i} \q^{\b}_{12} \d_{\g}^{\a} \, , \hspace{10mm} D_{(1) \a} \boldsymbol{x}_{12}^{\a \b} = - 4 \text{i} \q^{\b}_{12} \, , \label{Two-point building blocks 1 - differential identities}
\end{equation}
where $D_{(i) \a}$ acts on the superspace point $z_{i}$. From here we can now construct an operator analogous to the conformal inversion tensor acting on the space of symmetric traceless spin-tensors of arbitrary rank. Given a two-point function $\boldsymbol{x}$, we define the operator
\begin{equation} \label{Higher-spin inversion operators a}
	\cI_{\a(k) \b(k)}(\boldsymbol{x}) = \hat{\boldsymbol{x}}_{(\a_{1} (\b_{1}} \dots \hat{\boldsymbol{x}}_{ \a_{k}) \b_{k})}  \, ,
\end{equation}
along with its inverse
\begin{equation} \label{Higher-spin inversion operators b}
	\cI^{\a(k) \b(k)}(\boldsymbol{x}) = \hat{\boldsymbol{x}}^{(\a_{1} (\b_{1}} \dots \hat{\boldsymbol{x}}^{ \a_{k}) \b_{k})} \, .
\end{equation}
The spinor indices may be raised and lowered using the standard conventions as follows:
\begin{align}
	\cI_{\a(k)}{}^{\b(k)}(\boldsymbol{x}) &= \ve^{\b_{1} \g_{1}} \dots \ve^{\b_{k} \g_{k}} \, \cI_{\a(k) \g(k)}(\boldsymbol{x}) \, .
\end{align}
Now due to the property
\begin{equation}
	\cI_{\a(k) \b(k)}(-\boldsymbol{x}) = (-1)^{k} \cI_{\a(k) \b(k)}(\boldsymbol{x}) \, ,
\end{equation}
the following identity holds for products of inversion tensors:
\begin{align} \label{Higher-spin inversion operators - properties}
	\cI_{\a(k) \s(k)}(\boldsymbol{x}_{12}) \, \cI^{\s(k) \b(k)}(\boldsymbol{x}_{21}) &= \d_{(\a_{1}}^{(\b_{1}} \dots \d_{\a_{k})}^{\b_{k})} \, .
\end{align}
The objects \eqref{Higher-spin inversion operators a}, \eqref{Higher-spin inversion operators b} prove to be essential in the construction of correlation functions of primary operators with arbitrary spin. Indeed, the vector representation of the inversion tensor may be recovered in terms of the spinor two-point functions as follows:
\begin{equation}
	I_{m n}(x) = - \frac{1}{2} \, \text{Tr}( \g_{m} \, \hat{\boldsymbol{x}} \, \g_{n} \, \hat{\boldsymbol{x}} )|_{\theta = 0} \, .
\end{equation}
%


\subsection{Three-point building blocks}


Essential to the analysis of three-point correlation functions are three-point covariants/building blocks. Indeed, given three superspace points, $z_{1}, z_{2}, z_{3}$, one can define the objects, $\cZ_{k} = ( \boldsymbol{X}_{ij} , \Q_{ij} )$ as follows:
\begin{subequations} \label{Three-point building blocks 1}
	\begin{align}
		\boldsymbol{X}_{ij \, \a \b} &= -(\boldsymbol{x}_{ik}^{-1})_{\a \g}  \boldsymbol{x}_{ij}^{\g \d} (\boldsymbol{x}_{kj}^{-1})_{\d \b} \, , \hspace{5mm} \Q_{ij \, \a} = (\boldsymbol{x}_{ik}^{-1})_{\a \b} \q_{ki}^{\b} - (\boldsymbol{x}_{jk}^{-1})_{\a \b} \q_{kj}^{\b} \, , \label{Three-point building blocks} 
	\end{align}
\end{subequations}
where the labels $(i,j,k)$ are a cyclic permutation of $(1,2,3)$. These objects possess the important property $\boldsymbol{X}_{ij \, \a \b} = - \boldsymbol{X}_{ji \, \b \a}$. As a consequence, the three-point building blocks~\eqref{Three-point building blocks 1} possess many properties similar to those of the two-point building blocks
\begin{align} 
	\boldsymbol{X}_{ij}^{\a \s} \boldsymbol{X}^{}_{ji \, \s \b} = \boldsymbol{X}_{ij}^{2} \d_{\b}^{\a} \, , \hspace{5mm} \boldsymbol{X}_{ij}^{2} = - \frac{1}{2} \boldsymbol{X}_{ij}^{\a \b}  \boldsymbol{X}^{}_{ij \, \a \b} \, .
\end{align}
Hence, we find
\begin{equation}
	(\boldsymbol{X}_{ij}^{-1})^{\a \b} = - \frac{\boldsymbol{X}_{ij}^{ \b \a}}{\boldsymbol{X}_{ij}^{2}} \, .
\end{equation}
It is also useful to note that one may decompose $\boldsymbol{X}_{ij}$ into symmetric and anti-symmetric parts similar to \eqref{Two-point building blocks 1 - properties 1} as follows:
\begin{equation}
	\boldsymbol{X}_{ij  \, \a \b} = X_{ij  \, \a \b} - \frac{\text{i}}{2} \ve_{\a \b} \Q_{ij}^{2} \, , \hspace{10mm} X_{ij  \, \a \b} = X_{ij  \, \b \a} \, , \label{Three-point building blocks 1a - properties 3}
\end{equation}
where the symmetric spin-tensor, $X_{ij  \, \a \b}$, can be equivalently represented by the three-vector $X_{ij \, m} = - \frac{1}{2} (\g_{m})^{\a \b} X_{ij  \, \a \b}$. Since the building blocks possess the same properties up to cyclic permutations of the points, we will only examine the properties of $\boldsymbol{X}_{12}$ and $\Q_{12}$, as these objects appear most frequently in our analysis of correlation functions. One can compute
\begin{equation}
	\boldsymbol{X}_{12}^{2} = - \frac{1}{2} \boldsymbol{X}_{12}^{\a \b}  \boldsymbol{X}_{12 \, \a \b}^{} = \frac{\boldsymbol{x}_{12}^{2}}{\boldsymbol{x}_{13}^{2} \boldsymbol{x}_{23}^{2}} \, , \hspace{10mm}  \Q_{12}^{2} = \Q^{\a}_{12} \Q^{}_{12 \, \a} \, . \label{Three-point building blocks 2}
\end{equation}
The building block $\boldsymbol{X}_{12}$ also possesses the following superconformal transformation properties:
\begin{subequations}
	\begin{align}
		\tilde{\d} \boldsymbol{X}_{12 \, \a \b} &= \l_{\a}{}^{\g}(z_{3}) \boldsymbol{X}_{12 \, \g \b} + \boldsymbol{X}_{12 \, \a \g} \, \l^{\g}{}_{\b}(z_{3}) - \s(z_{3}) \boldsymbol{X}_{12 \, \a \b} \, , \label{Three-point building blocks 1a - transformation law 1} \\[2mm]
		\tilde{\d} \Q_{12 \, \a} &= \Big( \l_{\a}{}^{\b}(z_{3}) - \frac{1}{2} \, \d_{\a}{}^{\b} \s(z_{3}) \Big) \Q_{12 \, \b}  \, , \label{Three-point building blocks 1a - transformation law 2}
	\end{align}
\end{subequations}
and, therefore
\begin{equation}
	\tilde{\d} \boldsymbol{X}_{12}^{2} = - 2 \s(z_{3}) \boldsymbol{X}_{12}^{2} \, , \hspace{10mm} \tilde{\d} \Q_{12}^{2} = - \s(z_{3}) \, \Q_{12}^{2} \, , \label{Three-point building blocks 2 - transformation law 1}
\end{equation}
i.e. $(\boldsymbol{X}_{12}$, $\Q_{12})$ is superconformally covariant at $z_{3}$. As a consequence, one can identify the three-point superconformal invariant
\begin{equation}
	\boldsymbol{J} = \frac{\Q_{12}^{2}}{\sqrt{\boldsymbol{X}_{12}^{2}}} \hspace{5mm} \Longrightarrow \hspace{5mm} \tilde{\d} \boldsymbol{J} = 0 \, ,
\end{equation}
which proves to be invariant under permutations of the superspace points, i.e.
\begin{equation}
	\boldsymbol{J} = \frac{\Q_{12}^{2}}{\sqrt{\boldsymbol{X}_{12}^{2}}} = \frac{\Q_{31}^{2}}{\sqrt{\boldsymbol{X}_{31}^{2}}} = \frac{\Q_{23}^{2}}{\sqrt{\boldsymbol{X}_{23}^{2}}}  \, . \label{Superconformal invariants}
\end{equation}
Analogous to the two-point functions, it is also useful to introduce the normalised three-point building blocks, denoted by $\hat{\boldsymbol{X}}_{ij}$, $\hat{\Q}_{ij}$, 
\begin{align} \label{Normalised three-point building blocks}
	\hat{\boldsymbol{X}}_{ij \, \a \b} = \frac{\boldsymbol{X}_{ij \, \a \b}}{( \boldsymbol{X}_{ij}^{2})^{1/2}} \, , \hspace{10mm} \hat{\Q}_{ij}^{\a} = \frac{ \Q_{ij}^{\a} }{(\boldsymbol{X}_{ij}^{2})^{1/4}} \, ,
\end{align}
such that
\begin{align}
	\hat{\boldsymbol{X}}_{ij}^{\a \s} \hat{\boldsymbol{X}}^{}_{ji \, \s \b} = \d_{\b}^{\a} \, , \hspace{10mm} \boldsymbol{J} = \hat{\Q}_{ij}^{2} \, .
\end{align}
Compared with the standard three-point building blocks, \eqref{Three-point building blocks 1}, the objects \eqref{Normalised three-point building blocks} transform only with local Lorentz parameters. Now given an arbitrary three-point building block, $\boldsymbol{X}$, let us construct the following higher-spin inversion operator:
\begin{equation}
	\cI_{\a(k) \b(k)}(\boldsymbol{X}) = \hat{\boldsymbol{X}}_{ (\a_{1} (\b_{1}} \dots \hat{\boldsymbol{X}}_{\a_{k}) \b_{k})}  \, , \label{Inversion tensor identities - three point functions a}
\end{equation}
along with its inverse
\begin{equation}
	\cI^{\a(k) \b(k)}(\boldsymbol{X}) = \hat{\boldsymbol{X}}^{(\a_{1} (\b_{1}} \dots \hat{\boldsymbol{X}}^{ \a_{k}) \b_{k})} \, . \label{Inversion tensor identities - three point functions b}
\end{equation}
These operators possess properties similar to the two-point higher-spin inversion operators \eqref{Higher-spin inversion operators a}, \eqref{Higher-spin inversion operators b}, and are essential to the analysis of three-point correlation functions involving higher-spin primary superfields. In particular, one can prove the following useful identities involving $\boldsymbol{X}_{ij}$ and $\Q_{ij}$ at different superspace points:
\begin{subequations}
	\begin{align}
		\cI_{\a}{}^{\s}(\boldsymbol{x}_{13}) \, \cI_{\b}{}^{\g}(\boldsymbol{x}_{13}) \, \cI_{\s \g}(\boldsymbol{X}_{12}) &= \cI_{\a \b}(\boldsymbol{X}_{23})  \, , \label{Three-point building blocks 1a - properties 1}\\[2mm]
		\cI_{\a}{}^{\g}(\boldsymbol{x}_{13}) \, \hat{\Q}_{12 \, \g} &= \hat{\Q}^{I}_{23 \, \a} \, , \label{Three-point building blocks 1a - properties 2}
	\end{align}
\end{subequations}
where we have defined
\begin{equation}
\Q^{I}_{ij \a} = \cI_{\a \b}(-\boldsymbol{X}_{ij}) \, \Q^{\b}_{ij}\,.
\label{zh3}
\end{equation}
Note that $\boldsymbol{X}^{I}$ is defined in the same way, using (2.42) we have $\boldsymbol{X}^{I}_{\a \b} = \cI_{(\a \a') (\b \b')}(-\boldsymbol{X}) \,  \boldsymbol{X}^{\a' \b'} = -\boldsymbol{X}_{\a \b}$, as expected.
Using the inversion operators above, the identity \eqref{Three-point building blocks 1a - properties 1} (and cyclic permutations) admits the following generalisation to higher-spins
\begin{equation}
	\cI_{\a(k)}{}^{\s(k)}(\boldsymbol{x}_{13}) \, \cI_{\b(k)}{}^{\g(k)}(\boldsymbol{x}_{13}) \, \cI_{\s(k) \g(k)}(\boldsymbol{X}_{12}) = \cI_{\a(k) \b(k)}(\boldsymbol{X}_{23})  \, . \label{Inversion tensor identities - higher spin case}
\end{equation}
Due to the transformation properties \eqref{Three-point building blocks 1a - transformation law 1}, \eqref{Three-point building blocks 1a - transformation law 2} it is often useful to make the identifications $(\boldsymbol{X}_{1}, \Q_{1}) := (\boldsymbol{X}_{23}, \Q_{23})$, $(\boldsymbol{X}_{2}, \Q_{2}) := (\boldsymbol{X}_{31}, \Q_{31})$, $(\boldsymbol{X}_{3}, \Q_{3}) := (\boldsymbol{X}_{12}, \Q_{12})$, in which case we have e.g. $\boldsymbol{X}_{21} = - \boldsymbol{X}_{3}^{\text{T}}$; we will switch between these notations when convenient. Let us now introduce the following analogues of the covariant spinor derivative and supercharge operators involving the three-point objects:
\begin{equation}
	\cD_{(i) \a} = \frac{\partial}{\partial \Q^{\a}_{i}} + \text{i} (\g^{m})_{\a \b} \Q^{\b}_{i} \frac{\partial}{\partial X^{m}_{i}} \, , \hspace{5mm} \cQ_{(i) \a} = \text{i} \frac{\partial}{\partial \Q^{\a}_{i}} + (\g^{m})_{\a \b} \Q^{\b}_{i} \frac{\partial}{\partial X^{m}_{i}} \, , \label{Supercharge and spinor derivative analogues}
\end{equation}
which obey the standard commutation relations
\begin{equation}
	\big\{ \cD_{(i) \a} , \cD_{(i) \b} \big\} = \big\{ \cQ_{(i) \a} , \cQ_{(i) \b} \big\} = 2 \text{i} \, (\g^{m})_{\a \b} \frac{\partial}{\partial X^{m}_{i}} \, .
\end{equation}
Some useful identities involving~\eqref{Supercharge and spinor derivative analogues} are, e.g.
\begin{equation}
	\cD_{(3) \g} \boldsymbol{X}_{3 \, \a \b} = - 2 \text{i} \ve_{\g \b} \Q_{3 \, \a} \, , \hspace{5mm} \cQ_{(3) \g} \boldsymbol{X}_{3 \, \a \b} = - 2 \ve_{\g \a} \Q_{3 \, \b} \, . \label{Three-point building blocks 1a - differential identities 1}
\end{equation}
We must also account for the fact that correlation functions of primary superfields obey differential constraints as a result of superfield conservation equations. Using \eqref{Two-point building blocks 1 - differential identities} we obtain the following identities
\begin{subequations}
	\begin{align}
		D_{(1) \g} \boldsymbol{X}_{3 \, \a \b} &= 2 \text{i} (\boldsymbol{x}^{-1}_{13})_{\a \g} \Q_{3 \, \b} \, , \hspace{5mm} D_{(1) \a} \Q_{3 \, \b} = - (\boldsymbol{x}_{13}^{-1})_{\b \a} \, , \label{Three-point building blocks 1c - differential identities 1}\\[2mm]
		D_{(2) \g} \boldsymbol{X}_{3 \, \a \b} &= 2 \text{i} (\boldsymbol{x}^{-1}_{23})_{\b \g} \Q_{3 \, \b} \, , \hspace{5mm} D_{(2) \a} \Q_{3 \, \b} = (\boldsymbol{x}_{23}^{-1})_{\b \a} \, . \label{Three-point building blocks 1c - differential identities 2}
	\end{align}
\end{subequations}
Now given a function $f(\boldsymbol{X}_{3} , \Q_{3})$, there are the following differential identities which arise as a consequence of \eqref{Three-point building blocks 1a - differential identities 1}, \eqref{Three-point building blocks 1c - differential identities 1} and \eqref{Three-point building blocks 1c - differential identities 2}:
\begin{subequations}
	\begin{align}
		D_{(1) \g} f(\boldsymbol{X}_{3} , \Q_{3}) &= (\boldsymbol{x}_{13}^{-1})_{\a \g} \cD_{(3)}^{\a} f(\boldsymbol{X}_{3} , \Q_{3}) \, ,  \label{Three-point building blocks 1c - differential identities 3} \\[2mm]
		D_{(2) \g} f(\boldsymbol{X}_{3} , \Q_{3}) &= \text{i} (\boldsymbol{x}_{23}^{-1})_{\a \g} \cQ_{(3)}^{\a} f(\boldsymbol{X}_{3} , \Q_{3}) \, .  \label{Three-point building blocks 1c - differential identities 4}
	\end{align}
\end{subequations}
These will prove to be essential for imposing differential constraints on three-point correlation functions of primary superfields.

\section{General formalism for correlation functions of primary superfields}\label{section3}

In this section we develop a formalism to construct correlation functions of primary superfields in 3D superconformal field theories. We utilise a hybrid method which combines auxiliary spinors 
with the approach of~\cite{Park:1999cw, Buchbinder:2015qsa}.

\subsection{Two-point functions}\label{subsection3.1}

Let $\F_{\cA}$ be a primary superfield with dimension $\D$, where $\cA$ denotes a collection of Lorentz spinor indices. The two-point correlation function of $\F_{\cA}$ is fixed by superconformal symmetry to the form
\begin{equation} \label{Two-point correlation function}
	\langle \F_{\cA}(z_{1}) \, \F^{\cB}(z_{2}) \rangle = c \, \frac{\cI_{\cA}{}^{\cB}(\boldsymbol{x}_{12})}{(\boldsymbol{x}_{12}^{2})^{\D}} \, , 
\end{equation} 
where $\cI$ is an appropriate representation of the inversion tensor and $c$ is a constant real parameter. The denominator of the two-point function is determined by the conformal dimension of $\F_{\cA}$, which guarantees that the correlation function transforms with the appropriate weight under scale transformations.

\subsection{Three-point functions}\label{subsection3.2}

In this subsection we will review the various properties of three-point correlation functions in 3D $\cN=1$ superconformal field theory. First we present the superfield ansatz 
introduced by Park in \cite{Park:1999cw}. We then develop a new index free formalism utilising auxiliary spinors to simplify the overall form of three-point function, with the ultimate aim of constructing a generating function for arbitrary spins.

\subsubsection{Superfield ansatz}\label{subsubsection3.2.1}
Concerning three-point correlation functions, let $\F$, $\J$, $\P$ be primary superfields with scale dimensions $\D_{1}$, $\D_{2}$ and $\D_{3}$ respectively. The three-point function may be 
constructed using the general ansatz
\begin{align}
	\langle \F_{\cA_{1}}(z_{1}) \, \J_{\cA_{2}}(z_{2}) \, \P_{\cA_{3}}(z_{3}) \rangle = \frac{ \cI^{(1)}{}_{\cA_{1}}{}^{\cA'_{1}}(\boldsymbol{x}_{13}) \,  \cI^{(2)}{}_{\cA_{2}}{}^{\cA'_{2}}(\boldsymbol{x}_{23}) }{(\boldsymbol{x}_{13}^{2})^{\D_{1}} (\boldsymbol{x}_{23}^{2})^{\D_{2}}}
	\; \cH_{\cA'_{1} \cA'_{2} \cA_{3}}(\boldsymbol{X}_{12}, \Q_{12}) \, , \label{Three-point function - general ansatz}
\end{align} 
where the tensor $\cH_{\cA_{1} \cA_{2} \cA_{3}}$ encodes all information about the correlation function, and is related to the leading singular OPE coefficient \cite{Osborn:1993cr}. It is highly constrained by superconformal symmetry as follows:
\begin{enumerate}
	\item[\textbf{(i)}] Under scale transformations of $\mathbb{M}^{3|2}$, $z = ( x, \q ) \mapsto z' = ( \l^{-2} x, \l^{-1} \q )$, hence, the three-point covariants transform as $( \boldsymbol{X}, \Q) \mapsto ( \boldsymbol{X}', \Q') = ( \l^{2} \boldsymbol{X}, \l \Q )$. As a consequence, the correlation function transforms as 
	\begin{equation}
		\langle \F_{\cA_{1}}(z_{1}') \, \J_{\cA_{2}}(z_{2}') \, \P_{\cA_{3}}(z_{3}') \rangle = (\l^{2})^{\D_{1} + \D_{2} + \D_{3}} \langle \F_{\cA_{1}}(z_{1}) \, \J_{\cA_{2}}(z_{2}) \,  \P_{\cA_{3}}(z_{3}) \rangle \, ,
	\end{equation}
	which implies that $\cH$ obeys the scaling property
	\begin{equation}
		\cH_{\cA_{1} \cA_{2} \cA_{3}}( \l^{2} \boldsymbol{X}, \l \Q) = (\l^{2})^{\D_{3} - \D_{2} - \D_{1}} \, \cH_{\cA_{1} \cA_{2} \cA_{3}}(\boldsymbol{X}, \Q) \, , \hspace{5mm} \forall \l \in \mathbb{R} \, \backslash \, \{ 0 \} \, .
	\end{equation}
	This guarantees that the correlation function transforms correctly under scale transformations.
	
	\item[\textbf{(ii)}] If any of the fields $\F$, $\J$, $\P$ obey differential equations, such as conservation laws in the case of conserved currents, then the tensor $\cH$ is also constrained by differential equations which may be derived with the aid of identities \eqref{Three-point building blocks 1c - differential identities 3}, \eqref{Three-point building blocks 1c - differential identities 4}.
	
	\item[\textbf{(iii)}] If any (or all) of the operators $\F$, $\J$, $\P$ coincide, the correlation function possesses symmetries under permutations of spacetime points, e.g.
	\begin{equation}
		\langle \F_{\cA_{1}}(z_{1}) \, \F_{\cA_{2}}(z_{2}) \, \P_{\cA_{3}}(z_{3}) \rangle = (-1)^{\e(\F)} \langle \F_{\cA_{2}}(z_{2}) \, \F_{\cA_{1}}(z_{1}) \, \P_{\cA_{3}}(z_{3}) \rangle \, ,
	\end{equation}
	where $\e(\F)$ is the Grassmann parity of $\F$. As a consequence, the tensor $\cH$ obeys constraints which will be referred to as ``point-switch identities".
	
\end{enumerate}
The constraints above fix the functional form of $\cH$ (and therefore the correlation function) up to finitely many independent parameters. Hence, using the general formula \eqref{H ansatz}, the problem of computing three-point correlation functions is reduced to deriving the general structure of the tensor $\cH$ subject to the above constraints.

\subsubsection{A note on conserved three-point functions}\label{subsubsection3.2.2}

An important aspect of this construction is that depending on the way in which one constructs the general ansatz \eqref{H ansatz}, it can be difficult to impose conservation equations on one of the three fields due to a lack of useful identities such as \eqref{Three-point building blocks 1c - differential identities 1}, \eqref{Three-point building blocks 1c - differential identities 2}. For this reason it is useful to switch between the various representations of the three-point function. To illustrate this process more clearly, consider the following example; suppose we have obtained a solution for the correlation function $\langle \F_{\cA_{1}}(z_{1}) \, \J_{\cA_{2}}(z_{2}) \, \P_{\cA_{3}}(z_{3}) \rangle$, with the ansatz
\begin{equation} \label{H ansatz}
	\langle \F_{\cA_{1}}(z_{1}) \, \J_{\cA_{2}}(z_{2}) \, \P_{\cA_{3}}(z_{3}) \rangle = \frac{ \cI^{(1)}{}_{\cA_{1}}{}^{\cA'_{1}}(\boldsymbol{x}_{13}) \,  \cI^{(2)}{}_{\cA_{2}}{}^{\cA'_{2}}(\boldsymbol{x}_{23}) }{(\boldsymbol{x}_{13}^{2})^{\D_{1}} (\boldsymbol{x}_{23}^{2})^{\D_{2}}}
	\; \cH_{\cA'_{1} \cA'_{2} \cA_{3}}(\boldsymbol{X}_{12}, \Q_{12}) \, . 
\end{equation} 
All information about this correlation function is encoded in the tensor $\cH$, and one can impose conservation on $z_{1}$ and $z_{2}$ using the identities \eqref{Three-point building blocks 1c - differential identities 1}, \eqref{Three-point building blocks 1c - differential identities 2}, \eqref{Three-point building blocks 1c - differential identities 3}, \eqref{Three-point building blocks 1c - differential identities 4}. However, this particular formulation of the three-point function prevents us from imposing conservation on $z_{3}$ in a straightforward way. Let us now reformulate the ansatz with $\P$ at the front as follows:
\begin{equation} \label{Htilde ansatz}
	\langle \P_{\cA_{3}}(z_{3}) \, \J_{\cA_{2}}(z_{2}) \, \F_{\cA_{1}}(z_{1}) \rangle = \frac{ \cI^{(3)}{}_{\cA_{3}}{}^{\cA'_{3}}(\boldsymbol{x}_{31}) \,  \cI^{(2)}{}_{\cA_{2}}{}^{\cA'_{2}}(\boldsymbol{x}_{21}) }{(\boldsymbol{x}_{31}^{2})^{\D_{3}} (\boldsymbol{x}_{21}^{2})^{\D_{2}}}
	\; \tilde{\cH}_{\cA_{1} \cA'_{2} \cA'_{3} }(\boldsymbol{X}_{23}, \Q_{23}) \, . 
\end{equation} 
In this case, all information about this correlation function is now encoded in the tensor $\tilde{\cH}$, which has a completely different structure compared to $\cH$. Conservation on $\P$ can now be imposed by treating $z_{3}$ as the first point with the aid of identities analogous to \eqref{Three-point building blocks 1c - differential identities 3}, \eqref{Three-point building blocks 1c - differential identities 4}. We now require an equation relating the tensors $\cH$ and $\tilde{\cH}$, which correspond to different representations of the same correlation function. Equating the two ansatz above, we obtain the following:
\begin{align} \label{Htilde and H relation}
	\tilde{\cH}_{\cA_{1} \cA_{2}  \cA_{3} }(\boldsymbol{X}_{23}, \Q_{23}) &= (\boldsymbol{x}_{13}^{2})^{\D_{3} - \D_{1}} \bigg(\frac{\boldsymbol{x}_{21}^{2}}{\boldsymbol{x}_{23}^{2}} \bigg)^{\hspace{-1mm} \D_{2}} \, \cI^{(1)}{}_{\cA_{1}}{}^{\cA'_{1}}(\boldsymbol{x}_{13}) \, \cI^{(2)}{}_{\cA_{2}}{}^{\cB_{2}}(\boldsymbol{x}_{12}) \,  \cI^{(2)}{}_{\cB_{2}}{}^{\cA'_{2}}(\boldsymbol{x}_{23}) \nonumber \\[-2mm]
	& \hspace{50mm} \times \cI^{(3)}{}_{\cA_{3}}{}^{\cA'_{3}}(\boldsymbol{x}_{13}) \, \cH_{\cA'_{1} \cA'_{2} \cA'_{3}}(\boldsymbol{X}_{12}, \Q_{12}) \, ,
\end{align}
where we have ignored any signs due to Grassmann parity. Before we can simplify the above equation, we must understand how the inversion tensor acts on $\cH(\boldsymbol{X},\Q)$. Now let:
\begin{align}
	\cH_{ \cA_{1} \cA_{2} \cA_{3} }(\boldsymbol{X},\Q) &= \boldsymbol{X}^{\D_{3} - \D_{3}- \D_{1}} \hat{\cH}_{\cA_{1} \cA_{2} \cA_{3}}(\boldsymbol{X}, \Q) \, ,
\end{align}
where $\hat{\cH}_{\cA_{1} \cA_{2} \cA_{3}}(\boldsymbol{X}, \Q)$ is homogeneous degree 0 in $(\boldsymbol{X}, \Q)$, i.e.
\begin{align}
	\hat{\cH}_{\cA_{1} \cA_{2} \cA_{3}}(\l^{2} \boldsymbol{X}, \l \Q) &= \hat{\cH}_{\cA_{1} \cA_{2} \cA_{3}}(\boldsymbol{X}, \Q) \, .
\end{align}
The tensor $\hat{\cH}_{\cA_{1} \cA_{2} \cA_{3}}(\boldsymbol{X}, \Q)$ can be constructed from totally symmetric, homogeneous degree 0 combinations of $\ve$, $\boldsymbol{X}$ and $\Q$, compatible with the set of indices $\cA_{1}, \cA_{2}, \cA_{3}$, hence, we consider the following objects:
\begin{align}
	\ve_{\a \b} \, , \hspace{5mm} \hat{\boldsymbol{X}}_{\a \b} \, , \hspace{5mm} \hat{\Q}_{\a} \, , \hspace{5mm} (\hat{\boldsymbol{X}} \cdot \hat{\Q})_{\a} =  \hat{\boldsymbol{X}}_{\a \b} \hat{\Q}^{\b} \, , \hspace{5mm} \boldsymbol{J} = \hat{\Q}^{2} \, .
\end{align}
Now to simplify \eqref{Htilde and H relation}, consider
\begin{equation}
	\cI^{(1)}{}_{\cA_{1}}{}^{\cA'_{1}}(\boldsymbol{x}_{13}) \, \cI^{(2)}{}_{\cA_{2}}{}^{\cA'_{2}}(\boldsymbol{x}_{13}) \, \cI^{(3)}{}_{\cA_{3}}{}^{\cA'_{3}}(\boldsymbol{x}_{13}) \, \hat{\cH}_{\cA'_{1} \cA'_{2} \cA'_{3}}(\boldsymbol{X}_{12}, \Q_{12}) \, .
\end{equation}
Only combinations of the following fundamental products may appear in the result:
\begin{subequations}
	\begin{align}
		\cI_{\a}{}^{\a'}(\boldsymbol{x}_{13}) \, \cI_{\b}{}^{\b'}(\boldsymbol{x}_{13}) \, \ve_{\a' \b'} &= - \ve_{\a \b} \, , \\
		\cI_{\a}{}^{\a'}(\boldsymbol{x}_{13}) \, \cI_{\b}{}^{\b'}(\boldsymbol{x}_{13}) \, \hat{\boldsymbol{X}}_{12 \, \a' \b'} &= \hat{\boldsymbol{X}}_{23 \, \a \b} \, , \\
		\cI_{\a}{}^{\a'}(\boldsymbol{x}_{13}) \, \hat{\Q}_{12 \, \a'} &= \hat{\Q}^{I}_{23 \, \a} \, , \\
		\cI_{\a}{}^{\a'}(\boldsymbol{x}_{13}) \, (\hat{\boldsymbol{X}}_{12} \cdot \hat{\Q}_{12})_{\a'} &= - (\hat{\boldsymbol{X}}_{23} \cdot \hat{\Q}^{I}_{23})_{\a} \, ,
	\end{align}
\end{subequations}
where $\hat{\Q}^{I}_{ij}$ was defined in~\eqref{zh3}.
For correlation functions involving the superconformal invariant, $\boldsymbol{J}$, we must note that $\boldsymbol{J}^{I} = (\hat{\Q}^{I})^2 = - \boldsymbol{J}$. These identities are consequences of \eqref{Three-point building blocks 1a - properties 1}, \eqref{Three-point building blocks 1a - properties 2}. If we now denote the above transformations by $\cI_{13}$, it acts on $\hat{\cH}(\boldsymbol{X}_{12},\Q_{12})$ as follows:
\begin{subequations}
	\begin{align} \label{Inversion even objects}
		\hat{\boldsymbol{X}}_{12} \xrightarrow[]{\cI_{13}} \hat{\boldsymbol{X}}_{23} \, , \hspace{10mm} \hat{\Q}_{12} \xrightarrow[]{\cI_{13}} \hat{\Q}^{I}_{23} \, ,
	\end{align}
	\vspace{-10mm}
	\begin{align} \label{Inversion odd objects}
		\ve \xrightarrow[]{\cI_{13}} -\ve  \, , \hspace{10mm} \hat{\boldsymbol{X}}_{12} \cdot \hat{\Q}_{12} \xrightarrow[]{\cI_{13}} - \hat{\boldsymbol{X}}_{23} \cdot \hat{\Q}^{I}_{23} \, ,\hspace{10mm} \boldsymbol{J} \xrightarrow[]{\cI_{13}} - \boldsymbol{J}^{I} \, .
	\end{align}
\end{subequations}
Hence, due to their transformation properties under $\cI$, the objects \eqref{Inversion even objects} are classified as ``parity-even" as they are invariant under $\cI$, while the objects \eqref{Inversion odd objects} are classified as ``parity-odd", as they are pseudo-invariant under $\cI$. At this point it is convenient to partition our solution into ``even" and ``odd" sectors as follows:
\begin{equation}
		\cH_{\cA_{1} \cA_{2}  \cA_{3} }(\boldsymbol{X}, \Q) = \cH^{(+)}_{\cA_{1} \cA_{2}  \cA_{3} }(\boldsymbol{X}, \Q) + \cH^{(-)}_{\cA_{1} \cA_{2}  \cA_{3} }(\boldsymbol{X}, \Q) \, , 
\end{equation}
where $\cH^{(+)}$ contains all structures that are invariant under $\cI$, and $\cH^{(-)}$ contains all structures that are pseudo-invariant under $\cI$. With this choice of convention, as a consequence of \eqref{Three-point building blocks 1a - properties 1}, \eqref{Three-point building blocks 1a - properties 2}, the following relation holds:
\begin{align} \label{Hc and H relation}
	\hat{\cH}^{I \, (\pm)}_{\cA_{1} \cA_{2} \cA_{3}}(\boldsymbol{X}_{23}, \Q_{23}) &= \pm \, \cI^{(1)}{}_{\cA_{1}}{}^{\cA'_{1}}(\boldsymbol{x}_{13}) \, \cI^{(2)}{}_{\cA_{2}}{}^{\cA'_{2}}(\boldsymbol{x}_{13}) \nonumber \\
	& \hspace{20mm} \times \cI^{(3)}{}_{\cA_{3}}{}^{\cA'_{3}}(\boldsymbol{x}_{13}) \, \hat{\cH}^{(\pm)}_{\cA'_{1} \cA'_{2} \cA'_{3}}(\boldsymbol{X}_{12}, \Q_{12}) \, ,
\end{align}
where $\hat{\cH}^{I \, (\pm)}_{\cA_{1} \cA_{2} \cA_{3}}(\boldsymbol{X}, \Q) = \hat{\cH}^{(\pm)}_{\cA_{1} \cA_{2} \cA_{3}}(\boldsymbol{X}, \Q^{I})$. A result analogous to \eqref{Inversion even objects}, \eqref{Inversion odd objects} that follows from the properties of the inversion tensor acting on $(\boldsymbol{X}, \Q)$ is
\begin{subequations}
	\begin{align} \label{Inverison even objects - X}
		\hat{\boldsymbol{X}} \xrightarrow[]{\cI_{\boldsymbol{X}}}  - \hat{\boldsymbol{X}} \, , \hspace{10mm} \hat{\Q} \xrightarrow[]{\cI_{\boldsymbol{X}}} \hat{\Q}^{I} \, , 
	\end{align}
	\vspace{-10mm}
	\begin{align} \label{Inverison odd objects - X}
		\ve \xrightarrow[]{\cI_{\boldsymbol{X}}} -\ve  \, , \hspace{10mm} \hat{\boldsymbol{X}} \cdot \hat{\Q} \xrightarrow[]{\cI_{\boldsymbol{X}}} \hat{\boldsymbol{X}} \cdot \hat{\Q}^{I}\, ,\hspace{10mm} \boldsymbol{J} \xrightarrow[]{\cI_{\boldsymbol{X}}} - \boldsymbol{J}^{I} \, .
	\end{align}
\end{subequations}
Hence, to obtain the desired transformation properties as in \eqref{Inversion even objects}, \eqref{Inversion odd objects}, we consider $\cH(-\boldsymbol{X}, \Q)$ and obtain the formula
\begin{align} \label{H inversion}
	\cH^{I \, (\pm)}_{\cA_{1} \cA_{2} \cA_{3}}(\boldsymbol{X}, \Q) = \pm \, \cI^{(1)}{}_{\cA_{1}}{}^{\cA'_{1}}(\boldsymbol{X}) \, \cI^{(2)}{}_{\cA_{2}}{}^{\cA'_{2}}(\boldsymbol{X}) \, \cI^{(3)}{}_{\cA_{3}}{}^{\cA'_{3}}(\boldsymbol{X}) \, \cH^{(\pm)}_{\cA'_{1} \cA'_{2} \cA'_{3}}(-\boldsymbol{X}, \Q) \, ,
\end{align}
which is generally more simple to compute. After substituting \eqref{Hc and H relation} into \eqref{Htilde and H relation}, we obtain the following relation between $\cH$ and $\tilde{\cH}$:
\begin{equation} \label{Htilde and Hc relation}
	\tilde{\cH}^{(\pm)}_{\cA_{1} \cA_{2} \cA_{3} }(\boldsymbol{X},\Q) = \pm \, (\boldsymbol{X}^{2})^{\D_{1} - \D_{3}} \, \cI^{(2)}{}_{\cA_{2}}{}^{\cA'_{2}}(\boldsymbol{X}) \, \cH^{I \, (\pm)}_{\cA_{1} \cA'_{2} \cA_{3}}(\boldsymbol{X}, \Q) \, . 
\end{equation}
It is now apparent that $\cI$ acts as an intertwining operator between the various representations of the correlation function. Once $\tilde{\cH}$ is obtained we can then impose conservation on $\Pi$ as if it were located at the ``first point'', using identities analogous to \eqref{Three-point building blocks 1c - differential identities 3}, \eqref{Three-point building blocks 1c - differential identities 4}.  

If we now consider the correlation function of three conserved primary superfields $\mathbf{J}^{}_{\a(I)}$, $\mathbf{J}'_{\b(J)}$, $\mathbf{J}''_{\g(K)}$, where $I=2s_{1}$, $J=2s_{2}$, $K=2s_{3}$, then the general ansatz is
\begin{align} \label{Conserved correlator ansatz}
	\langle \, \mathbf{J}^{}_{\a(I)}(z_{1}) \, \mathbf{J}'_{\b(J)}(z_{2}) \, \mathbf{J}''_{\g(K)}(z_{3}) \rangle = \frac{ \cI_{\a(I)}{}^{\a'(I)}(\boldsymbol{x}_{13}) \,  \cI_{\b(J)}{}^{\b'(J)}(\boldsymbol{x}_{23}) }{(\boldsymbol{x}_{13}^{2})^{\D_{1}} (\boldsymbol{x}_{23}^{2})^{\D_{2}}}
	\; \cH_{\a'(I) \b'(J) \g(K)}(\boldsymbol{X}_{12}, \Q_{12}) \, ,
\end{align} 
where $\D_{i} = s_{i} + 1$. The constraints on $\cH$ are then as follows:
\begin{enumerate}
	\item[\textbf{(i)}] {\bf Homogeneity:}
	\begin{equation}
		\cH_{\a(I) \b(J) \g(K)}(\l^{2} \boldsymbol{X}, \l \Q) = (\l^{2})^{\D_{3} - \D_{2} - \D_{1}} \, \cH_{\a(I) \b(J) \g(K)}(\boldsymbol{X}, \Q) \, ,
	\end{equation}
	It is often convenient to introduce $\hat{\cH}_{\a(I) \b(J) \g(K)}(\boldsymbol{X}, \Q)$, such that
	\begin{align}
		\cH_{\a(I) \b(J) \g(K)}(\boldsymbol{X},\Q) &= \boldsymbol{X}^{\D_{3} - \D_{3}- \D_{1}} \hat{\cH}_{\a(I) \b(J) \g(K)}(\boldsymbol{X}, \Q) \, ,
	\end{align}
	where $\hat{\cH}_{\a(I) \b(J) \g(K)}(\boldsymbol{X}, \Q)$ is homogeneous degree 0 in $(\boldsymbol{X}, \Q)$, i.e.
	\begin{align}
		\hat{\cH}_{\a(I) \b(J) \g(K)}(\l^{2} \boldsymbol{X}, \l \Q) &= \hat{\cH}_{\a(I) \b(J) \g(K)}(\boldsymbol{X}, \Q) \, .
	\end{align}
	
	\item[\textbf{(ii)}] {\bf Differential constraints:} \\
	After application of the identities \eqref{Three-point building blocks 1c - differential identities 3}, \eqref{Three-point building blocks 1c - differential identities 4} we obtain the following constraints:
	\begin{subequations}
		\begin{align}
			\text{Conservation at $z_{1}$:} && \cD^{\a} \cH_{\a \a(I - 1) \b(J) \g(K)}(\boldsymbol{X}, \Q) &= 0 \, , \\
			\text{Conservation at $z_{2}$:} && \cQ^{\b} \cH_{\a(I) \b \b(J-1) \g(K)}(\boldsymbol{X}, \Q) &= 0 \, , \\
			\text{Conservation at $z_{3}$:} && \cQ^{\g} \tilde{\cH}_{\a(I) \b(J) \g \g(K-1)  }(\boldsymbol{X}, \Q) &= 0 \, ,
		\end{align}
	\end{subequations}
	where
	\begin{equation}
		\tilde{\cH}^{(\pm)}_{\a(I) \b(J) \g(K) }(\boldsymbol{X}, \Q) = (\boldsymbol{X}^{2})^{\D_{1} - \D_{3}} \, \cI_{\b(J)}{}^{\b'(J)}(\boldsymbol{X}) \, \cH^{I \, (\pm)}_{\a(I) \b'(J) \g(K)}(\boldsymbol{X}, \Q) \, . 
	\end{equation}

	\item[\textbf{(iii)}] {\bf Point-switch symmetries:} \\
	If the fields $\mathbf{J}$ and $\mathbf{J}'$ coincide, then we obtain the following point-switch identity
	\begin{equation}
		\cH_{\a(I) \b(I) \g(K)}(\boldsymbol{X}, \Q) = (-1)^{\e(\mathbf{J})} \cH_{\b(I) \a(I) \g(K)}(-\boldsymbol{X}^{\text{T}}, -\Q) \, ,
	\end{equation}
	where $\e(\mathbf{J})$ is the Grassmann parity of $\mathbf{J}$. Likewise, if the fields $\mathbf{J}$ and $\mathbf{J}''$ coincide, then we obtain the constraint
	\begin{equation}
		\tilde{\cH}_{\a(I) \b(J) \g(I) }(\boldsymbol{X}, \Q) = (-1)^{\e(\mathbf{J})} \cH_{\g(I) \b(J) \a(I)}(-\boldsymbol{X}^{\text{T}}, -\Q) \, .
	\end{equation}
\end{enumerate}
In practice, imposing these constraints on correlation functions involving higher-spin supercurrents quickly becomes unwieldy using the tensor formalism, particularly due to the sheer number of possible tensor structures for a given set of superspins. Hence, in the next subsections we will develop an index-free formalism to handle the computations efficiently, using the same approach as \cite{Buchbinder:2022mys}.

\subsubsection{Auxiliary spinor formalism}\label{subsubsection3.2.3}

Suppose we must analyse the constraints on a general spin-tensor $\cH_{\cA_{1} \cA_{2} \cA_{3}}(\boldsymbol{X}, \Q)$, where $\cA_{1} = \{ \a_{1}, ... , \a_{I} \}, \cA_{2} = \{ \b_{1}, ... , \b_{J} \}, \cA_{3} = \{ \g_{1}, ... , \g_{K} \}$ represent sets of totally symmetric spinor indices associated with the fields at points $z_{1}$, $z_{2}$ and $z_{3}$ respectively. We introduce sets of commuting auxiliary spinors for each point; $u$ at $z_{1}$, $v$ at $z_{2}$, and $w$ at $z_{3}$, where the spinors satisfy 
\begin{align}
u^2 &= \varepsilon_{\a \b} \, u^{\a} u^{\b}=0\,,  &
v^{2} &= \varepsilon_{\a \b} \, v^{\a} v^{\b}=0\,, & w^{2} &= \varepsilon_{\a \b} \, w^{\a} w^{\b}=0\,. 
\label{extra1}
\end{align}
Now if we define the objects
\begin{subequations}
	\begin{align}
		\boldsymbol{u}^{\cA_{1}} &\equiv \boldsymbol{u}^{\a(I)} = u^{\a_{1}} \dots u^{\a_{I}} \, , \\
		\boldsymbol{v}^{\cA_{2}} &\equiv \boldsymbol{v}^{\b(J)} = v^{\b_{1}} \dots v^{\b_{J}} \, , \\
		\boldsymbol{w}^{\cA_{3}} &\equiv \boldsymbol{w}^{\g(K)} = w^{\g_{1}} \dots w^{\g_{K}} \, ,
	\end{align}
\end{subequations}
then the generating polynomial for $\cH$ is constructed as follows:
\begin{equation} \label{H - generating polynomial}
	\cH(\boldsymbol{X}, \Q; u,v,w) = \,\cH_{ \cA_{1} \cA_{2} \cA_{3} }(\boldsymbol{X}, \Q) \, \boldsymbol{u}^{\cA_{1}} \boldsymbol{v}^{\cA_{2}} \boldsymbol{w}^{\cA_{3}} \, . \\
\end{equation}
%
There is a one-to-one mapping between the space of symmetric traceless spin tensors and the polynomials constructed using the above method. Indeed, the tensor $\cH$ is extracted from the polynomial by acting on it with the following partial derivative operators:
\begin{subequations}
	\begin{align}
		\frac{\pa}{\pa \boldsymbol{u}^{\cA_{1}} } &\equiv \frac{\pa}{\pa \boldsymbol{u}^{\a(I)}} = \frac{1}{I!} \frac{\pa}{\pa u^{\a_{1}} } \dots \frac{\pa}{\pa u^{\a_{I}}}  \, , \\
		\frac{\pa}{\pa \boldsymbol{v}^{\cA_{2}} } &\equiv \frac{\pa}{\pa \boldsymbol{v}^{\b(J)}} = \frac{1}{J!} \frac{\pa}{\pa v^{\b_{1}} } \dots \frac{\pa}{\pa v^{\b_{J}}} \, , \\
		\frac{\pa}{\pa \boldsymbol{w}^{\cA_{3}} } &\equiv \frac{\pa}{\pa \boldsymbol{w}^{\g(K)}} = \frac{1}{K!} \frac{\pa}{\pa w^{\g_{1}} } \dots \frac{\pa}{\pa w^{\g_{K}}} \, . 
	\end{align}
\end{subequations}
The tensor $\cH$ is then extracted from the polynomial as follows:
\begin{equation}
	\cH_{\cA_{1} \cA_{2} \cA_{3}}(\boldsymbol{X}, \Q) = \frac{\pa}{ \pa \boldsymbol{u}^{\cA_{1}} } \frac{\pa}{ \pa \boldsymbol{v}^{\cA_{2}}} \frac{\pa}{ \pa \boldsymbol{w}^{\cA_{3}} } \, \cH(\boldsymbol{X}, \Q; u, v, w) \, .
\end{equation}
Auxiliary spinors are widely used 
in the construction of correlation functions throughout the literature (see e.g.~\cite{Giombi:2011rz, Costa:2011mg, Stanev:2012nq, Zhiboedov:2012bm, Nizami:2013tpa, Elkhidir:2014woa}), 
however, usually the entire correlator is contracted with auxiliary variables and as a result one produces a polynomial 
depending on all three superspace points and the auxiliary spinors. In contrast, this approach contracts the auxiliary spinors with the tensor $\cH_{ \cA_{1} \cA_{2} \cA_{3} }(\boldsymbol{X}, \Q)$, 
which depends only on $\boldsymbol{X}$, $\Q$. As a result, it is straightforward to impose constraints on the correlation function as $\cH$ does not depend on any of the superspace points explicitly. 

The full three-point function may be translated into the auxiliary spinor formalism; recalling that $I = 2s_{1}$, $J = 2s_{2}$, $K = 2s_{3} $, first we define:
\begin{subequations}
	\begin{align}
		\mathbf{J}^{}_{s_{1}}(z_{1}; u) & = \mathbf{J}_{\a(I)}(z_{1}) \, \boldsymbol{u}^{\a(I)} \, , & \mathbf{J}'_{s_{2}}(z_{2}; v) &= \mathbf{J}_{\b(J)}(z_{2}) \, \boldsymbol{v}^{\a(J)} \, ,
	\end{align}
	\vspace{-10mm}
	\begin{align}
		\mathbf{J}''_{s_{3}}(z_{3}; w) &= \mathbf{J}_{\g(K)}(z_{3}) \, \boldsymbol{w}^{\g(K)} \, .
	\end{align}
\end{subequations}
The general ansatz for the three-point function is as follows:
\begin{align}
	\langle \, \mathbf{J}^{}_{s_{1}}(z_{1}; u) \, \mathbf{J}'_{s_{2}}(z_{2}; v) \, \mathbf{J}''_{s_{3}}(z_{3}; w) \rangle = \frac{ \cI^{(I)}(\boldsymbol{x}_{13}; u, \tilde{u}) \,  \cI^{(J)}(\boldsymbol{x}_{23}; v, \tilde{v}) }{(\boldsymbol{x}_{13}^{2})^{\D_{1}} (\boldsymbol{x}_{23}^{2})^{\D_{2}}}
	\; \cH(\boldsymbol{X}_{12}, \Q_{12}; \tilde{u},\tilde{v},w) \, ,
\end{align} 
where 
\begin{equation}
	\cI^{(s)}(\boldsymbol{x}; u,\tilde{u}) \equiv \cI^{(s)}_{\boldsymbol{x}}(u,\tilde{u}) = \boldsymbol{u}^{\a(s)} \cI_{\a(s)}{}^{\a'(s)}(\boldsymbol{x}) \, \frac{\pa}{\pa \tilde{\boldsymbol{u}}^{\a'(s)}} \, ,
\end{equation}
is the inversion operator acting on polynomials degree $s$ in $\tilde{u}$, and $\D_{i} = s_{i} + 1$. After converting the constraints summarised in the previous subsection into the auxiliary spinor formalism, we obtain:
\begin{enumerate}
	\item[\textbf{(i)}] {\bf Homogeneity:}
	\begin{equation}
		\cH(\l^{2} \boldsymbol{X}, \l \Q ; u(I), v(J), w(K)) = (\l^{2})^{\D_{3} - \D_{2} - \D_{1}} \, \cH(\boldsymbol{X}, \Q; u(I), v(J), w(K)) \, ,
	\end{equation}
	where we have used the notation $u(I)$, $v(J)$, $w(K)$ to keep track of the homogeneity of the auxiliary spinors $u$, $v$ and $w$.
	\item[\textbf{(ii)}] {\bf Differential constraints:} \\
	First, define the following three differential operators:
	\begin{align}
		D_{1} = \cD^{\a} \frac{\pa}{\pa u^{\a}} \, , && D_{2} = \cQ^{\a} \frac{\pa}{\pa v^{\a}} \, , && D_{3} = \cQ^{\a} \frac{\pa}{\pa w^{\a}} \, .
	\end{align}
	Conservation on all three points may be imposed using the following constraints:
	\begin{subequations} \label{Conservation equations}
		\begin{align}
			\text{Conservation at $z_{1}$:} && D_{1} \, \cH(\boldsymbol{X}, \Q; u(I), v(J), w(K)) &= 0 \, , \\[1mm]
			\text{Conservation at $z_{2}$:} && D_{2} \, \cH(\boldsymbol{X}, \Q; u(I), v(J), w(K)) &= 0 \, , \\[1mm]
			\text{Conservation at $z_{3}$:} && D_{3} \, \tilde{\cH}(\boldsymbol{X}, \Q; u(I), v(J), w(K)) &= 0 \, ,
		\end{align}
	\end{subequations}
	where, in the auxiliary spinor formalism, $\tilde{\cH} = \tilde{\cH}^{(+)} + \tilde{\cH}^{(-)}$ is computed as follows:
	\begin{equation}
		\tilde{\cH}^{(\pm)}(\boldsymbol{X}, \Q; u(I), v(J), w(K) ) = \pm (\boldsymbol{X}^{2})^{\D_{1} - \D_{3}} \cI^{(J)}_{\boldsymbol{X}}(v,\tilde{v}) \, \cH^{I \, (\pm)}(\boldsymbol{X}, \Q; u(I), \tilde{v}(J), w(K)) \, , 
	\end{equation}
	where $\cI^{(s)}_{\boldsymbol{X}}(v,\tilde{v}) \equiv \cI^{(s)}(\boldsymbol{X}; v,\tilde{v})$.
	\item[\textbf{(iii)}] {\bf Point switch symmetries:} \\
	If the fields $\mathbf{J}$ and $\mathbf{J}'$ coincide (hence $I = J$), then we obtain the following point-switch constraint
	\begin{equation} \label{Point switch A}
		\cH(\boldsymbol{X}, \Q; u(I), v(I), w(K)) = (-1)^{\e(\mathbf{J})} \cH(- \boldsymbol{X}^{\text{T}}, - \Q; v(I), u(I), w(K)) \, ,
	\end{equation}
	where, again, $\e(\mathbf{J})$ is the Grassmann parity of $\mathbf{J}$. Similarly, if the fields $\mathbf{J}$ and $\mathbf{J}''$ coincide (hence $I = K$) then we obtain the constraint
	\begin{equation} \label{Point switch B}
		\tilde{\cH}(\boldsymbol{X}, \Q; u(I), v(J), w(I)) = (-1)^{\e(\mathbf{J})} \cH(- \boldsymbol{X}^{\text{T}}, - \Q; w(I), v(J), u(I)) \, .
	\end{equation}
\end{enumerate}
To find an explicit solution for the polynomial \eqref{H - generating polynomial}, one must now consider all possible scalar combinations of $\boldsymbol{X}$, $\Q$, $\ve$, $u$, $v$ and $w$ with the appropriate homogeneity. Hence, let us introduce the following structures: \\[2mm]
\textbf{Bosonic:}
\begin{subequations} \label{Basis scalar structures 1}
	\begin{align}
		P_{1} &= \ve_{\a \b} v^{\a} w^{\b} \, , & P_{2} &= \ve_{\a \b} w^{\a} u^{\b} \, , & P_{3} &= \ve_{\a \b} u^{\a} v^{\b} \, , \\
		\mathbb{Q}_{1} &= \hat{\boldsymbol{X}}_{\a \b} v^{\a} w^{\b} \, , & \mathbb{Q}_{2} &= \hat{\boldsymbol{X}}_{\a \b} w^{\a} u^{\b} \, , & \mathbb{Q}_{3} &= \hat{\boldsymbol{X}}_{\a \b} u^{\a} v^{\b} \, , \\
		\mathbb{Z}_{1} &= \hat{\boldsymbol{X}}_{\a \b} u^{\a} u^{\b} \, , & \mathbb{Z}_{2} &= \hat{\boldsymbol{X}}_{\a \b} v^{\a} v^{\b} \, , & \mathbb{Z}_{3} &= \hat{\boldsymbol{X}}_{\a \b} w^{\a} w^{\b} \, ,
	\end{align}
\end{subequations}
\textbf{Fermionic:}
\begin{subequations} \label{Basis scalar structures 2}
	\begin{align}
		R_{1} &= \ve_{\a \b} u^{\a} \hat{\Q}^{\b} \, , & R_{2} &= \ve_{\a \b} v^{\a} \hat{\Q}^{\b} \, , & R_{3} &= \ve_{\a \b} w^{\a} \hat{\Q}^{\b} \, , \\
		\mathbb{S}_{1} &= \hat{\boldsymbol{X}}_{\a \b} u^{\a} \hat{\Q}^{\b} \, , & \mathbb{S}_{2} &= \hat{\boldsymbol{X}}_{\a \b} v^{\a} \hat{\Q}^{\b} \, , & \mathbb{S}_{3} &= \hat{\boldsymbol{X}}_{\a \b} w^{\a} \hat{\Q}^{\b} \, .
	\end{align}
\end{subequations}
A general solution for $\cH(\boldsymbol{X}, \Q)$ is comprised of all possible combinations of $P_{i}, \mathbb{Q}_{i}, \mathbb{Z}_{i}, R_{i}, \mathbb{S}_{i}$ and $\boldsymbol{J}$ which possess the correct homogeneity in $u$, $v$ and $w$. Comparing with \eqref{Inversion even objects}, \eqref{Inversion odd objects}, we can identify the objects $P_{i}$, $\mathbb{S}_{i}$ and $\boldsymbol{J}$ as being ``parity-odd" due to their transformation properties under inversions. 

For the subsequent analysis of conserved three-point functions, due to the property \eqref{Three-point building blocks 1a - properties 3}, and the fact that in $\cN=1$ theories $\Q^{3} = 0 \implies \boldsymbol{X}^{2} = X^{2}$, it is generally more convenient to construct the polynomial in terms of the symmetric spin-tensor, $X_{\a \b}$, rather than $\boldsymbol{X}_{\a \b}$, resulting in the polynomial $\cH(X, \Q)$. Hence, we expand $\mathbb{Q}_{i}, \mathbb{Z}_{i}, \mathbb{S}_{i}$ as follows:
\begin{align}
	\mathbb{Q}_{i} = Q_{i} - \frac{\text{i}}{2} \, P_{i} \, \boldsymbol{J} \, , && \mathbb{Z}_{i} &= Z_{i} \, ,
	&& \mathbb{S}_{i} = S_{i} \, ,
\end{align}
where we have defined
\begin{subequations}
	\begin{align}
		Q_{1} &= \hat{X}_{\a \b} v^{\a} w^{\b} \, , & Q_{2} &= \hat{X}_{\a \b} w^{\a} u^{\b} \, , & Q_{3} &= \hat{X}_{\a \b} u^{\a} v^{\b} \, , \\
		Z_{1} &= \hat{X}_{\a \b} u^{\a} u^{\b} \, , & Z_{2} &= \hat{X}_{\a \b} v^{\a} v^{\b} \, , & Z_{3} &= \hat{X}_{\a \b} w^{\a} w^{\b} \, , \\
		S_{1} &= \hat{X}_{\a \b} u^{\a} \hat{\Q}^{\b} \, , & S_{2} &= \hat{X}_{\a \b} v^{\a} \hat{\Q}^{\b} \, , & S_{3} &= \hat{X}_{\a \b} w^{\a} \hat{\Q}^{\b} \, .
	\end{align}
\end{subequations}
The polynomial $\cH(X, \Q)$ is now constructed from all possible combinations of $P_{i}$, $Q_{i}$, $Z_{i}$, $R_{i}$, $S_{i}$ 
and $\boldsymbol{J}$. Once a general solution for $\cH(X, \Q)$ is obtained, one can convert back to ``covariant form", $\cH(\boldsymbol{X}, \Q)$, by making the replacements
\begin{align}
	Q_{i} \rightarrow \mathbb{Q}_{i} + \frac{\text{i}}{2} \, P_{i} \, \boldsymbol{J} \, , && Z_{i} \rightarrow \mathbb{Z}_{i} \, ,
	&& S_{i} \rightarrow \mathbb{S}_{i} \, .
\end{align}

\subsubsection{Generating function method}\label{subsubsection3.2.4}
In general, it is a non-trivial technical problem to come up with an exhaustive list of possible solutions for $\cH(X,\Q;u,v,w)$ for a given set of superspins, however, this process can be simplified by introducing generating functions for the polynomial $\cH(X,\Q; u, v, w)$. First we introduce the function $\cF(X)$, defined as follows:
\begin{align} \label{Generating function 1}
	\cF(X) &= X^{\d} P_{1}^{k_{1}} P_{2}^{k_{2}} P_{3}^{k_{3}} Q_{1}^{l_{1}} Q_{2}^{l_{2}} Q_{3}^{l_{3}} Z_{1}^{m_{1}} Z_{2}^{m_{2}} Z_{3}^{m_{3}}
\end{align}
where, typically, $\d = \D_{3} - \D_{2} - \D_{1}$. The generating functions for Grassmann-even and Grassmann-odd correlators in $\cN=1$ theories are then defined as follows:
\begin{align} \label{Generating function 2}
	\cG(X,\Q \, | \, \G) &= \begin{cases} 
		\cF(X) \, \boldsymbol{J}^{\s} \, , & \text{Bosonic} \\
		\cF(X) \, R_{1}^{p_{1}} R_{2}^{p_{2}} R_{3}^{p_{3}} S_{1}^{q_{1}} S_{2}^{q_{2}} S_{3}^{q_{3}} \, , & \text{Fermionic}
	\end{cases}
\end{align}
Here the non-negative integers, $ \G = \{ k_{i}, l_{i}, m_{i}, p_{i}, q_{i}, \s\}$, $i=1,2,3$, are  constrained; for overall bosonic correlation functions they are solutions to the following linear system
\begin{subequations} \label{Diophantine equations 1}
	\begin{align}
		k_{2} + k_{3} + l_{2} + l_{3} + 2m_{1} &= I \, , \\
		k_{1} + k_{3} + l_{1} + l_{3} + 2m_{2} &= J \, , \\
		k_{1} + k_{2} + l_{1} + l_{2} + 2m_{3} &= K \, ,
	\end{align}
\end{subequations}
with $\s = 0,1$. Likewise, for overall fermionic correlation functions, the integers $\G$ are solutions to the following system
\begin{subequations} \label{Diophantine equations 2}
	\begin{align}
		k_{2} + k_{3} + l_{2} + l_{3} + 2m_{1} + p_{1} + q_{1} &= I \, , \\
		k_{1} + k_{3} + l_{1} + l_{3} + 2m_{2} + p_{2} + q_{2} &= J \, , \\
		k_{1} + k_{2} + l_{1} + l_{2} + 2m_{3} + p_{3} + q_{3} &= K \, , \\
		p_{1} + p_{2} + p_{3} + q_{1} + q_{2} + q_{3} &= 1 \, ,
	\end{align}
\end{subequations}
where $I = 2s_{1}$, $J = 2s_{2}$, $K = 2s_{3}$ specify the spin-structure of the correlation function. These equations are obtained by comparing the homogeneity of the auxiliary spinors $u$, $v$, $w$ in the generating functions \eqref{Generating function 2}, against the index structure of the tensor $\cH$. The solutions correspond to a linearly dependent basis of structures in which the polynomial $\cH$ can be decomposed. Using \textit{Mathematica} it is straightforward to generate all possible solutions to \eqref{Diophantine equations 1}, \eqref{Diophantine equations 2} for fixed values of the superspins. 

Now let us assume there exists a finite number of solutions $\G_{i}$, $i = 1, ..., N$ to \eqref{Diophantine equations 1}, \eqref{Diophantine equations 2} for a given choice of $I,J,K$. The set of solutions $\G = \{ \G_{i} \}$ may be partitioned into ``even" and ``odd" sets $\G^{+}$ and $\G^{-}$ respectively by counting the number of pseudo-invariant basis structures present in a particular solution. Therefore we define:
\begin{align}
	\G^{+} = \G|_{ \, k_{1} + k_{2} + k_{3} + q_{1} + q_{2} + q_{3} + \s \, ( \hspace{-1mm}\bmod 2 ) = 0} \, , && \G^{-} = \G|_{ \, k_{1} + k_{2} + k_{3} + q_{1} + q_{2} + q_{3} + \s\, ( \hspace{-1mm} \bmod 2 ) = 1} \, .
\end{align}
Hence, the even solutions are those such that $k_{1} + k_{2} + k_{3} + q_{1} + q_{2} + q_{3} + \s = \text{even}$ (i.e contains an even number of parity-odd building blocks), while the odd solutions are those such that $k_{1} + k_{2} + k_{3} + q_{1} + q_{2} + q_{3} + \s= \text{odd}$ (contains an odd number of parity-odd building blocks). Let $|\G^{+}| = N^{+}$ and $|\G^{-}| = N^{-}$, with $N = N^{+} + N^{-}$, then the most general ansatz for the polynomial $\cH$ in \eqref{H - generating polynomial} is as follows:
\begin{equation}  \label{H decomposition}
	\cH(X, \Q; u, v, w) = \cH^{(+)}(X, \Q; u, v, w) + \cH^{(-)}(X, \Q; u, v, w) \, ,
\end{equation}
where
\begin{subequations}
	\begin{align}
		\cH^{(+)}(X, \Q; u, v, w) &= \sum_{i=1}^{N^{+}} A_{i} \, \cG(X, \Q \, | \,\G^{+}_{i}) \, , \\
		\cH^{(-)}(X, \Q; u, v, w) &= \sum_{i=1}^{N^{-}} B_{i} \, \cG(X, \Q \, | \, \G^{-}_{i}) \, ,
	\end{align}
\end{subequations}
and $A_{i}$ and $B_{i}$ are real constants. Using this method one can generate all the possible structures for a given set of superspins $(s_{1}, s_{2}, s_{3} )$, however, at this stage we must recall that the solutions generated using this approach are linearly dependent. To form a linearly independent set of solutions we must systematically take into account the following non-linear relations between the primitive structures: 
\begin{subequations}
	\begin{align} \label{Linear dependence 1}
		Z_{2} Z_{3} + P_{1}^{2} - Q_{1}^{2} &= 0 \, , \\
		Z_{1} Z_{3} + P_{2}^{2} - Q_{2}^{2} &= 0 \, , \\
		Z_{1} Z_{2} + P_{3}^{2} - Q_{3}^{2} &= 0 \, ,
	\end{align}
\end{subequations}
\vspace{-8mm}
\begin{subequations}
	\begin{align} \label{Linear dependence 2}
		P_{1} Z_{1} + P_{2} Q_{3} + P_{3} Q_{2} &= 0 \, , & Q_{1} Z_{1} - Q_{2} Q_{3} - P_{2} P_{3} &= 0 \, , \\
		P_{2} Z_{2} + P_{1} Q_{3} + P_{3} Q_{1} &= 0 \, , & Q_{2} Z_{2} - Q_{1} Q_{3} - P_{1} P_{3} &= 0 \, , \\
		P_{3} Z_{3} + P_{1} Q_{2} + P_{2} Q_{1} &= 0 \, , & Q_{3} Z_{3} - Q_{1} Q_{2} - P_{1} P_{2} &= 0 \, .
	\end{align}
\end{subequations}
These allow elimination of the combinations $Z_{i} Z_{j}$, $Z_{i} P_{i}$, $Z_{i} Q_{i}$. There is also another relation involving triple products:
\begin{align} \label{Linear dependence 3}
	P_{1} P_{2} P_{3} + P_{1} Q_{2} Q_{3} + P_{2} Q_{1} Q_{3} + P_{3} Q_{1} Q_{2} &= 0 \, ,
\end{align}
which allows elimination of $P_{1} P_{2} P_{3}$. The relations above are identical to those appearing in the 3D CFT case \cite{Buchbinder:2022mys}, however, they must be supplemented by relations involving the fermionic structures:
\begin{subequations}
	\begin{align} \label{Linear dependence 4}
		P_{1} R_{1} - Q_{2} S_{2} + Q_{3} S_{3} &= 0 \, , & P_{1} S_{1} - Q_{2} R_{2} + Q_{3} R_{3} &= 0 \, , \\
		P_{2} R_{2} - Q_{3} S_{3} + Q_{1} S_{1} &= 0 \, , & P_{2} S_{2} - Q_{3} R_{3} + Q_{1} R_{1} &= 0 \, , \\
		P_{3} R_{3} - Q_{1} S_{1} + Q_{2} S_{2} &= 0 \, , & P_{3} S_{3} - Q_{1} R_{1} + Q_{2} R_{2} &= 0 \, ,
	\end{align}
\end{subequations}
\vspace{-8mm}
\begin{subequations}
	\begin{align} \label{Linear dependence 5}
		Z_{1} R_{2} - Q_{3} R_{1} + P_{3} S_{1} &= 0 \, , & Z_{2} R_{1} - Q_{3} R_{2} - P_{3} S_{2} &= 0 \, , \\
		Z_{2} R_{3} - Q_{1} R_{2} + P_{1} S_{2} &= 0 \, , & Z_{3} R_{2} - Q_{1} R_{3} - P_{1} S_{3} &= 0 \, , \\
		Z_{3} R_{1} - Q_{2} R_{3} + P_{2} S_{3} &= 0 \, , & Z_{1} R_{3} - Q_{2} R_{1} - P_{2} S_{1} &= 0 \, ,
	\end{align}
\end{subequations}
\vspace{-8mm}
\begin{subequations}
	\begin{align} \label{Linear dependence 6}
		Z_{1} S_{2} - Q_{3} S_{1} + P_{3} R_{1} &= 0 \, , & Z_{2} S_{1} - Q_{3} S_{2} - P_{3} R_{2} &= 0 \, , \\
		Z_{2} S_{3} - Q_{1} S_{2} + P_{1} R_{2} &= 0 \, , & Z_{3} S_{2} - Q_{1} S_{3} - P_{1} R_{3} &= 0 \, , \\
		Z_{3} S_{1} - Q_{2} S_{3} + P_{2} R_{3} &= 0 \, , & Z_{1} S_{3} - Q_{2} S_{1} - P_{2} R_{1} &= 0 \, .
	\end{align}
\end{subequations}
These allow for elimination of the products $P_{i} R_{i}$, $P_{i} S_{i}$, $Z_{i} R_{j}$, $Z_{i} S_{j}$. As a consequence of \eqref{Linear dependence 4}, the following also hold:
\begin{subequations}
	\begin{align}
		P_{1} R_{1} + P_{2} R_{2} + P_{3} R_{3} &= 0 \, , \\
		P_{1} S_{1} + P_{2} S_{2} + P_{3} S_{3} &= 0 \, .	
	\end{align}
\end{subequations}
Applying the above relations to a set of linearly dependent polynomial structures significantly reduces the number of structures to consider for a given three-point function, since we are now restricted to linearly independent contributions. This process is relatively straightforward to implement using Mathematica's pattern matching functions.

Now that we have taken care of linear-dependence, it now remains to impose conservation on all three points in addition to the various point-switch symmetries; introducing the objects $P_{i}, Q_{i}, Z_{i}, R_{i}, S_{i}$ proves to streamline this analysis significantly. First let us consider conservation;
to impose conservation on $z_{1}$, (for either sector) we compute
\begin{align}
	D_{1} \cH(X, \Q; u,v,w) &= D_{1} \Bigg\{ \sum_{i=1}^{N} c_{i} \, \cG(X, \Q \, | \, \G_{i}) \Bigg\} \nonumber \\
	&= \sum_{i=1}^{N} c_{i} \, D_{1} \cG(X, \Q \, | \, \G_{i}) \, .
\end{align}
We then solve for the coefficient $c_{i}$ such that the result vanishes. To impose the superfield conservation equations, the identities \eqref{Derivative identities} are essential. The same approach applies for conservation on $z_{2}$.

Next, to impose conservation on $z_{3}$ we must first obtain an explicit expression for $\tilde{\cH}(\boldsymbol{X},\Q)$ in terms of $\cH(\boldsymbol{X},\Q)$, that is, we must compute (e.g. for the even sector)
\begin{equation}
	\tilde{\cH}(\boldsymbol{X}, \Q; u(I), v(J), w(K) ) = (\boldsymbol{X}^{2})^{\D_{1} - \D_{3}} \cI^{(J)}_{\boldsymbol{X}}(v,\tilde{v}) \, \cH^{I}(\boldsymbol{X}, \Q; u(I), \tilde{v}(J), w(K)) \, .
\end{equation}
Recall that any solution for $\cH(\boldsymbol{X}, \Q)$ can be written in terms of the structures \eqref{Basis scalar structures 1}, \eqref{Basis scalar structures 2}; given the transformation properties \eqref{Hc and H relation}, and \eqref{Htilde and Hc relation}, the computation of $\cH^{I}(\boldsymbol{X}, \Q)$ from $\cH(\boldsymbol{X}, \Q)$ is equivalent to the following replacements:
\begin{subequations} \label{Inversion transformation 1}
	\begin{align} 
		P_{1} &\rightarrow - P_{1} \, , & P_{2} &\rightarrow -P_{2} \, , & P_{3} &\rightarrow -P_{3} \, , \\
		R_{1} &\rightarrow - \mathbb{S}_{1} \, , & R_{2} &\rightarrow - \mathbb{S}_{2} \, , & R_{3} &\rightarrow - \mathbb{S}_{3} \, , \\
		\mathbb{S}_{1} &\rightarrow R_{1} \, , & \mathbb{S}_{2} &\rightarrow R_{2} \, , & \mathbb{S}_{3} &\rightarrow R_{3} \, .
	\end{align}
\end{subequations}
Now to compute $\tilde{\cH}(\boldsymbol{X},\Q)$ from $\cH^{I}(\boldsymbol{X}, \Q)$, we make use of the fact that $P_{1}$, $P_{3}$, $\mathbb{Q}_{1}$, $\mathbb{Q}_{3}$, $\mathbb{Z}_{2}$, $R_{2}$, and $\mathbb{S}_{2}$ are the only objects with $\tilde{v}$ dependence, and apply the identities
\begin{subequations} \label{Inversion transformation 2}
	\begin{align}
		\cI_{\boldsymbol{X}}(v,\tilde{v}) \, P_{1} &= - \mathbb{Q}_{1} \, , & \cI_{\boldsymbol{X}}(v,\tilde{v}) \, P_{3} &= \mathbb{Q}_{3} + \text{i} P_{3} \boldsymbol{J} \, , \\
		\cI_{\boldsymbol{X}}(v,\tilde{v}) \, \mathbb{Q}_{1} &= - P_{1} + \text{i} \, \mathbb{Q}_{1} \boldsymbol{J} \, , & \cI_{\boldsymbol{X}}(v,\tilde{v}) \, \mathbb{Q}_{3} &= P_{3} \, , \\
		\cI_{\boldsymbol{X}}(v,\tilde{v}) \, R_{2} &= -\mathbb{S}_{2} \, ,  & \cI_{\boldsymbol{X}}(v,\tilde{v}) \, \mathbb{S}_{2} &= - R_{2} \, ,
	\end{align}
\vspace{-8mm}
	\begin{align}
		\cI^{(2)}_{\boldsymbol{X}}(v,\tilde{v}) \, \mathbb{Z}_{2} = - \mathbb{Z}_{2} \,.
	\end{align}
\end{subequations}
Hence, given a solution for the polynomial $\cH(\boldsymbol{X}, \Q)$, the computation of $\tilde{\cH}(\boldsymbol{X}, \Q)$ is now equivalent to the following replacements of the basis structures \eqref{Basis scalar structures 1}, \eqref{Basis scalar structures 2}:
\begin{subequations} \label{Htilde structure replacements}
	\begin{align} 
		P_{1} &\rightarrow \mathbb{Q}_{1} \, , & P_{2} &\rightarrow - P_{2} \, , & P_{3} &\rightarrow - \mathbb{Q}_{3} - \text{i} P_{3} \boldsymbol{J} \, , \\
		\mathbb{Q}_{1} &\rightarrow - P_{1} + \text{i} \, \mathbb{Q}_{1} \boldsymbol{J} \, , & \mathbb{Q}_{2} &\rightarrow \mathbb{Q}_{2} \, , & \mathbb{Q}_{3} &\rightarrow P_{3} \, , \\
		\mathbb{Z}_{1} &\rightarrow \mathbb{Z}_{1} \, , & \mathbb{Z}_{2} &\rightarrow - \mathbb{Z}_{2} \, , & \mathbb{Z}_{3} &\rightarrow \mathbb{Z}_{3} \, \\
		R_{1} &\rightarrow - \mathbb{S}_{1} \, , & R_{2} &\rightarrow R_{2} \, , & R_{3} &\rightarrow - \mathbb{S}_{3} \, , \\
		\mathbb{S}_{1} &\rightarrow R_{1} \, , & \mathbb{S}_{2} &\rightarrow - \mathbb{S}_{2} \, , & \mathbb{S}_{3} &\rightarrow R_{3} \, .
	\end{align}
\end{subequations}
These rules are obtained by combining \eqref{Inversion transformation 1}, \eqref{Inversion transformation 2}. Conservation on $z_{3}$ can now be imposed using the operator $D_{3}$.

It now remains to find out how point-switch symmetries act on the basis structures; this analysis is more simple when working with $\cH(X,\Q)$, instead of $\cH(\boldsymbol{X},\Q)$. For permutation of superspace points $z_{1}$ and $z_{2}$, we have $X \rightarrow - X$, $\Q \rightarrow - \Q$, $u \leftrightarrow v$. This results in the following replacement rules for the basis objects \eqref{Basis scalar structures 1}, \eqref{Basis scalar structures 2}:
\begin{subequations} \label{Point switch A - basis}
	\begin{align} 
		P_{1} &\rightarrow - P_{2} \, , & P_{2} &\rightarrow - P_{1} \, , & P_{3} &\rightarrow - P_{3} \, , \\
		Q_{1} &\rightarrow - Q_{2} \, , & Q_{2} &\rightarrow - Q_{1} \, , & Q_{3} &\rightarrow - Q_{3} \, , \\
		Z_{1} &\rightarrow - Z_{2} \, , & Z_{2} &\rightarrow - Z_{1} \, , & Z_{3} &\rightarrow - Z_{3} \, , \\
		R_{1} &\rightarrow - R_{2} \, , & R_{2} &\rightarrow - R_{1} \, , & R_{3} &\rightarrow - R_{3} \, , \\
		S_{1} &\rightarrow S_{2} \, , & S_{2} &\rightarrow S_{1} \, , & S_{3} &\rightarrow S_{3} \, .
	\end{align}
\end{subequations}
Likewise, for permutation of superspace points $z_{1}$ and $z_{3}$ we have  $X \rightarrow - X$, $\Q \rightarrow - \Q$, $u \leftrightarrow w$, resulting in the following replacements:
\begin{subequations} \label{Point switch B - basis}
	\begin{align} 
		P_{1} &\rightarrow - P_{3} \, , & P_{2} &\rightarrow - P_{2} \, , & P_{3} &\rightarrow - P_{1} \, , \\
		Q_{1} &\rightarrow - Q_{3} \, , & Q_{2} &\rightarrow - Q_{2} \, , & Q_{3} &\rightarrow - Q_{1} \, , \\
		Z_{1} &\rightarrow - Z_{3} \, , & Z_{2} &\rightarrow - Z_{2} \, , & Z_{3} &\rightarrow - Z_{1} \, , \\
		R_{1} &\rightarrow - R_{3} \, , & R_{2} &\rightarrow - R_{2} \, , & R_{3} &\rightarrow - R_{1} \, , \\
		S_{1} &\rightarrow S_{3} \, , & S_{2} &\rightarrow S_{2} \, , & S_{3} &\rightarrow S_{1} \, .
	\end{align}
\end{subequations}
We have now developed all the formalism necessary to analyse the structure of three-point correlation functions in 3D $\cN=1$ SCFT. To summarise, in the remaining sections of this paper we will analyse the three-point functions of conserved higher-spin supercurrents (for both integer and half-integer superspin) using the following method:
\begin{enumerate}
	\item For a given set of superspins, we construct all possible (linearly dependent) structures for the polynomial $\cH(X, \Q; u,v,w)$, which is governed by the solutions to \eqref{Diophantine equations 1}, \eqref{Diophantine equations 2}. The solutions are sorted into even and odd sectors.
	\item We systematically apply the linear dependence relations \eqref{Linear dependence 1}, \eqref{Linear dependence 2}, \eqref{Linear dependence 3}, \eqref{Linear dependence 4}, \eqref{Linear dependence 5}, \eqref{Linear dependence 6} to the set of all polynomial structures. This is sufficient to form the most general linearly independent ansatz for the correlation function.
	\item Using the method outlined in subsection \ref{subsubsection3.2.3}, we impose the superfield conservation equations on the correlation function, resulting in the differential contraints \eqref{Conservation equations} on $\cH$. The result of each computation is a large polynomial in the basis structures \eqref{Basis scalar structures 1}, \eqref{Basis scalar structures 2}. The linear dependence relations are systematically applied to this polynomial again to ensure that it is composed of only linearly independent terms. The coefficients are read off the structures, resulting in algebraic constraint relations on the coefficients $A_{i}, B_{i}$. This process significantly reduces the number of structures in the three-point function.
	\item Once the general form of the polynomial $\cH(X,\Q; u,v,w)$ (associated with the conserved three-point function $\langle \mathbf{J}^{}_{s_{1}} \mathbf{J}'_{s_{2}} \mathbf{J}''_{s_{3}} \rangle$) is obtained for a given set of superspins $(s_{1},s_{2}, s_{3})$, we then impose any symmetries under permutation of superspace points, that is, \eqref{Point switch A} and \eqref{Point switch B} (if applicable). In certain cases, imposing these constraints can eliminate the remaining structures. The solution is then converted into covariant form $\cH(\boldsymbol{X},\Q; u,v,w)$.
\end{enumerate}
The computations are done completely analytically with the use of Mathematica and the Grassmann package. By using pattern matching functions, the calculations are carried out purely amongst the basis structures \eqref{Basis scalar structures 1}, \eqref{Basis scalar structures 2}, as a result we do not have to fix superspace points to certain values. The only chosen parameters are the spins. Due to computational limitations we could carry out the analysis up to $s_{i} = 20$ (some steps of the calculations involve millions of terms), however, with more optimisation and sufficient computational resources this approach should hold for arbitrary superspins. 
Since there are an enormous number of possible three-point functions with $s_{i} \leq 20$, we present the final results (in the form of Mathematica outputs) for $\cH(\boldsymbol{X},\Q; u,v,w)$ for some particularly interesting examples, as the solutions and coefficient constraints become cumbersome to present beyond cases involving low superspins. We are primarily interested in counting the number of independent polynomial structures after imposing all the constraints.



\section{Three-point functions of conserved supercurrents}\label{section4}
 
In the next subsections we analyse the structure of three-point correlation functions involving conserved higher-spin supercurrents. As a test of our approach we begin with an analysis of three-point functions involving currents with low superspins, such as the supercurrent and flavour current multiplets. 

\subsection{Supercurrent and flavour current correlators}\label{subsection4.1}

The most important examples of conserved supercurrents in 3D $\cN=1$ superconformal field theories are the supercurrent and flavour current multiplets. The supercurrent multiplet is described by the spin-tensor superfield, $J_{\a(3)}(z)$, with scale dimension $\Delta_{J} = 5/2$. It satisfies $D^{\a_{1}} J_{\a_{1} \a_{2} \a_{3}}(z) = 0$ and contains the energy-momentum tensor, $T_{\a(4)}(x) = D_{(\a_{1}} J_{\a_{2} \a_{3} \a_{4})}(z) |_{\q = 0} $, and the supersymmetry current, $Q_{\a(3)}(x) = J_{\a(3)}(z) |_{\q = 0} $, as its independent component fields. Likewise, the flavour current multiplet is described by a spinor superfield, $L_{\a}(z)$, with scale dimension $\Delta_{L} = 3/2$. It satisfies the superfield conservation equation $D^{\a} L_{\a}(z) = 0$, and contains a conserved vector current $V_{\a(2)}(x) = D_{(\a_{1}} L_{\a_{2})}(z) |_{\q = 0} $. Three-point functions of these supercurrents were originally studied in \cite{Buchbinder:2015qsa,Buchbinder:2021gwu} (for analysis of three-point functions of the component currents in 3D/4D CFT see \cite{Buchbinder:2022cqp,Buchbinder:2022mys}), here we present the solutions for them using our formalism. The possible three-point functions involving the supercurrent and flavour current multiplets are: 
\begin{align} \label{Low-superspin component correlators}
	\langle L_{\a}(z_{1}) \, L_{\b}(z_{2}) \, L_{\g}(z_{3}) \rangle \, , &&  \langle L_{\a}(z_{1}) \, L_{\b}(z_{2}) \, J_{\g(3)}(z_{3}) \rangle \, , \\
	\langle J_{\a(3)}(z_{1}) \, J_{\b(3)}(z_{2}) \, L_{\a}(z_{3}) \rangle \, , &&  \langle J_{\a(3)}(z_{1}) \, J_{\b(3)}(z_{2}) \, J_{\g(3)}(z_{3}) \rangle \, .
\end{align}
We note that in all cases the correlation functions are overall Grassmann-odd, hence, it's expected that each of them are fixed up to a single parity-even solution after imposing conservation on all three points. The analysis of these three-point functions is relatively straightforward using our computational approach.

\newpage
\noindent\textbf{Correlation function} $\langle L L L \rangle$\textbf{:}

Let us first consider $\langle L L L \rangle$; within the framework of our formalism we study the three-point function $\langle \mathbf{J}^{}_{1/2} \mathbf{J}'_{1/2} \mathbf{J}''_{1/2} \rangle$. The general ansatz for this correlation function, according to \eqref{Conserved correlator ansatz} is
\begin{align}
	\langle \mathbf{J}^{}_{\a}(z_{1}) \, \mathbf{J}'_{\b}(z_{2}) \, \mathbf{J}''_{\g}(z_{3}) \rangle = \frac{ \cI_{\a}{}^{\a'}(\boldsymbol{x}_{13}) \,  \cI_{\b}{}^{\b'}(\boldsymbol{x}_{23}) }{(\boldsymbol{x}_{13}^{2})^{3/2} (\boldsymbol{x}_{23}^{2})^{3/2}}
	\; \cH_{\a' \b' \g}(\boldsymbol{X}_{12}, \Q_{12}) \, .
\end{align} 
Using the formalism outlined in subsection \ref{subsection3.2}, all information about this correlation function is encoded in the following polynomial:
\begin{align}
	\cH(\boldsymbol{X}, \Q; u(1), v(1), w(1)) = \cH_{ \a \b \g }(\boldsymbol{X}, \Q) \, \boldsymbol{u}^{\a}  \boldsymbol{v}^{\b}  \boldsymbol{w}^{\g} \, .
\end{align}
Using Mathematica we solve \eqref{Diophantine equations 2} for the chosen spins and substitute each solution into the generating function \eqref{Generating function 2}. This provides us with the following list of linearly dependent polynomial structures for the polynomial $\cH(X,\Q;u,v,w)$ in the even and odd sectors respectively:
\begin{flalign*}
	\hspace{5mm} \includegraphics[width=0.95\textwidth]{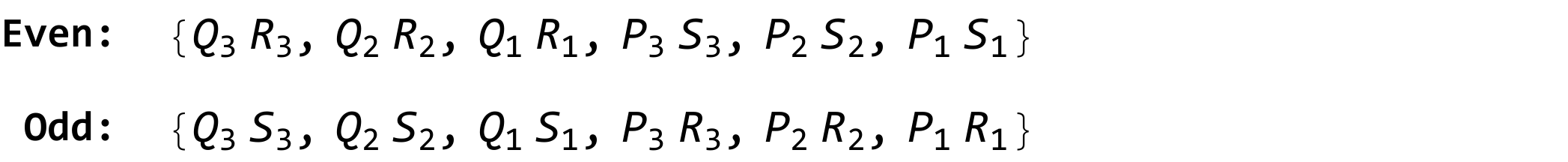} &&
\end{flalign*} 
After systematic application of the linear dependence relations \eqref{Linear dependence 1}-\eqref{Linear dependence 6} we obtain the following linearly independent sets:
\begin{flalign*}
	\hspace{5mm} \includegraphics[width=0.95\textwidth]{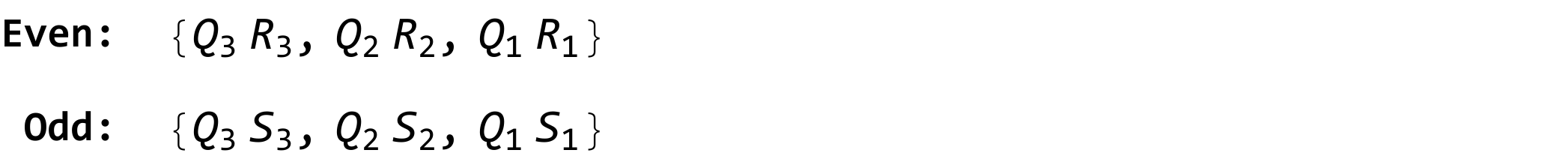} &&
\end{flalign*} 
Next, we impose conservation on all three points, where we obtain the following constraints on the coefficients $A_{i}$ and $B_{i}$:
\begin{flalign*}
	\hspace{5mm} \includegraphics[width=0.95\textwidth]{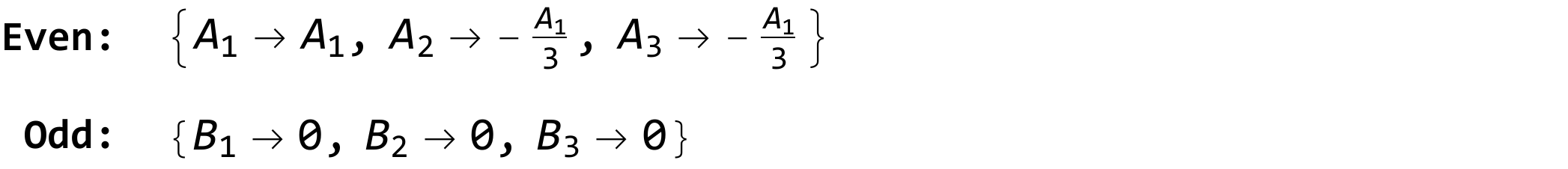} &&
\end{flalign*} 
and the explicit solution for $\cH(\boldsymbol{X}, \Q; u,v,w)$
\begin{flalign*}
	\hspace{5mm} \includegraphics[width=0.95\textwidth]{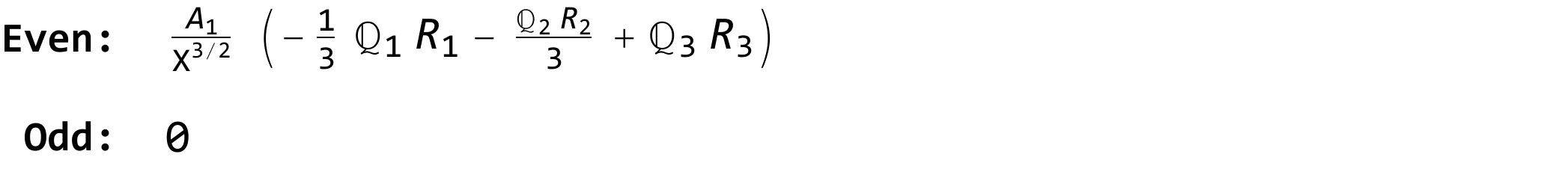} &&
\end{flalign*} 
Hence, the three-point function is fixed up to a single parity-even polynomial structure. 
After imposing symmetries under permutation of spacetime points, 
e.g. $\mathbf{J}=\mathbf{J}'=\mathbf{J}''$, the remaining structure vanishes.
This vanishing result is not surprising because it corresponds to the contribution proportional to the symmetric invariant tensor of the flavour symmetry group. 
In four dimensions this contribution is related to the chiral anomaly which does not exist in three dimensions.
The correlator $\langle \mathbf{J}_{1/2} \, \mathbf{J}_{1/2} \, \mathbf{J}_{1/2} \rangle$ has, however, a non-vanishing contribution proportional to the totally antisymmetric structure constants. 
In our analysis in this paper any possible ``antisymmetric" contributions are ignored when we impose the point-switch identities.  
The most general form of three-point function of flavour current multiplets was found explicitly in~\cite{Buchbinder:2015qsa, Buchbinder:2021gwu} and we will not discuss it here. 

\vspace{2mm}

\noindent
\textbf{Correlation function} $\langle L L J \rangle$\textbf{:}

The next example to consider is the mixed correlator $\langle L L J \rangle$; to study this case we may examine the correlation function $\langle \mathbf{J}^{}_{1/2} \mathbf{J}'_{1/2} \mathbf{J}''_{3/2} \rangle$. Using the general formula, the ansatz for this three-point function is
\begin{align}
	\langle \mathbf{J}^{}_{\a}(z_{1}) \, \mathbf{J}'_{\b}(z_{2}) \, \mathbf{J}''_{\g(3)}(z_{3}) \rangle = \frac{ \cI_{\a}{}^{\a'}(\boldsymbol{x}_{13}) \,  \cI_{\b}{}^{\b'}(\boldsymbol{x}_{23}) }{(\boldsymbol{x}_{13}^{2})^{3/2} (\boldsymbol{x}_{23}^{2})^{3/2}}
	\; \cH_{\a' \b' \g(3)}(\boldsymbol{X}_{12}, \Q_{12}) \, .
\end{align} 
Using the formalism outlined in \ref{subsection3.2}, all information about this correlation function is encoded in the following polynomial:
\begin{align}
	\cH(\boldsymbol{X}, \Q; u(1), v(1), w(3)) = \cH_{ \a \b \g(3) }(\boldsymbol{X}, \Q) \, \boldsymbol{u}^{\a}  \boldsymbol{v}^{\b} \boldsymbol{w}^{\g(3)} \, .
\end{align}
After solving \eqref{Diophantine equations 2}, we obtain the following list of polynomial structures for $\cH(X,\Q;u,v,w)$ in the even and odd sectors respectively:
\begin{flalign*}
	\hspace{5mm} \includegraphics[width=0.95\textwidth]{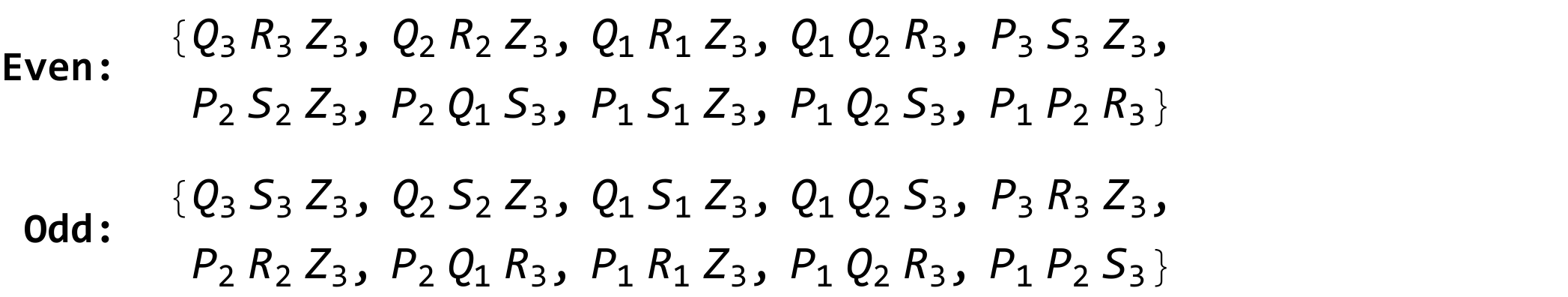} &&
\end{flalign*} 
After systematic application of the linear dependence relations \eqref{Linear dependence 1}-\eqref{Linear dependence 6} we obtain the following linearly independent sets:
\begin{flalign*}
	\hspace{5mm} \includegraphics[width=0.95\textwidth]{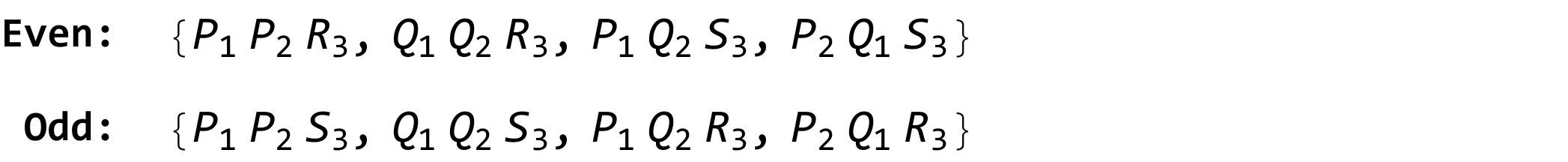} &&
\end{flalign*} 
Next, we impose conservation on all three points; we obtain the following constraints on the coefficients $A_{i}$ and $B_{i}$:
\begin{flalign*}
	\hspace{5mm} \includegraphics[width=0.95\textwidth]{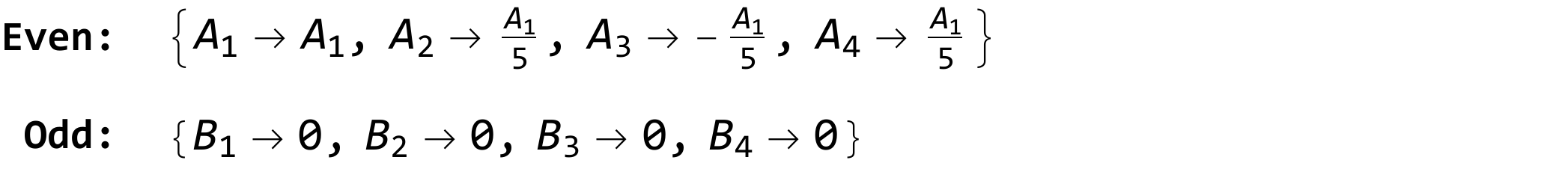} &&
\end{flalign*} 
and the explicit solution for $\cH(\boldsymbol{X}, \Q; u,v,w)$
\begin{flalign*}
	\hspace{5mm} \includegraphics[width=0.95\textwidth]{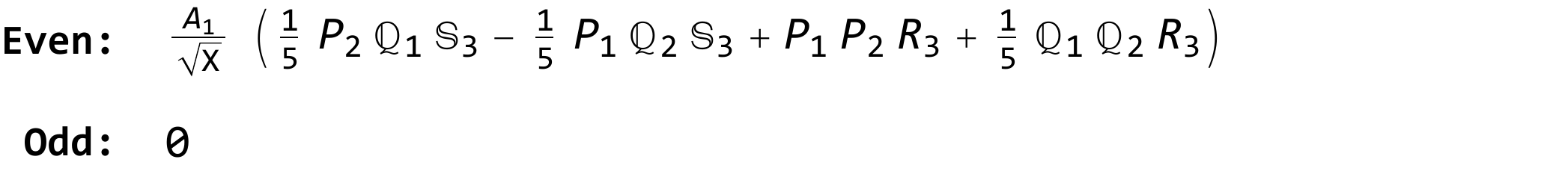} &&
\end{flalign*} 
Hence, after conservation, the three-point function is fixed up to a single even structure. This structure is also compatible with the symmetry $\mathbf{J}=\mathbf{J}'$, therefore $\langle L L J \rangle$ is fixed up to a single structure. \\[5mm]
\noindent
\textbf{Correlation function} $\langle J J L \rangle$\textbf{:}

The next example to consider is the mixed correlator $\langle J J L \rangle$; to study this case we may examine the correlation function $\langle \mathbf{J}^{}_{3/2} \mathbf{J}'_{3/2} \mathbf{J}''_{1/2} \rangle$. Using the general formula, the ansatz for this three-point function is
\begin{align}
	\langle \mathbf{J}^{}_{\a(3)}(z_{1}) \, \mathbf{J}'_{\b(3)}(z_{2}) \, \mathbf{J}''_{\g}(z_{3}) \rangle = \frac{ \cI_{\a(3)}{}^{\a'(3)}(\boldsymbol{x}_{13}) \,  \cI_{\b(3)}{}^{\b'(3)}(\boldsymbol{x}_{23}) }{(\boldsymbol{x}_{13}^{2})^{5/2} (\boldsymbol{x}_{23}^{2})^{5/2}}
	\; \cH_{\a'(3) \b'(3) \g}(\boldsymbol{X}_{12}, \Q_{12}) \, .
\end{align} 
Using the formalism outlined in \ref{subsection3.2}, all information about this correlation function is encoded in the following polynomial:
\begin{align}
	\cH(\boldsymbol{X}, \Q; u(3), v(3), w(1)) = \cH_{ \a(3) \b(3) \g }(\boldsymbol{X}, \Q) \, \boldsymbol{u}^{\a(3)}  \boldsymbol{v}^{\b(3)} \boldsymbol{w}^{\g} \, .
\end{align}
After solving \eqref{Diophantine equations 2}, we obtain the following list of (linearly dependent) polynomial structures in the even and odd sectors respectively:

\begin{flalign*}
	\hspace{5mm} \includegraphics[width=0.95\textwidth]{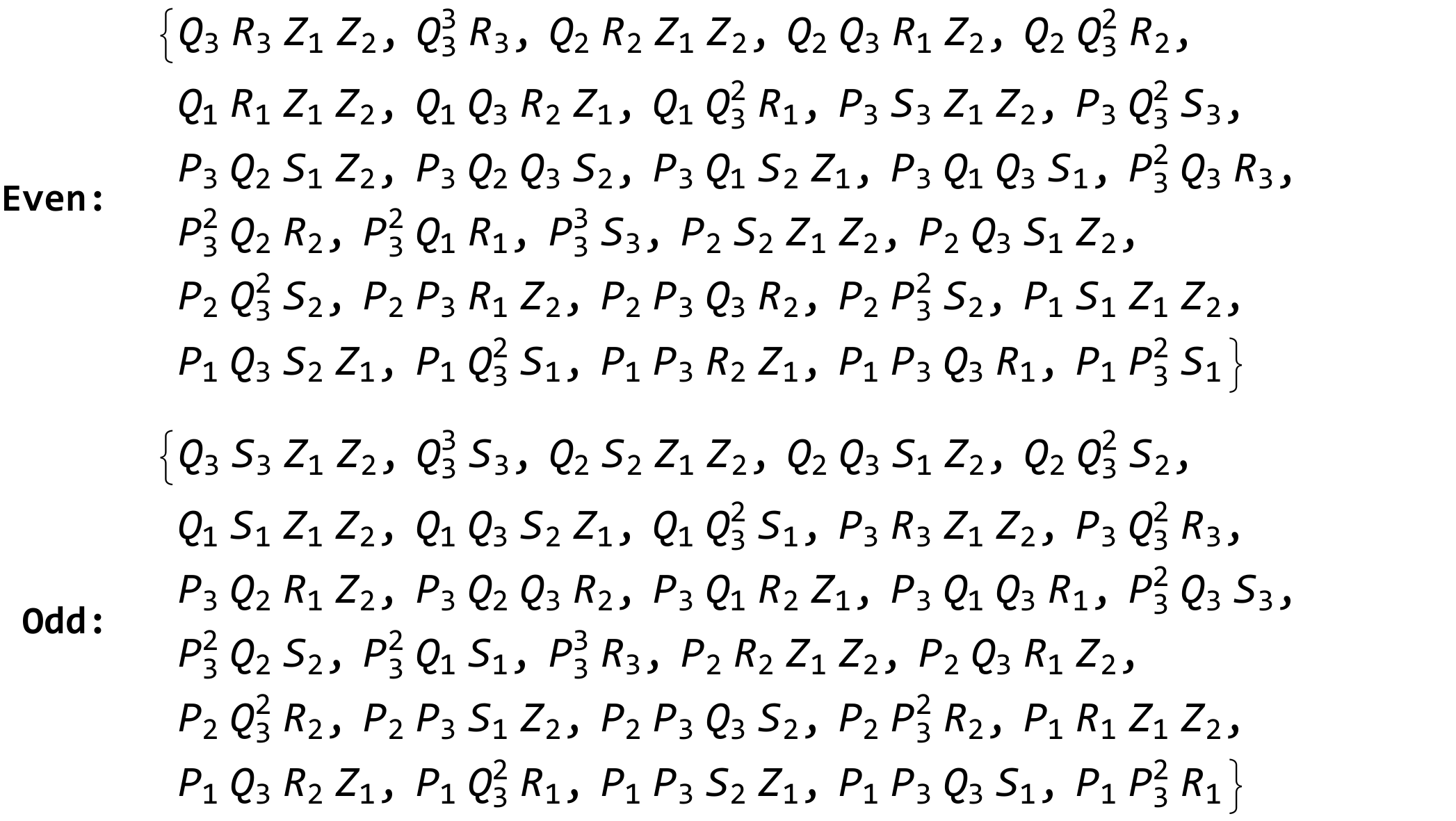} &&
\end{flalign*} 
After systematic application of the linear dependence relations \eqref{Linear dependence 1}-\eqref{Linear dependence 6} we obtain the following linearly independent sets:
\begin{flalign*}
	\hspace{5mm} \includegraphics[width=0.95\textwidth]{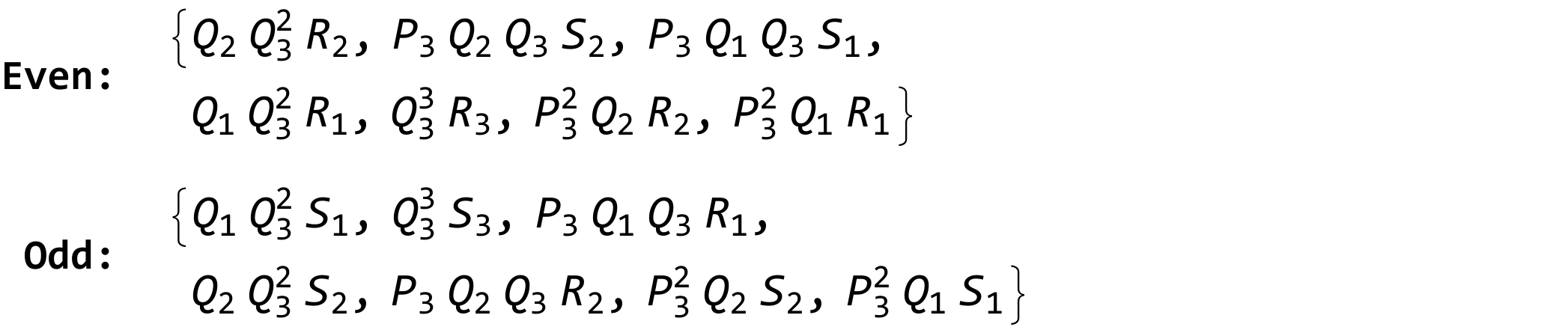} &&
\end{flalign*} 
Next, we impose conservation on all three points; we obtain the following constraints on the coefficients $A_{i}$ and $B_{i}$:
\begin{flalign*}
	\hspace{5mm} \includegraphics[width=0.95\textwidth]{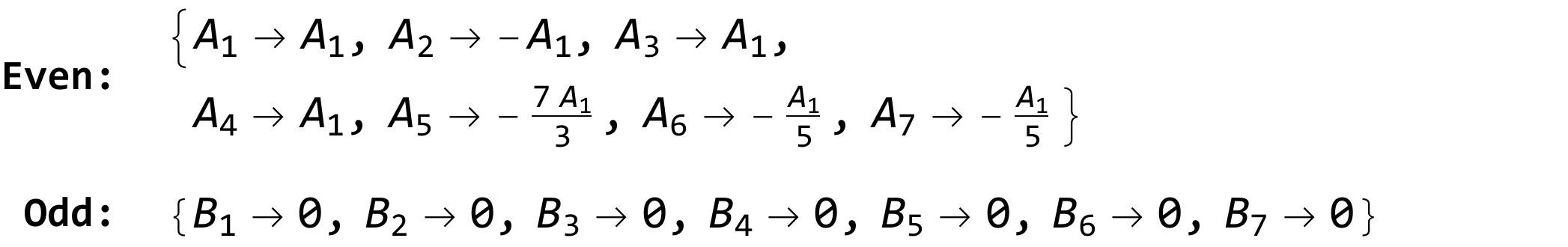} &&
\end{flalign*} 
and the explicit solution for $\cH(\boldsymbol{X}, \Q; u,v,w)$
\begin{flalign*}
	\hspace{5mm} \includegraphics[width=0.95\textwidth]{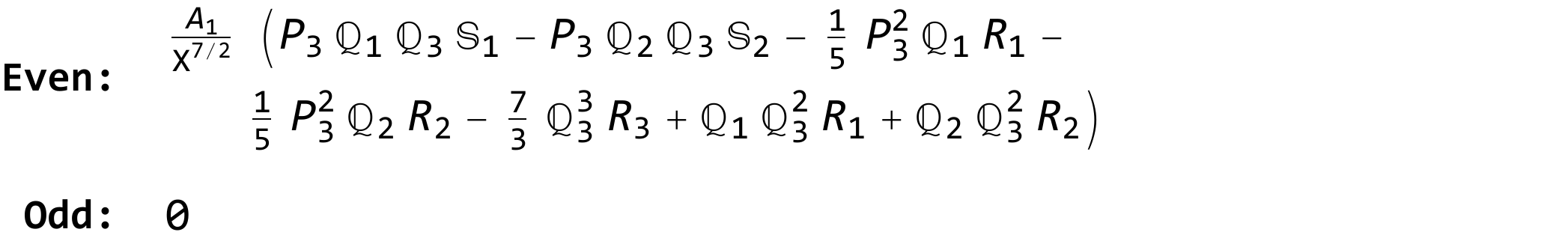} &&
\end{flalign*} 
Hence, after imposing conservation on all three points, the three-point function is fixed up to a single even structure. This structure is not compatible with the symmetry property $\mathbf{J} = \mathbf{J}'$, hence, $\langle J J L \rangle = 0$. \\[5mm]
\noindent
\textbf{Correlation function} $\langle J J J \rangle$\textbf{:}

The last example to consider is the three-point function of the supercurrent, $\langle J J J \rangle$. To study it we may examine the correlation function $\langle \mathbf{J}^{}_{3/2} \mathbf{J}'_{3/2} \mathbf{J}''_{3/2} \rangle$. Using the general formula, the ansatz for this three-point function is
\begin{align}
	\langle \mathbf{J}^{}_{\a(3)}(z_{1}) \, \mathbf{J}'_{\b(3)}(z_{2}) \, \mathbf{J}''_{\g(3)}(z_{3}) \rangle = \frac{ \cI_{\a(3)}{}^{\a'(3)}(\boldsymbol{x}_{13}) \,  \cI_{\b(3)}{}^{\b'(3)}(\boldsymbol{x}_{23}) }{(\boldsymbol{x}_{13}^{2})^{5/2} (\boldsymbol{x}_{23}^{2})^{5/2}}
	\; \cH_{\a'(3) \b'(3) \g(3)}(\boldsymbol{X}_{12}, \Q_{12}) \, .
\end{align} 
Using the formalism outlined in \ref{subsection3.2}, all information about this correlation function is encoded in the following polynomial:
\begin{align}
	\cH(\boldsymbol{X}, \Q; u(3), v(3), w(3)) = \cH_{ \a(3) \b(3) \g(3) }(\boldsymbol{X}, \Q) \, \boldsymbol{u}^{\a(3)}  \boldsymbol{v}^{\b(3)} \boldsymbol{w}^{\g(3)} \, .
\end{align}
In this case there are a vast number of linearly dependent structures to consider and the list is too large to present, however, after application of the linear dependence relations \eqref{Linear dependence 1}-\eqref{Linear dependence 6} we obtain the following linearly independent structures:
\begin{flalign*}
	\hspace{5mm} \includegraphics[width=0.95\textwidth]{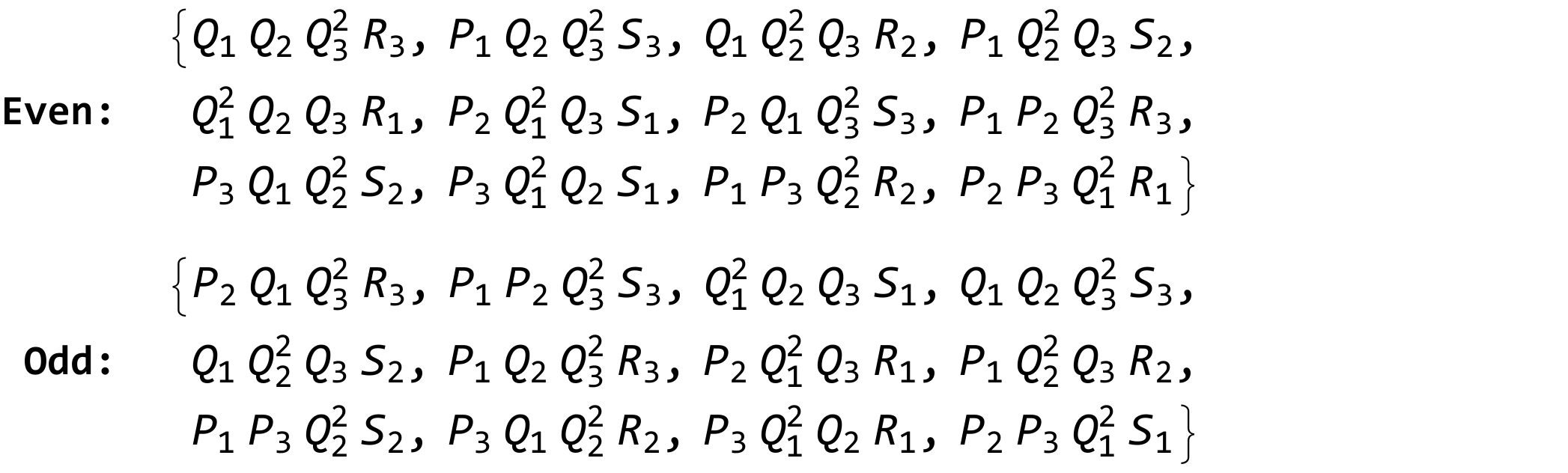} &&
\end{flalign*} 
%
Next, we impose conservation on all three points and obtain the following constraints on the coefficients $A_{i}$ and $B_{i}$:
\begin{flalign*}
	\hspace{5mm} \includegraphics[width=0.95\textwidth]{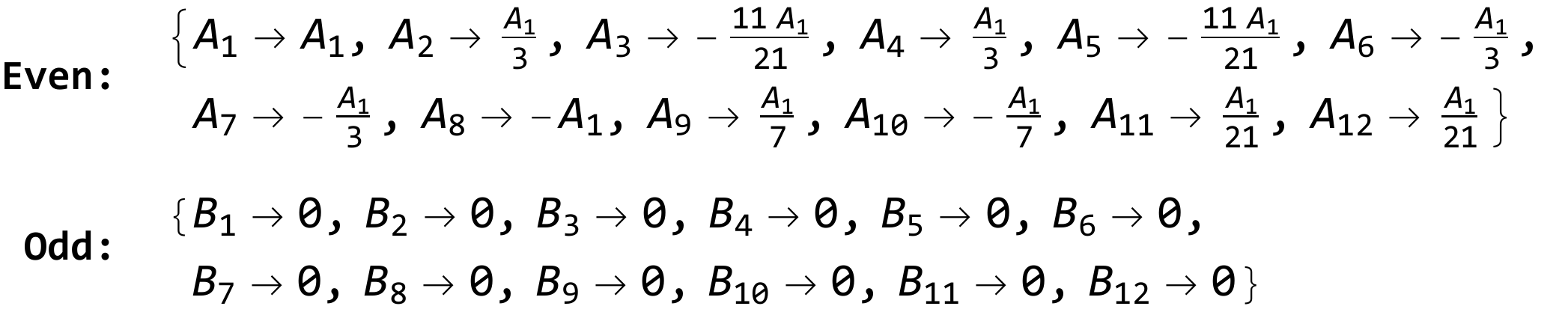} &&
\end{flalign*} 
and the explicit solution for $\cH(\boldsymbol{X}, \Q; u,v,w)$
\begin{flalign*}
	\hspace{5mm} \includegraphics[width=0.95\textwidth]{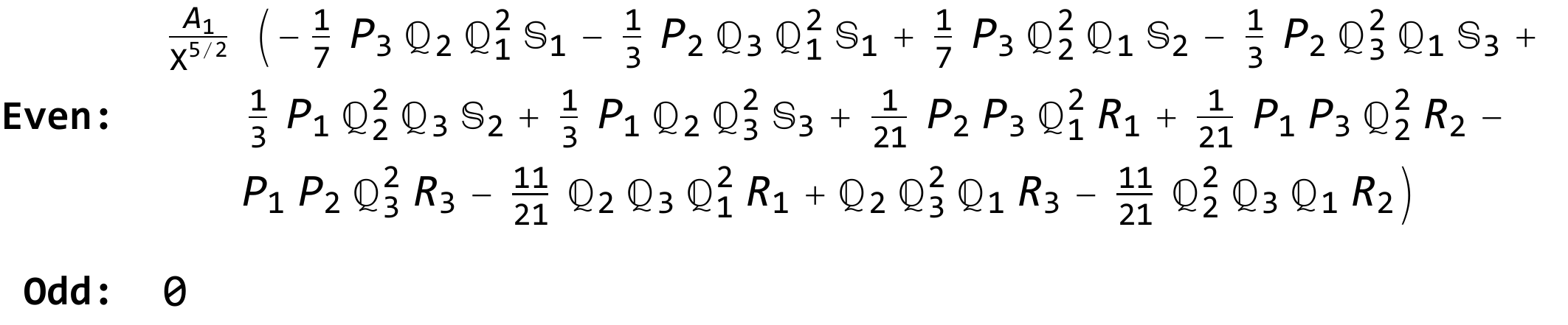} &&
\end{flalign*} 
Hence the three-point function $\langle \mathbf{J}^{}_{3/2} \mathbf{J}'_{3/2} \mathbf{J}''_{3/2} \rangle$ is fixed up to a single parity-even structure. The remaining polynomial structures are also compatible with the symmetry property $\mathbf{J}=\mathbf{J}'=\mathbf{J}''$, hence, the supercurrent three-point function $\langle J J J \rangle$ is fixed up to a single parity-even structure. In terms of the number of independent structures, these results are consistent with \cite{Buchbinder:2015qsa}.



\subsection{General structure of \texorpdfstring{$\langle \mathbf{J}^{}_{s_{1}} \mathbf{J}'_{s_{2}} \mathbf{J}''_{s_{3}} \rangle$
}{< J J' J'' >}}
\label{subsection4.2}


We performed a comprehensive analysis of the general structure of the three-point correlation function 
$\langle \mathbf{J}^{}_{s_{1}} \mathbf{J}'_{s_{2}} \mathbf{J}''_{s_{3}} \rangle$ using our computational approach. 
Due to computational limitations we were able to carry out this analysis for $s_{i} \leq 20$, however, the pattern in the solutions is very clear and we propose that the results 
stated in this section hold for arbitrary superspins. We also want to emphasise that for given $(s_1, s_2, s_3)$ our method produces a result which can be presented in an explicit form even for relatively
high superspins (see examples below). With a sufficiently powerful computer one can extend our results to larger values of $s_i$.

Based on our analysis we found that the general structure of the three-point correlation function 
$\langle \mathbf{J}^{}_{s_{1}} \mathbf{J}'_{s_{2}} \mathbf{J}''_{s_{3}} \rangle$ is constrained to the following form:
%
\begin{equation}
	\langle \mathbf{J}^{}_{s_{1}} \mathbf{J}'_{s_{2}} \mathbf{J}''_{s_{3}} \rangle = a \, \langle \mathbf{J}^{}_{s_{1}} \mathbf{J}'_{s_{2}} \mathbf{J}''_{s_{3}} \rangle_{E} + b \, \langle \mathbf{J}^{}_{s_{1}} \mathbf{J}'_{s_{2}} \mathbf{J}''_{s_{3}} \rangle_{O} \, .
\end{equation}
One of our main conclusions is that the odd structure, $\langle  \mathbf{J}^{}_{s_{1}}  \mathbf{J}'_{s_{2}}  \mathbf{J}''_{s_{3}} \rangle_{O}$ does not appear in correlators that are overall 
Grassmann-odd (or fermionic). The existence of the odd solution in the Grassmann-even (bosonic) correlators depend on the following superspin triangle inequalities:
\begin{align} \label{Triangle inequalities}
	s_{1} &\leq s_{2} + s_{3} \, , & s_{2} &\leq s_{1} + s_{3} \, , & s_{3} &\leq s_{1} + s_{2} \, .
\end{align}
When the triangle inequalities are simultaneously satisfied, there is one even solution and one odd solution, however, if any of the above inequalities are not satisfied then the odd solution is incompatible 
with current conservation.
Further, if any of the $ \mathbf{J}$, $ \mathbf{J}'$, $ \mathbf{J}''$ coincide then the resulting point-switch symmetries can kill off the remaining structures. 

Before we discuss in more detail Grassmann-even and Grassmann-odd correlators and present explicit examples we would like to make some general comments. 
In particular, we observe that if the triangle inequalities are simultaneously satisfied, each polynomial structure in the three-point functions can be written as a product of at most 5 of the $P_{i}$, $Q_{i}$, with the $Z_{i}$ 
completely eliminated. Another useful observation is that the triangle inequalities can be encoded in a discriminant, $\s$, which we define as follows:
\begin{align} \label{Discriminant}
	\s(s_{1}, s_{2}, s_{3}) = q_{1} q_{2} q_{3} \, , \hspace{10mm} q_{i} = s_{i} - s_{j} - s_{k} - 1 \, ,
\end{align}
where $(i,j,k)$ is a cyclic permutation of $(1,2,3)$. For $\s(s_{1}, s_{2}, s_{3}) < 0$, there is one even solution and one odd solution, while for $\s(s_{1}, s_{2}, s_{3}) \geq 0$ there is a single even solution. 
Also recall that the correlation function can be encoded in a tensor $\cH$, which is a function of two three-point covariants, $X$ and $\Q$. There are three different (equivalent) representations of a given three-point function, 
call them $\cH^{(i)}$, where the superscript $i$ denotes which point we set to act as the ``third point" in the ansatz \eqref{H ansatz}. As shown in subsection \ref{subsubsection3.2.1}, 
the representations are related by the intertwining operator, $\cI$. Since the dimensions of the conserved supercurrents $\Delta_i$ are related to the superspins as $\Delta_i= s_i+1$ 
it follows that each $\cH^{(i)}$ is homogeneous of degree $q_{i}$. Then it follows that the odd structure survives if and only if $\forall i$, $q_{i} < 0$. In other words, 
each $\cH^{(i)}$ must be a rational function of $X$ and $\Q$ with homogeneity $q_{i} < 0$. The discriminant \eqref{Discriminant} simply encodes information about whether the $\cH^{(i)}$ 
are simultaneously of negative homogeneity. 



\subsubsection{Grassmann-even correlators}

The complete classification of results for Grassmann-even conserved three-point functions, including cases where there is a point-switch symmetry, is as follows:
\begin{itemize}
	\item In all cases we have examined ($s_{i} \leq 20$) there is one even solution and one odd solution, however, the odd solution vanishes if the superspin triangle inequalities are not satisfied. 
	\item $\langle \mathbf{J}^{}_{s_{1}} \mathbf{J}^{}_{s_{1}} \mathbf{J}'_{s_{2}} \rangle$: 
	Note that in this case $s_2$ must  be an integer. 
	For $s_{2}$ even, the solutions survive the point-switch symmetry for arbitrary $s_{1}$ (integer or half-integer). For $s_{2}$ odd the point-switch symmetry is not satisfied and the three-point function vanishes. 
	\item $\langle \mathbf{J}_{s} \, \mathbf{J}_{s} \, \mathbf{J}_{s} \rangle$: in this case $s$ is restricted to integer values. For $s$ even the solutions are compatible with the point-switch symmetries. 
\end{itemize}
The number of linearly independent structures grows rapidly with the superspins, therefore we only present results for some low superspin cases after imposing conservation on all three points.

\vspace{4mm}


\noindent
\textbf{Correlation function} $\langle \mathbf{J}^{}_{1} \mathbf{J}'_{1} \mathbf{J}''_{1} \rangle$\textbf{:}
\begin{flalign*}
	\hspace{5mm} \includegraphics[width=0.97\textwidth]{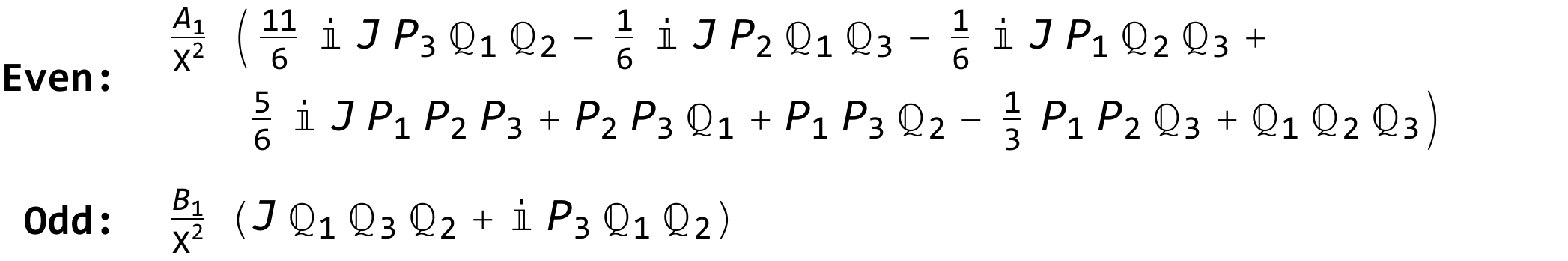} &&
\end{flalign*}

\noindent
\textbf{Correlation function} $\langle \mathbf{J}^{}_{1/2} \mathbf{J}'_{1/2} \mathbf{J}''_{2} \rangle$\textbf{:}
\begin{flalign*}
	\hspace{5mm} \includegraphics[width=0.97\textwidth]{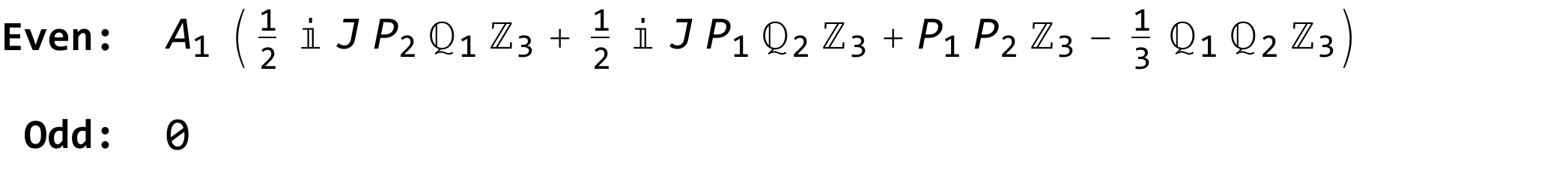} &&
\end{flalign*} 
This three-point function was initially studied in \cite{Nizami:2013tpa}, where it was shown that a parity odd solution could arise. However, it was proven later in \cite{Buchbinder:2021qlb} that such a structure cannot be consistent with the superfield conservation equations. The approach we have developed also confirms that a parity-odd solution cannot exist; this is further supported by the fact that the superspin triangle inequalities are not satisfied for this three-point function.

\vspace{4mm}

\noindent
\textbf{Correlation function} $\langle \mathbf{J}^{}_{1/2} \mathbf{J}'_{1/2} \mathbf{J}''_{3} \rangle$\textbf{:}
\begin{flalign*}
	\hspace{5mm} \includegraphics[width=0.97\textwidth]{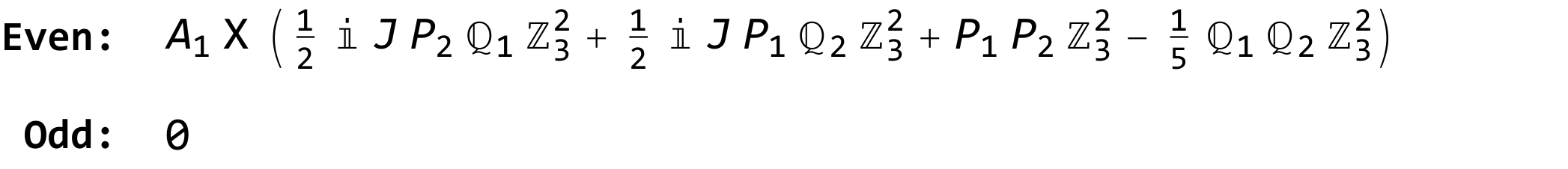} &&
\end{flalign*} 
This is another case where the superspin triangle inequalities are not satisfied, hence, the odd structure vanishes as expected.

\vspace{4mm}

\noindent
\textbf{Correlation function} $\langle \mathbf{J}^{}_{1/2} \mathbf{J}'_{3/2} \mathbf{J}''_{2} \rangle$\textbf{:}
\begin{flalign*}
	\hspace{5mm} \includegraphics[width=0.97\textwidth]{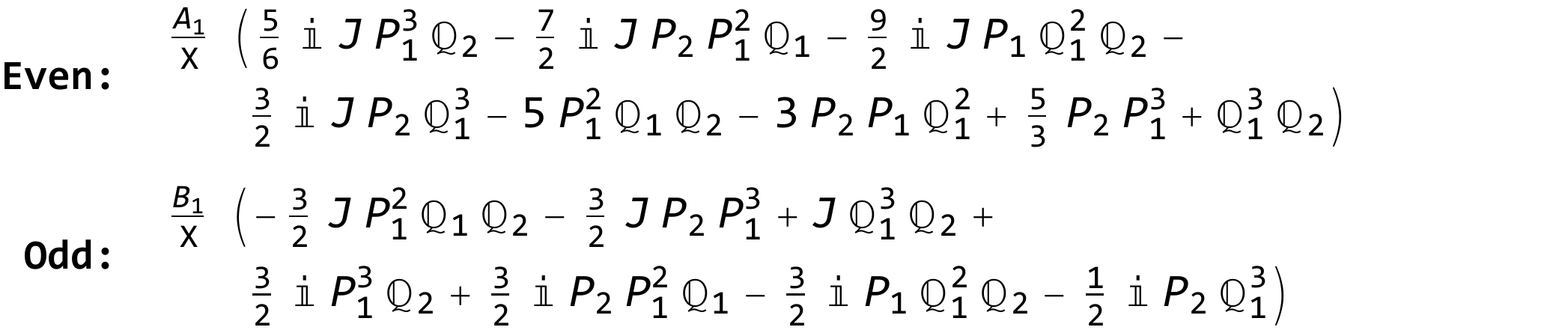} &&
\end{flalign*}

\noindent
\textbf{Correlation function} $\langle \mathbf{J}^{}_{3/2} \mathbf{J}'_{3/2} \mathbf{J}''_{2} \rangle$\textbf{:}
\begin{flalign*}
	\hspace{5mm} \includegraphics[width=0.97\textwidth]{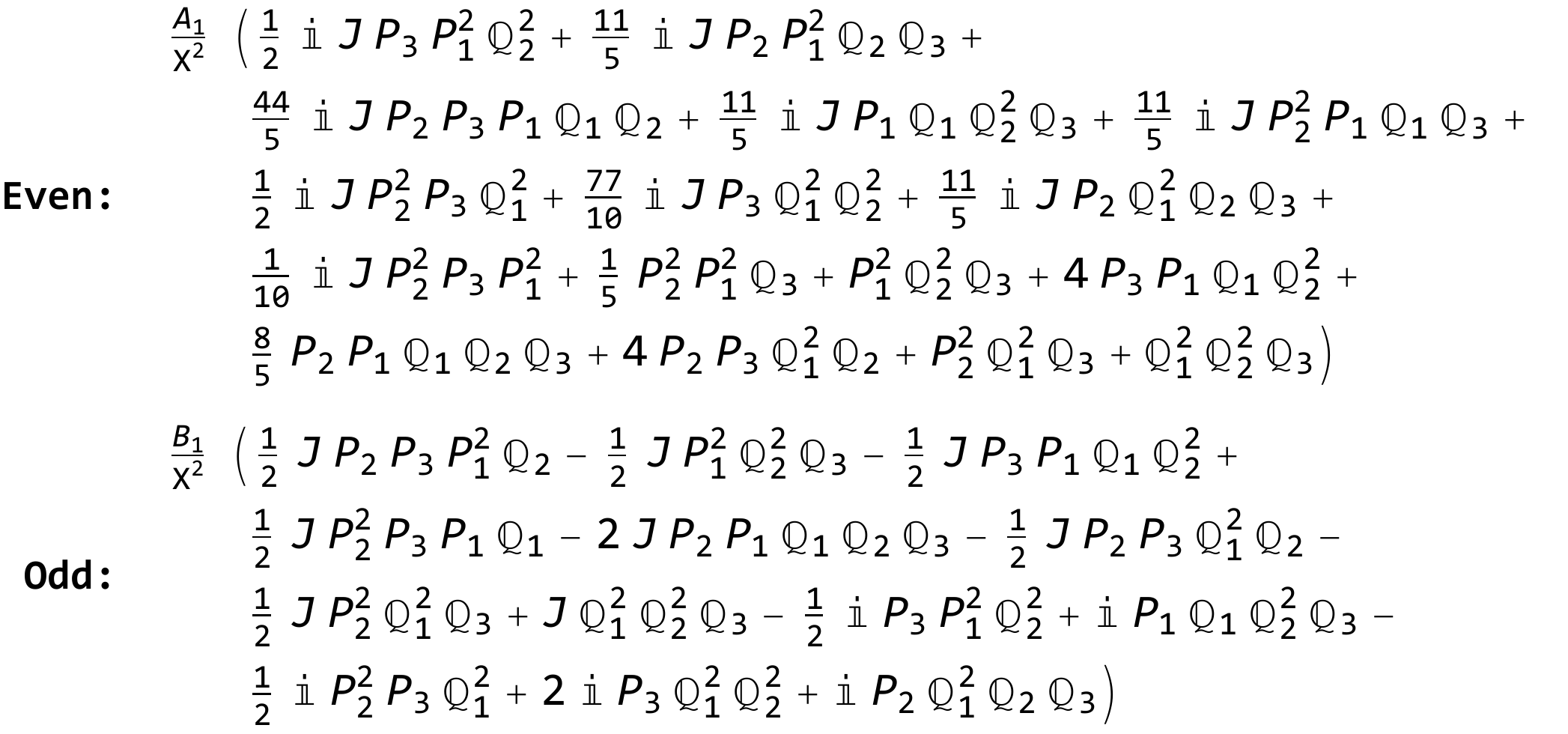} &&
\end{flalign*}

\noindent
\textbf{Correlation function} $\langle \mathbf{J}^{}_{2} \mathbf{J}'_{2} \mathbf{J}''_{2} \rangle$\textbf{:}
\begin{flalign*}
	\hspace{5mm} \includegraphics[width=0.97\textwidth]{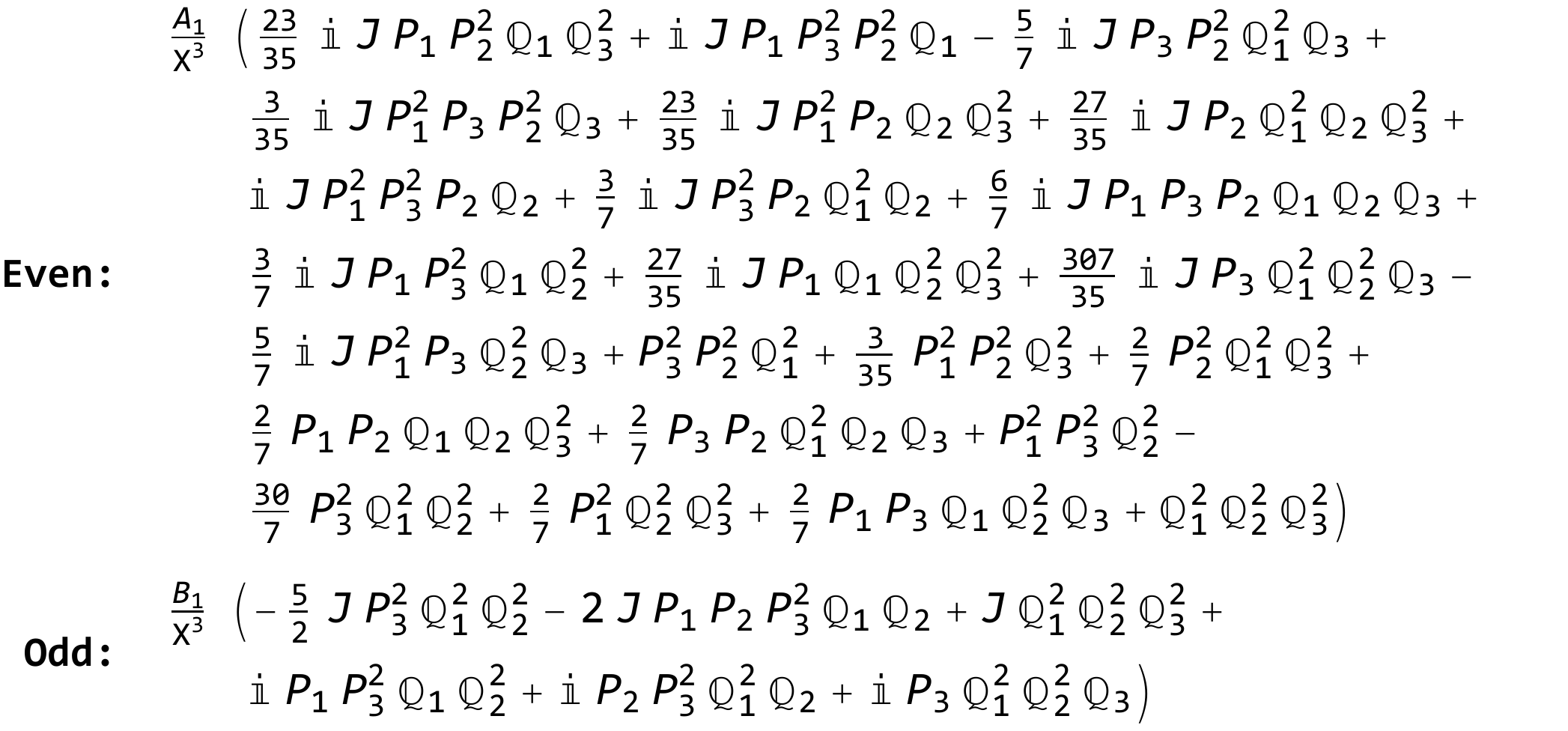} &&
\end{flalign*}
This three-point function has been studied explicitly using a tensor formalism in \cite{Buchbinder:2021qlb}, where it was shown that a parity-odd solution could arise in the three-point function. The approach we have developed can compute this correlator in seconds.

\vspace{4mm}

\noindent
\textbf{Correlation function} $\langle \mathbf{J}^{}_{1} \mathbf{J}'_{2} \mathbf{J}''_{4} \rangle$\textbf{:}
\begin{flalign*}
	\hspace{5mm} \includegraphics[width=0.97\textwidth]{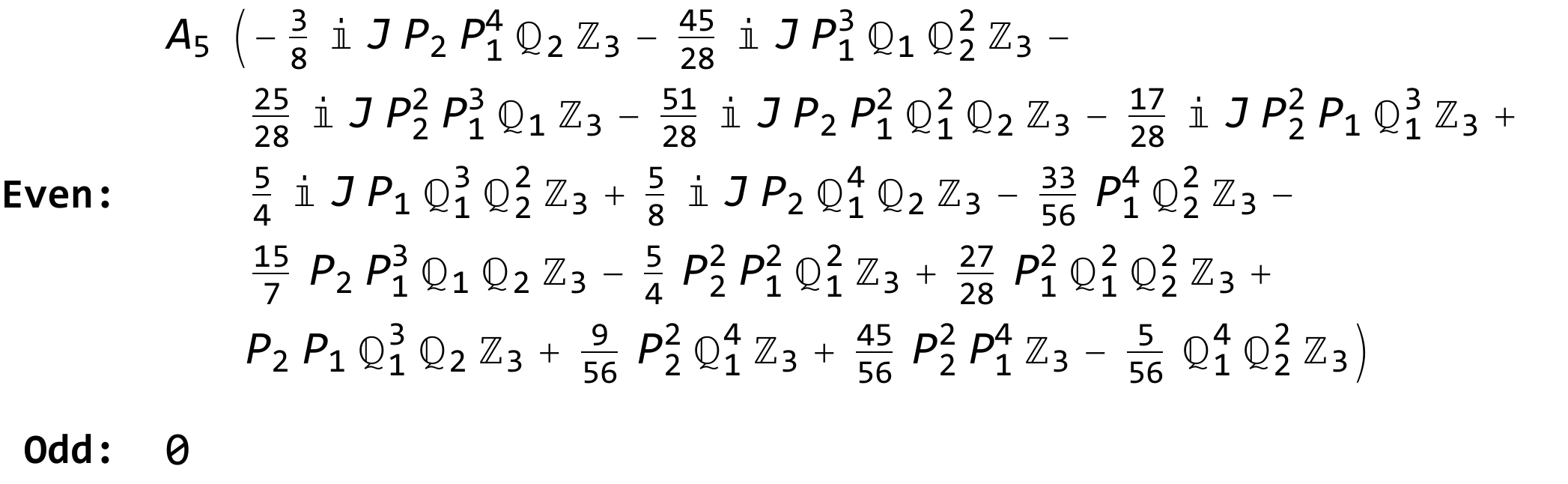} &&
\end{flalign*}
In this case we note that the superspin triangle inequalities are not satisfied and therefore the odd solution vanishes after current conservation.

\vspace{4mm}

\noindent
\textbf{Correlation function} $\langle \mathbf{J}^{}_{2} \mathbf{J}'_{2} \mathbf{J}''_{4} \rangle$\textbf{:}
\begin{flalign*}
	\hspace{5mm} \includegraphics[width=0.97\textwidth]{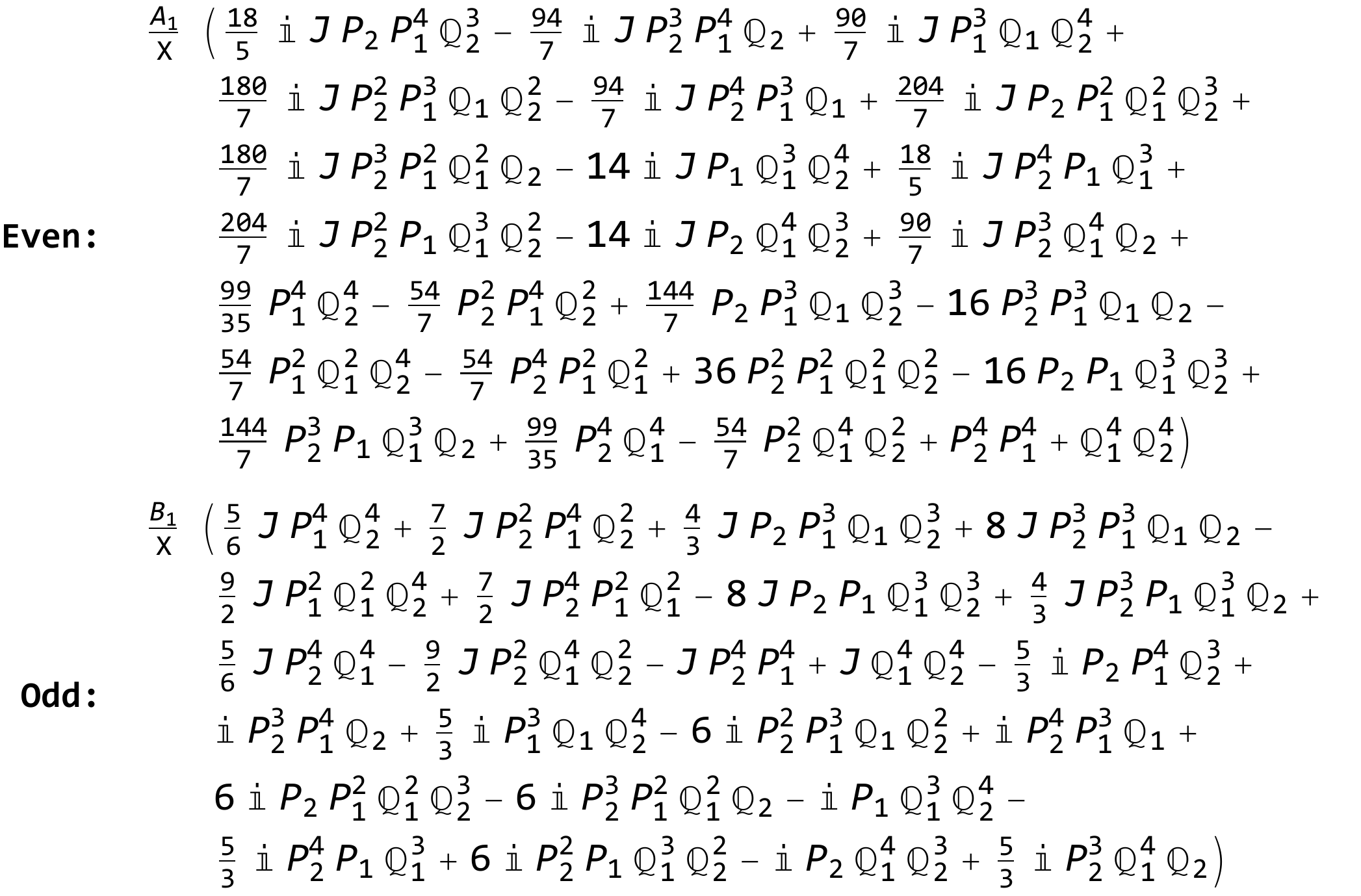} &&
\end{flalign*}

\noindent
\textbf{Correlation function} $\langle \mathbf{J}^{}_{4} \mathbf{J}'_{4} \mathbf{J}''_{4} \rangle$\textbf{:}
\begin{flalign*}
	\hspace{5mm} \includegraphics[width=0.97\textwidth]{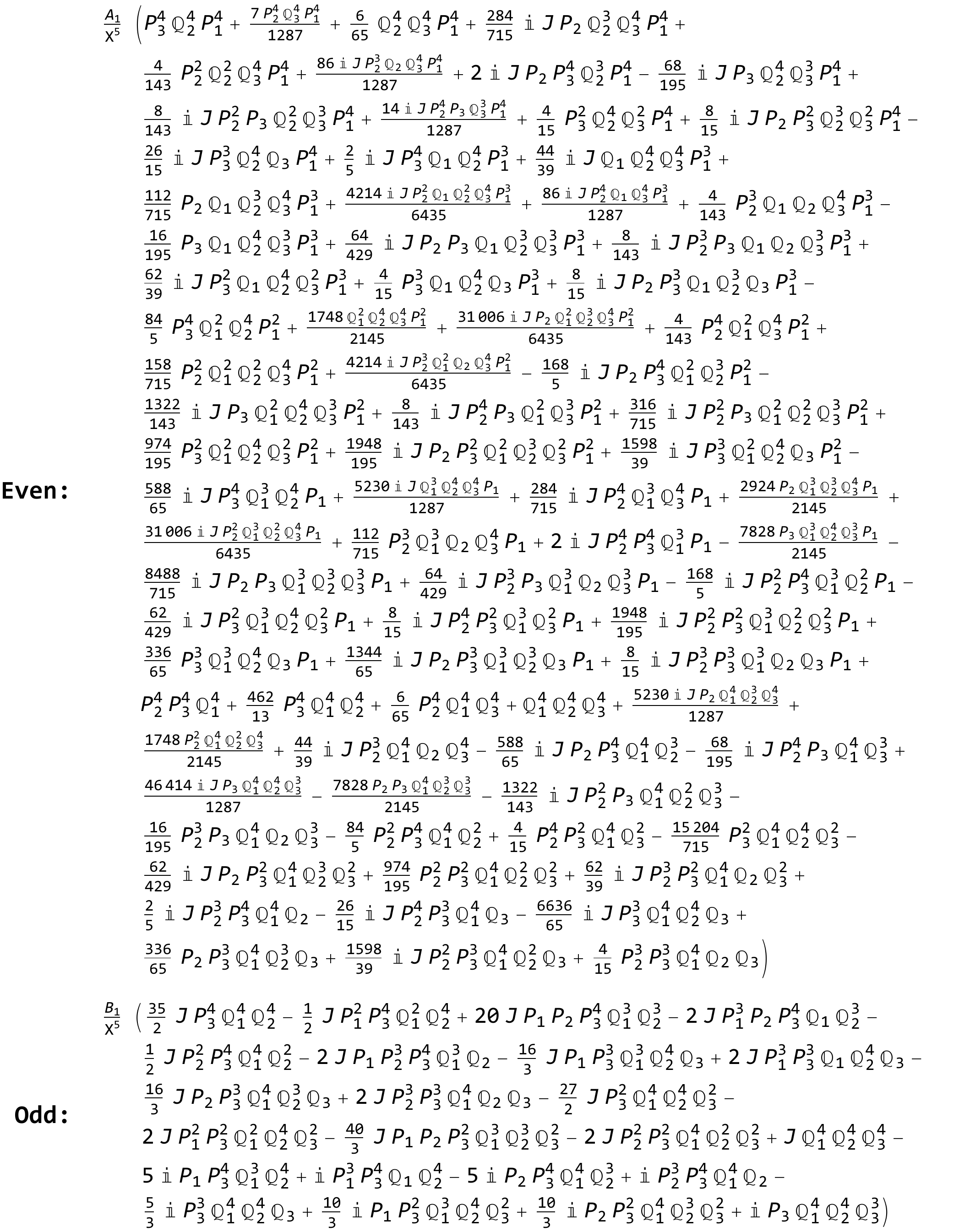} &&
\end{flalign*}

\subsubsection{Grassmann-odd correlators}

The classification of results for Grassmann-odd three-point functions, including cases where there is a point-switch symmetry, is as follows:
\begin{itemize}
	\item In all cases we have examined ($s_{i} \leq 20$), the three-point functions are fixed up to a single parity-even solution after conservation on all three points. In general, any parity-odd structures are incompatible with conservation.
	\item $\langle \mathbf{J}^{}_{s_{1}} \mathbf{J}^{}_{s_{1}} \mathbf{J}'_{s_{2}} \rangle$: 
	Note that in this case $s_2$ must be half-integer.
	For $s_{1} \neq s_{2}$, the classification is as follows:
	\begin{itemize}
		\item Let $s_{2} = 2k+\tfrac{1}{2}$, $k \in \mathbb{Z}_{\geq 0}$; for arbitrary $s_{1}$ (integer or half-integer) the point-switch symmetry is not satisfied and therefore the three-point function vanishes in general.
		\item Let $s_{2} = 2k+\tfrac{3}{2}$, $k \in \mathbb{Z}_{\geq 0}$; for arbitrary $s_{1}$ (integer or half-integer) the point-switch symmetry is satisfied and therefore the three-point function is fixed up to a single parity-even structure.
	\end{itemize}
	
	\item $\langle \mathbf{J}_{s} \, \mathbf{J}_{s} \, \mathbf{J}_{s} \rangle$: for $s = 2k + \tfrac{3}{2}$, $k \in \mathbb{Z}_{\geq 0}$ the solution is compatible with the point-switch symmetry. For $s = 2k+\tfrac{1}{2}$, $k \in \mathbb{Z}_{\geq 0}$ the three-point function vanishes. 
		
\end{itemize}
We now present results after imposing conservation on all three points.

\vspace{2mm}

\noindent
\textbf{Correlation function} $\langle \mathbf{J}^{}_{1/2} \mathbf{J}'_{3/2} \mathbf{J}''_{5/2} \rangle$\textbf{:}
\begin{flalign*}
	\hspace{5mm} \includegraphics[width=0.97\textwidth]{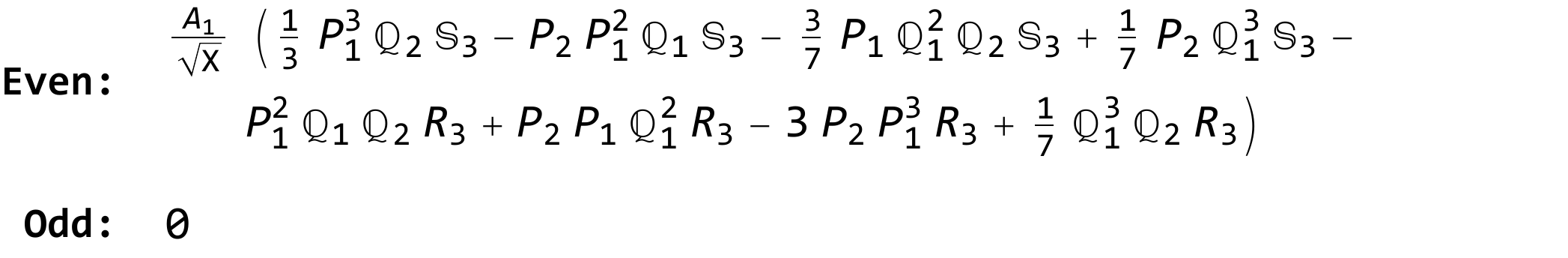} &&
\end{flalign*}

\noindent
\textbf{Correlation function} $\langle \mathbf{J}^{}_{2} \mathbf{J}'_{2} \mathbf{J}''_{1/2} \rangle$\textbf{:}
\begin{flalign*}
	\hspace{5mm} \includegraphics[width=0.97\textwidth]{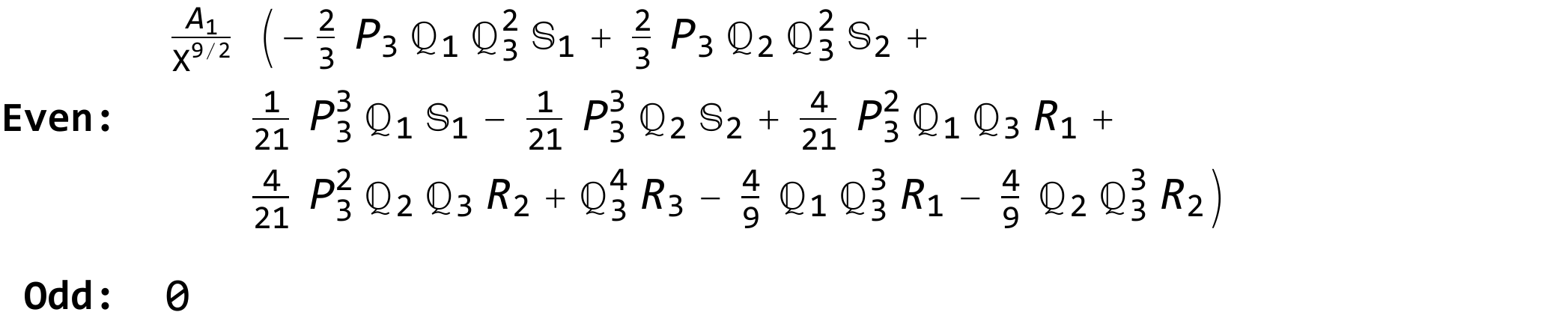} &&
\end{flalign*}
In this instance we note that the superspin triangle inequalities are not satisfied and therefore the odd solution vanishes after current conservation.

\vspace{4mm}

\noindent
\textbf{Correlation function} $\langle \mathbf{J}^{}_{2} \mathbf{J}'_{2} \mathbf{J}''_{3/2} \rangle$\textbf{:}
\begin{flalign*}
	\hspace{5mm} \includegraphics[width=0.97\textwidth]{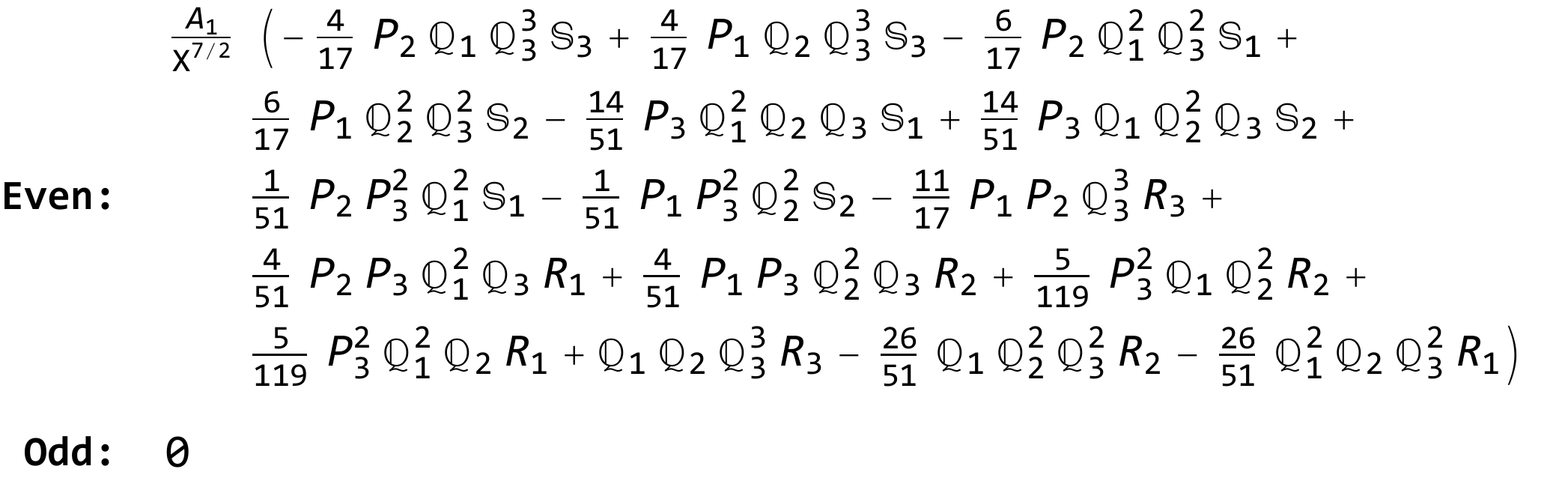} &&
\end{flalign*}

\noindent
\textbf{Correlation function} $\langle \mathbf{J}^{}_{3/2} \mathbf{J}'_{3/2} \mathbf{J}''_{5/2} \rangle$\textbf{:}
\begin{flalign*}
	\hspace{5mm} \includegraphics[width=0.97\textwidth]{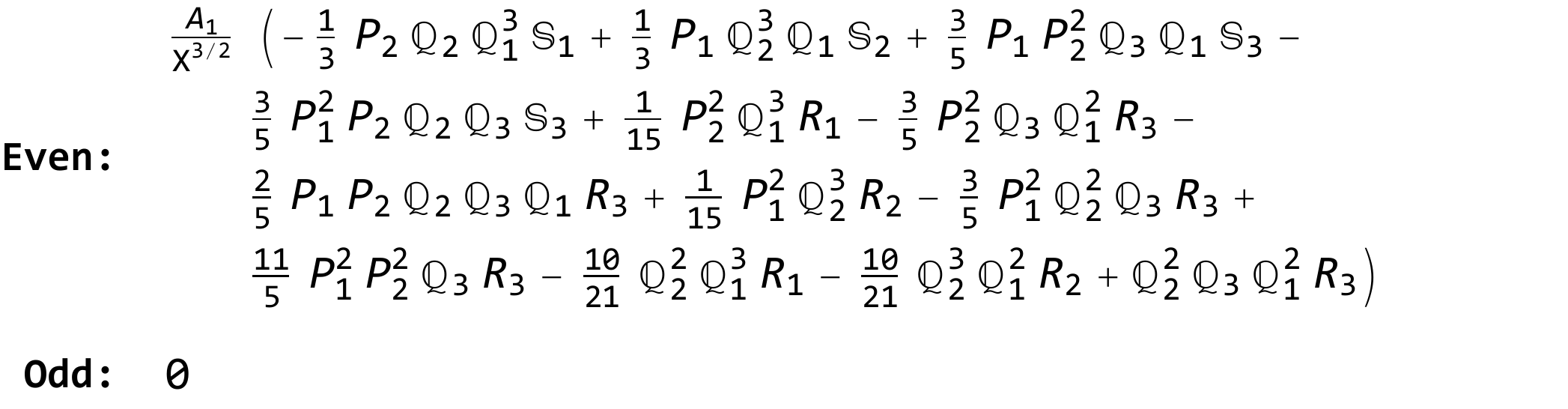} &&
\end{flalign*}

\noindent
\textbf{Correlation function} $\langle \mathbf{J}^{}_{3/2} \mathbf{J}'_{3/2} \mathbf{J}''_{7/2} \rangle$\textbf{:}
\begin{flalign*}
	\hspace{5mm} \includegraphics[width=0.97\textwidth]{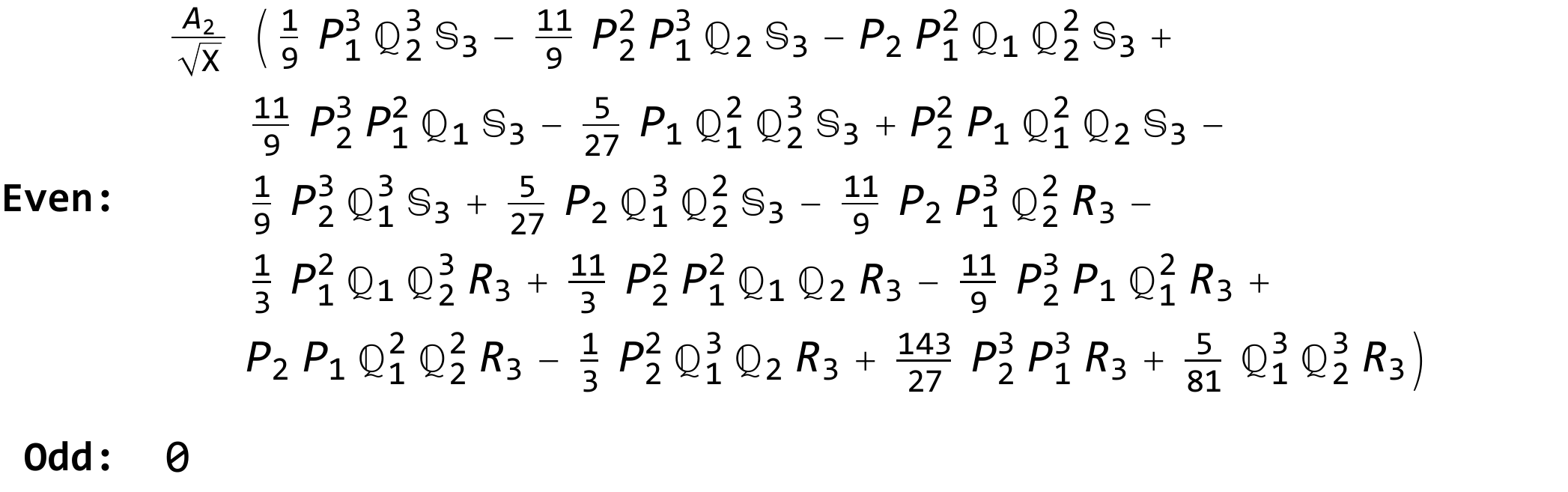} &&
\end{flalign*}

\noindent
\textbf{Correlation function} $\langle \mathbf{J}^{}_{2} \mathbf{J}'_{2} \mathbf{J}''_{7/2} \rangle$\textbf{:}
\begin{flalign*}
	\hspace{5mm} \includegraphics[width=0.97\textwidth]{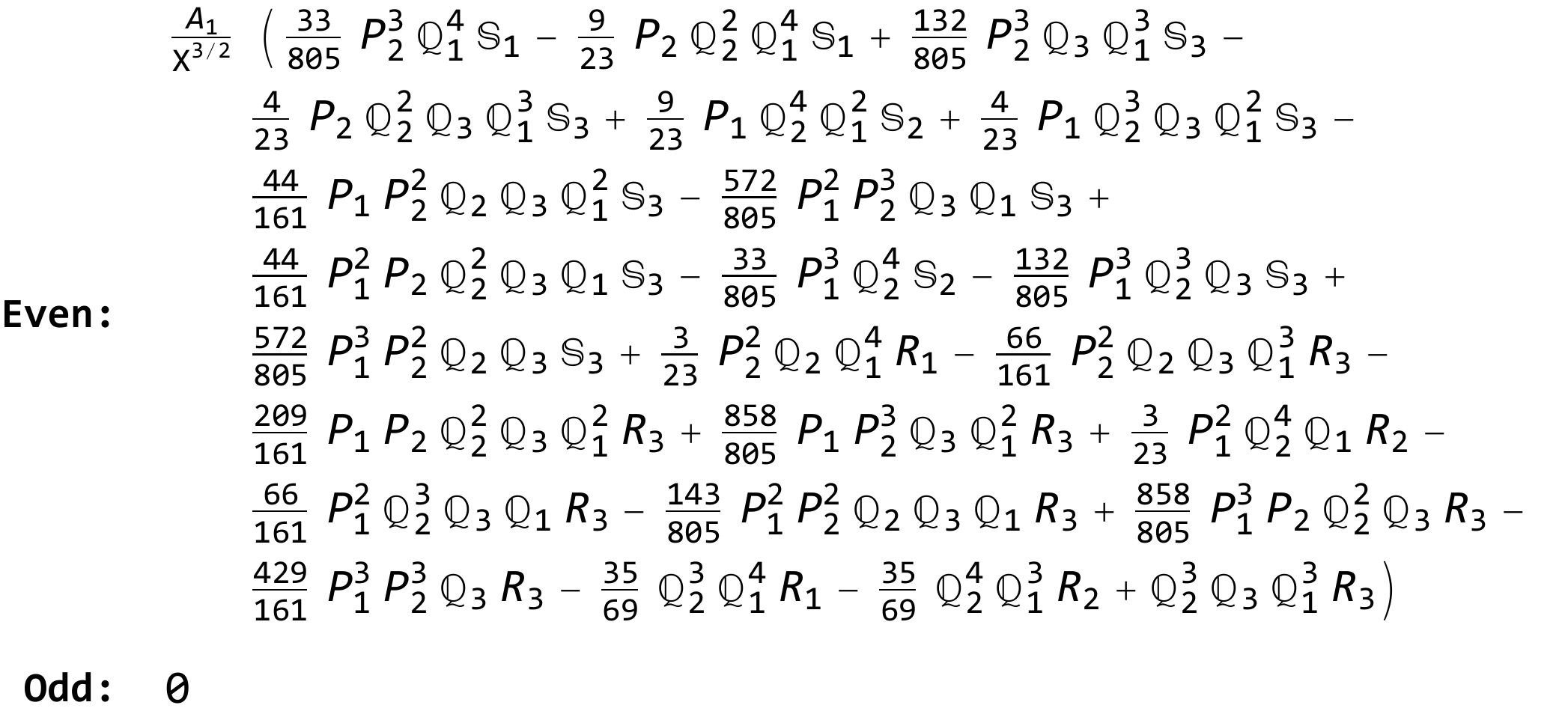} &&
\end{flalign*}

\noindent
\textbf{Correlation function} $\langle \mathbf{J}^{}_{5/2} \mathbf{J}'_{5/2} \mathbf{J}''_{5/2} \rangle$\textbf{:}
\begin{flalign*}
	\hspace{5mm} \includegraphics[width=0.97\textwidth]{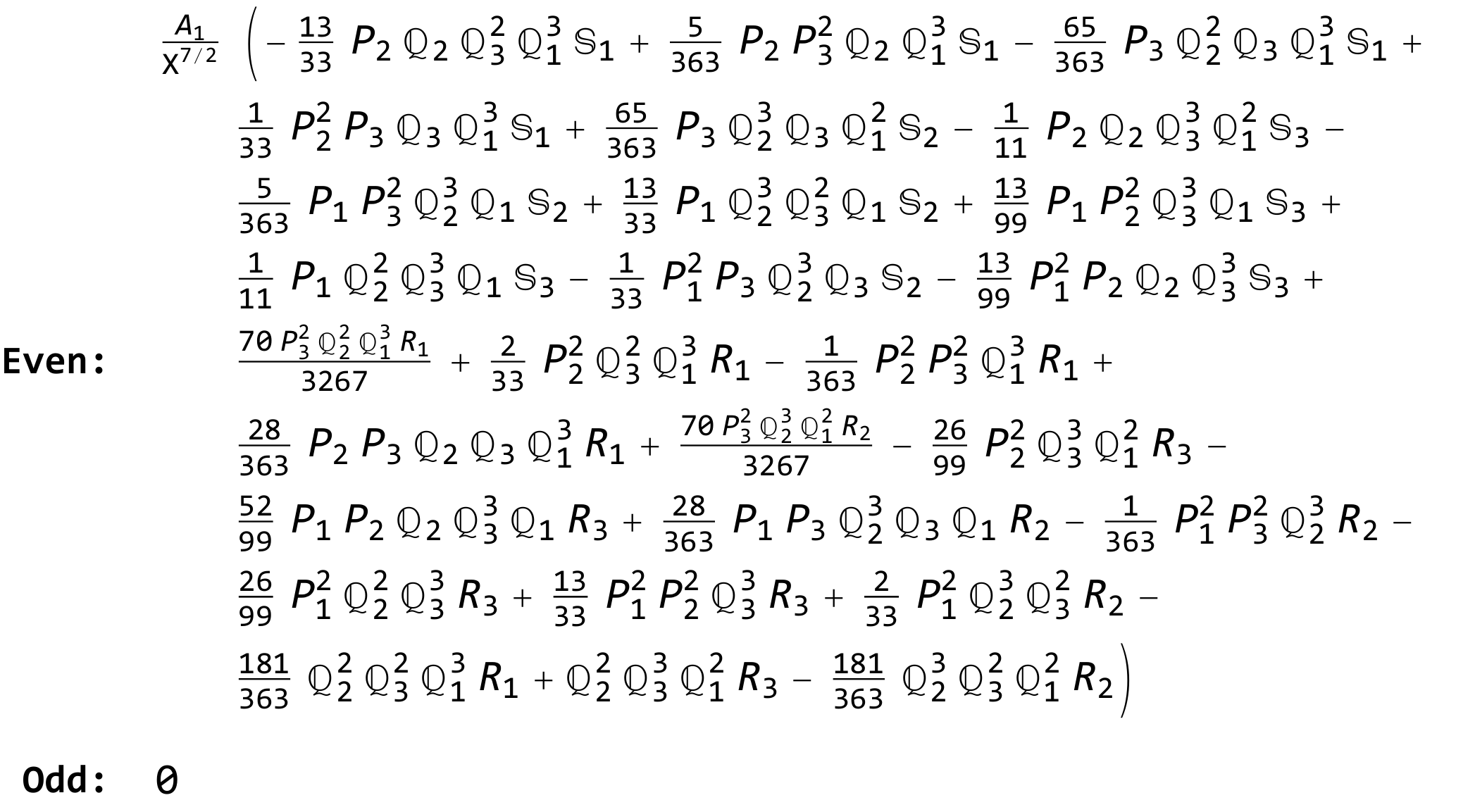} &&
\end{flalign*}

\noindent
\textbf{Correlation function} $\langle \mathbf{J}^{}_{7/2} \mathbf{J}'_{7/2} \mathbf{J}''_{7/2} \rangle$\textbf{:}
\begin{flalign*}
	\hspace{5mm} \includegraphics[width=0.97\textwidth]{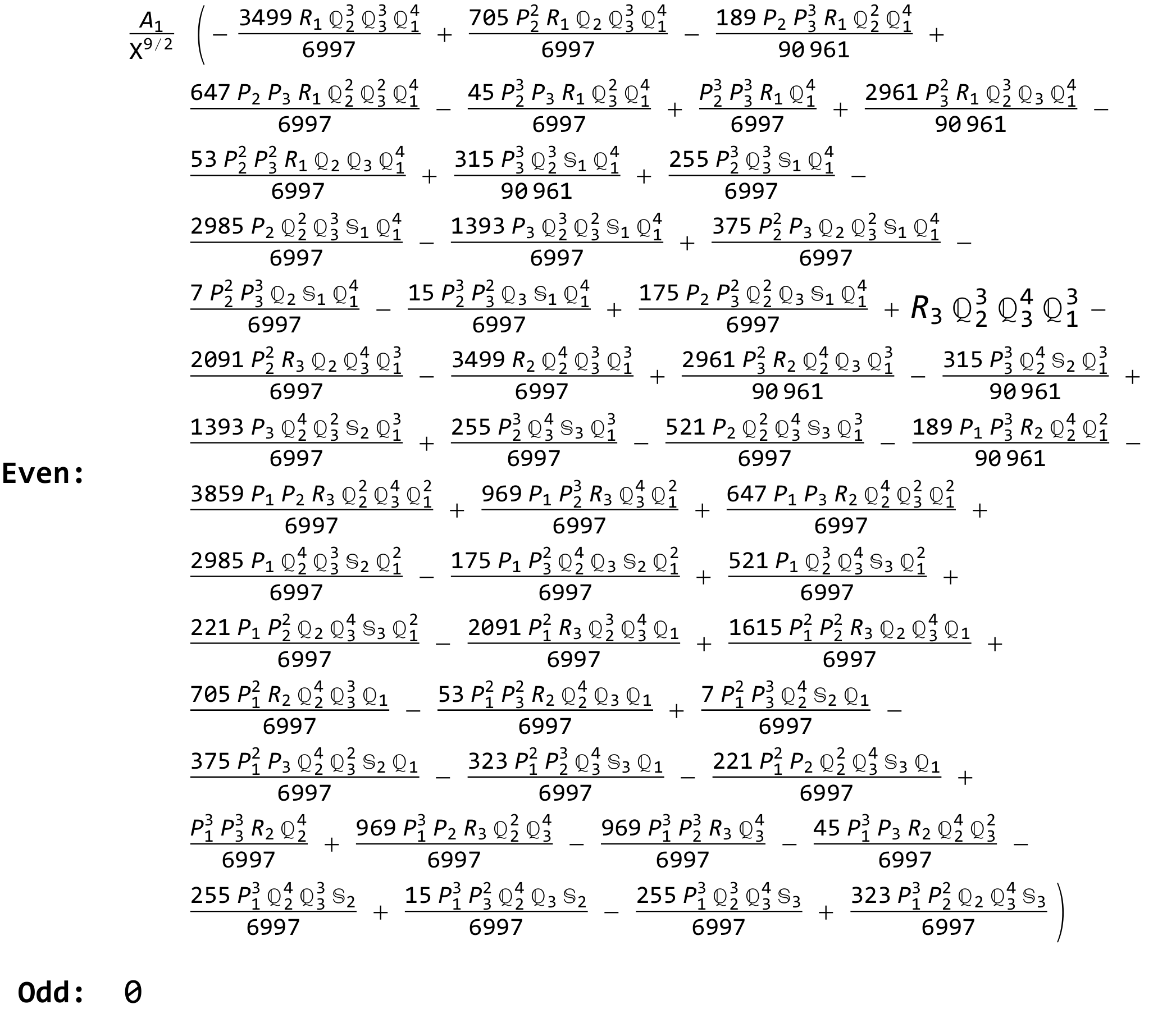} &&
\end{flalign*}

\section{Three-point functions of scalar superfields}\label{section5}

For completeness, in this section we analyse three-point correlation functions involving scalar superfields and conserved supercurrents. Some of the three-point functions contain parity-odd solutions, with their existence depending on both triangle inequalities and the weights of the scalars. We found that the following general results hold:
\begin{subequations}
	\begin{align}
		\langle \cO \, \cO' \, \mathbf{J}_{s} \rangle &= a \, \langle \cO \, \cO' \, \mathbf{J}_{s} \rangle_{E} \, , \\[2mm]
		\langle \mathbf{J}^{}_{s_{1}} \mathbf{J}'_{s_{2}} \, \cO \rangle &= a \, \langle \mathbf{J}^{}_{s_{1}} \mathbf{J}'_{s_{2}} \, \cO \rangle_{E} + b \, \langle \mathbf{J}^{}_{s_{1}} \mathbf{J}'_{s_{2}} \, \cO \rangle_{O} \, .
	\end{align}
\end{subequations}
The correlation functions are analysed using the same methods as in the previous sections; the full classification of results (for cases where there is a point-switch symmetry), is summarised below:
\begin{itemize}
	\item $\langle \cO \, \cO' \, \mathbf{J}_{s} \rangle$: in general there are solutions only for $\D_{\cO} = \D_{\cO'}$. For the Grassmann-even case the solution satisfies the point-switch symmetry $\cO = \cO'$ only for even $s$. For the Grassmann-odd case the solution satisfies the point-switch symmetry only for $s = 2k+\tfrac{3}{2}$, $k \in \mathbb{Z}_{\geq 0}$.
	\item $\langle \mathbf{J}^{}_{s_{1}} \mathbf{J}'_{s_{2}} \, \cO \rangle$: for $s_{1} \neq s_{2}$, there is a single even solution for $\Delta_{\cO}=1$, otherwise the three-point function vanishes. For $s_{1} = s_{2}$ there is one even and one odd solution and the point-switch symmetries are satisfied.
\end{itemize}
We now present explicit solutions for the above cases.

\vspace{4mm}

\noindent
\textbf{Correlation function} $\langle \cO \, \cO' \, \mathbf{J}_{1/2} \rangle$\textbf{:}\\
For $\delta_{1} = \delta_{2} = \delta$, there is a single even solution compatible with conservation
\begin{flalign*}
	\hspace{5mm} \includegraphics[width=0.97\textwidth]{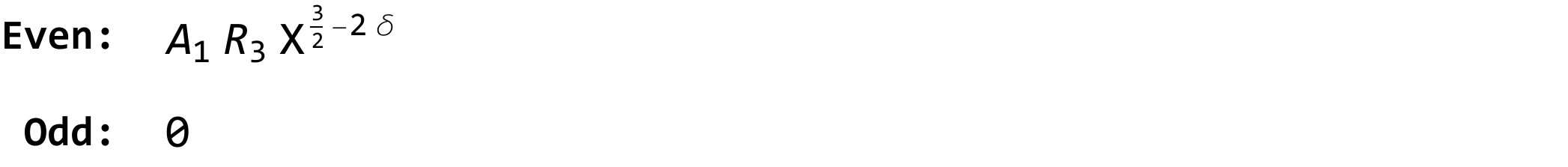} &&
\end{flalign*}

\noindent
\textbf{Correlation function} $\langle \cO \, \cO' \, \mathbf{J}_{1} \rangle$\textbf{:} \\
For $\delta_{1} = \delta_{2} = \delta$, there is a single even solution compatible with conservation
\begin{flalign*}
	\hspace{5mm} \includegraphics[width=0.97\textwidth]{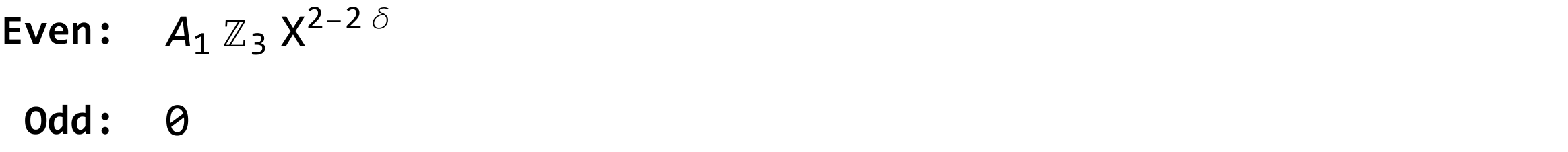} &&
\end{flalign*}

\noindent
\textbf{Correlation function} $\langle \cO \, \cO' \, \mathbf{J}_{3/2} \rangle$\textbf{:}\\
For $\delta_{1} = \delta_{2} = \delta$, there is a single even solution compatible with conservation
\begin{flalign*}
	\hspace{5mm} \includegraphics[width=0.97\textwidth]{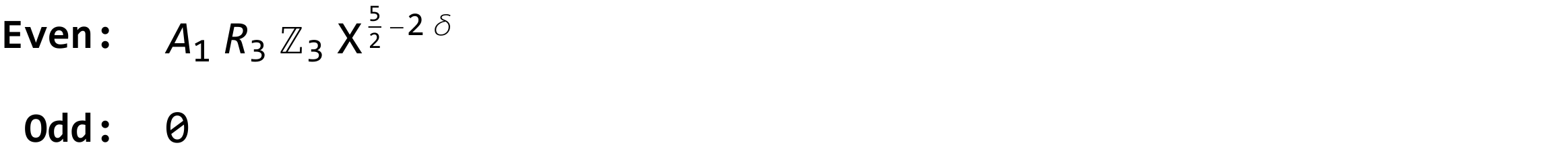} &&
\end{flalign*}

\noindent
\textbf{Correlation function} $\langle \cO \, \cO' \, \mathbf{J}_{2} \rangle$\textbf{:}\\
For $\delta_{1} = \delta_{2} = \delta$, there is a single even solution compatible with conservation
\begin{flalign*}
	\hspace{5mm} \includegraphics[width=0.97\textwidth]{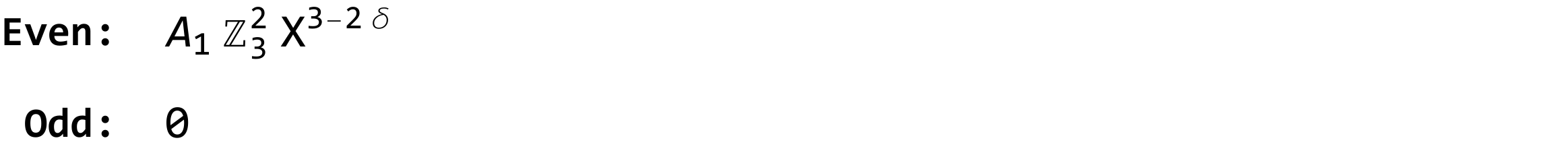} &&
\end{flalign*}

\noindent
\textbf{Correlation function} $\langle \mathbf{J}^{}_{1/2} \mathbf{J}'_{1/2} \cO \rangle$\textbf{:}\\
In this case, the superspin triangle inequalities are satisfied and there is one even and one odd solution for arbitrary $\d$:
\begin{flalign*}
	\hspace{5mm} \includegraphics[width=0.97\textwidth]{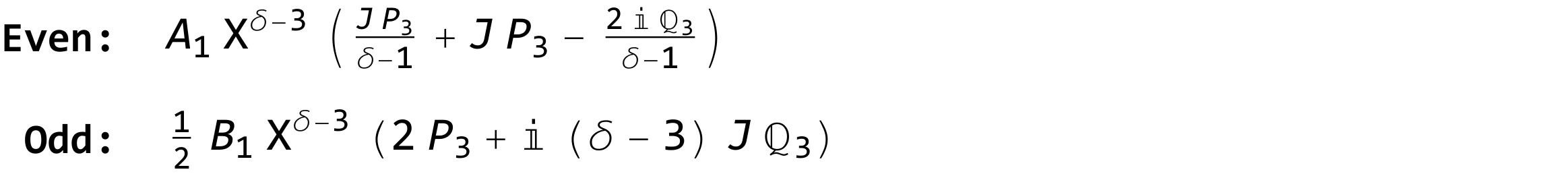} &&
\end{flalign*}

\noindent
\textbf{Correlation function} $\langle \mathbf{J}^{}_{1/2} \mathbf{J}'_{3/2} \cO \rangle$\textbf{:}\\
In this case there is a solution only for $\d = 1$:
\begin{flalign*}
	\hspace{5mm} \includegraphics[width=0.97\textwidth]{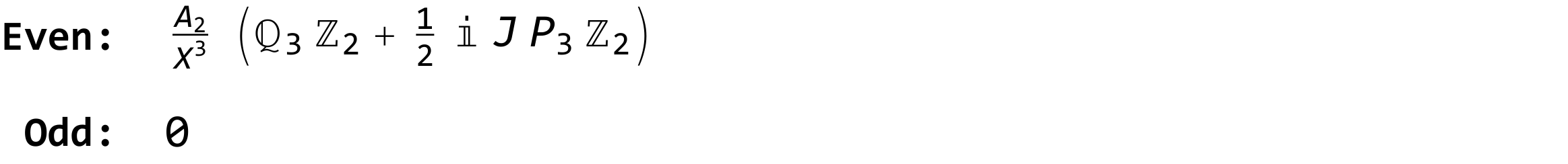} &&
\end{flalign*}

\noindent
\textbf{Correlation function} $\langle \mathbf{J}^{}_{3/2} \mathbf{J}'_{3/2} \cO \rangle$\textbf{:} \\
In this case, the superspin triangle inequalities are satisfied and there is one even and one odd solution for arbitrary $\d$:
\begin{flalign*}
	\hspace{5mm} \includegraphics[width=0.97\textwidth]{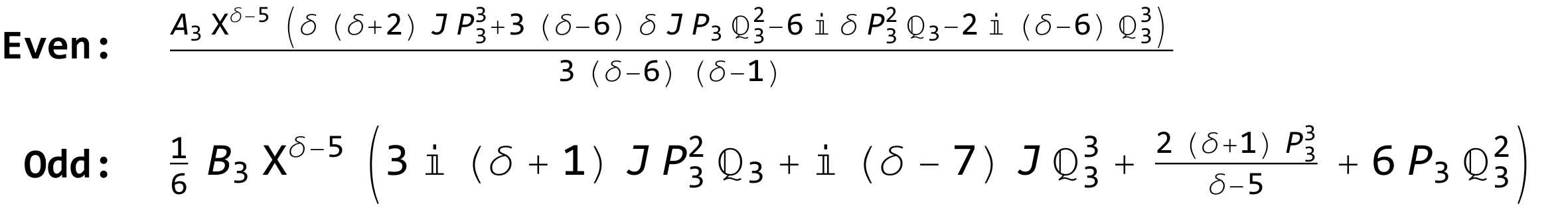} &&
\end{flalign*}

\noindent
\textbf{Correlation function} $\langle \mathbf{J}^{}_{1} \mathbf{J}'_{2} \cO \rangle$\textbf{:} \\
In this case there is a solution only for $\d = 1$:
\begin{flalign*}
	\hspace{5mm} \includegraphics[width=0.97\textwidth]{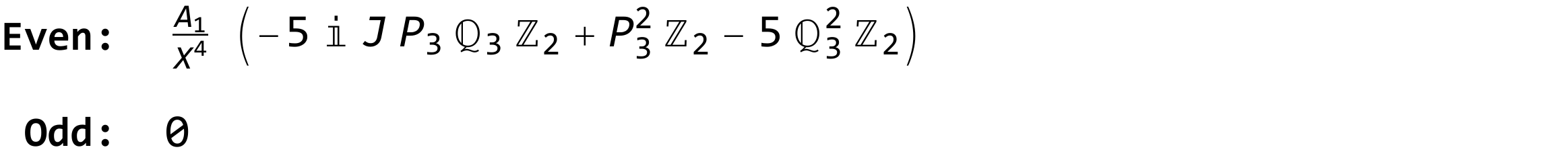} &&
\end{flalign*}

\noindent
\textbf{Correlation function} $\langle \mathbf{J}^{}_{2} \mathbf{J}'_{2} \cO \rangle$\textbf{:} \\
In this case, the superspin triangle inequalities are satisfied and there is one even and one odd solution for arbitrary $\d$:
\begin{flalign*}
	\hspace{5mm} \includegraphics[width=0.97\textwidth]{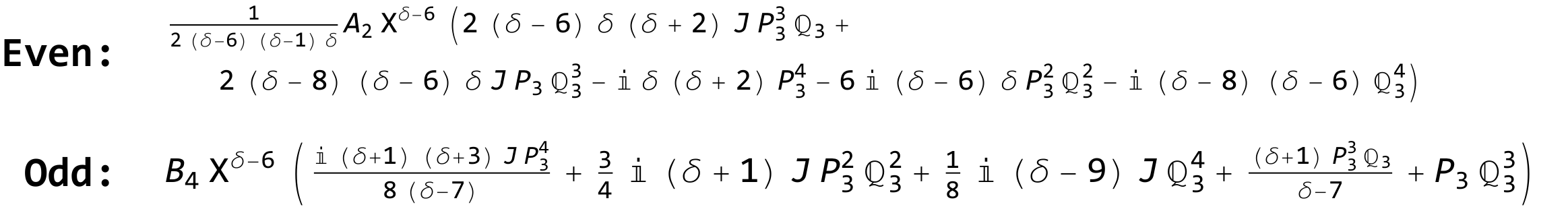} &&
\end{flalign*}
%

\section{Conclusion}\label{section6}

The purpose of this paper was to develop a formalism to determine the general structure of three-point correlation functions of conserved supercurrents for arbitrary 
superspins in three-dimensional superconformal field theory. Our method produces explicit results up to $s_{i} = 20$ and is limited only by computer power. 
We found that the main difference in the general structure of the three-point function $\langle \mathbf{J}^{}_{s_{1}} \mathbf{J}'_{s_{2}} \mathbf{J}''_{s_{3}} \rangle$
is whether it is Grassmann-odd or Grassmann-even in superspace. If $\langle \mathbf{J}^{}_{s_{1}} \mathbf{J}'_{s_{2}} \mathbf{J}''_{s_{3}} \rangle$ is Grassmann-odd
(that is the sum of the superspins is half-integer) then the correlator is fixed up to a single parity-even contribution. If $\langle \mathbf{J}^{}_{s_{1}} \mathbf{J}'_{s_{2}} \mathbf{J}''_{s_{3}} \rangle$ 
is Grassmann-even (that is the sum of the superspins is an integer) then it is fixed up to one even solution and one odd solution; the existence of the latter, however, depends on whether the triangle 
inequalities are satisfied. 
The pattern of the number of independent structures is clear and we have sufficient evidence to propose that our classification of results holds in general. 

There are various possible directions to extend our results. An open question is whether it is possible to find generating functions for arbitrary 
superspins that encapsulate the results in this paper, similar to the ones found in non-supersymmetric 
theories \cite{Stanev:2012nq,Zhiboedov:2012bm, Didenko:2012tv,Didenko:2013bj}. It would also be interesting to apply our methods to superconformal theories in higher dimensions 
(see~\cite{Buchbinder:2021izb,Buchbinder:2021kjk,Buchbinder:2022kmj} for recent progress) and to $\cN$-extended superconformal theories. 
Correlation functions of higher-spin currents in conformal theories with extended supersymmetry have practically not been studied, however recent progress has been reported in~\cite{Jain:2022izp}. 
An important difference compared to the $\cN=1$ case is that conserved currents can carry indices of the $R$-symmetry group. 
Concerning the study of three-point functions in four dimensions, in~\cite{Buchbinder:2022kmj} a method was introduced to study three-point functions of 
conserved supercurrents $J_{\alpha (r) \dot\alpha (r)}$ for arbitrary superspins in 4D $\cN=1$ superconformal field theories. Explicit solutions were constructed for three-point functions involving 
higher-spin supercurrents and flavour current multiplets. The method of~\cite{Buchbinder:2022kmj} was an extension of the one used in \cite{Buchbinder:2021kjk} where 
the classification problem was solved for generic three-point functions of conserved fermionic currents $S_{\a(k)}$ of arbitrary rank in 4D $\cN=1$ SCFT. We believe that the 
formalism developed in the present paper will generalise directly to 4D $\cN=1$ theories and will allow us to extend the results of~\cite{Buchbinder:2022kmj}.
We leave these considerations for a future study.

\section*{Acknowledgements}
The authors are grateful to Sergei Kuzenko and Jessica Hutomo for valuable discussions. We also acknowledge the use of Matthew Headrick’s \textit{Grassmann} Mathematica package for computations with fermionic variables. The work of E.I.B. is supported in part by the Australian Research Council, project No. DP200101944. The work of B.S. is supported by the \textit{Bruce and Betty Green Postgraduate Research Scholarship} under the Australian Government Research Training Program. 



\appendix

\section{3D conventions and notation}\label{AppA}

For the Minkowski metric we use the ``mostly plus'' convention: $\eta_{mn} = \text{diag}(-1,1,1)$. Spinor indices are then raised and lowered with the $\text{SL}(2,\mathbb{R})$ invariant anti-symmetric $\varepsilon$-tensor
\begin{subequations}
	\begin{align}
		\ve_{\a \b} = 
		\begingroup
		\setlength\arraycolsep{4pt}
		\begin{pmatrix}
			\, 0 & -1 \, \\
			\, 1 & 0 \,
		\end{pmatrix}
		\endgroup 
		\, , & \hspace{6mm}
		\ve^{\a \b} =
		\begingroup
		\setlength\arraycolsep{4pt}
		\begin{pmatrix}
			\, 0 & 1 \, \\
			\, -1 & 0 \,
		\end{pmatrix}
		\endgroup 
		\, , \hspace{6mm}
		\ve_{\a \g} \ve^{\g \b} = \d_{\a}{}^{\b} \, , \\[4mm]
		& \hspace{-8mm} \f_{\a} = \ve_{\a \b} \, \f^{\b} \, , \hspace{12mm} \f^{\a} = \ve^{\a \b} \, \f_{\b} \, .
	\end{align}
\end{subequations}
The $\g$-matrices are chosen to be real, and are expressed in terms of the Pauli matrices, $\s$, as follows:
\begin{subequations}
	\begin{align}
		(\g_{0})_{\a}{}^{\b} = - \text{i} \s_{2} = 
		\begingroup
		\setlength\arraycolsep{4pt}
		\begin{pmatrix}
			\, 0 & -1 \, \\
			\, 1 & 0 \,
		\end{pmatrix}
		\endgroup 
		\, , & \hspace{8mm}
		(\g_{1})_{\a}{}^{\b} = \s_{3} = 
		\begingroup
		\setlength\arraycolsep{4pt}
		\begin{pmatrix}
			\, 1 & 0 \, \\
			\, 0 & -1 \,
		\end{pmatrix}
		\endgroup 
		\, , \\[3mm]
		(\g_{2})_{\a}{}^{\b} = - \s_{1} &= 
		\begingroup
		\setlength\arraycolsep{4pt}
		\begin{pmatrix}
			\, 0 & -1 \, \\
			\, -1 & 0 \,
		\end{pmatrix}
		\endgroup 
		\, ,
	\end{align}
	\vspace{1mm}
	\begin{equation}
		(\g_{m})_{\a \b} = \ve_{\b \d} (\g_{m})_{\a}{}^{\d} \, , \hspace{10mm} (\g_{m})^{\a \b} = \ve^{\a \d} (\g_{m})_{\d}{}^{\b} \, .
	\end{equation}
\end{subequations}
The $\g$-matrices are traceless and symmetric
\begin{equation}
	(\g_{m})^{\a}{}_{\a} = 0 \, , \hspace{10mm} (\g_{m})_{\a \b} = (\g_{m})_{\b \a} \, ,
\end{equation} 
and also satisfy the Clifford algebra
\begin{equation}
	\g_{m} \g_{n} + \g_{n} \g_{m} = 2 \eta_{mn} \, .
\end{equation}
For products of $\g$-matrices we make use of the identities
\begin{subequations}
	\begin{align}
		(\g_{m})_{\a}{}^{\r} (\g_{n})_{\r}{}^{\b} &= \eta_{mn} \d_{\a}{}^{\b} + \e_{mnp} (\g^{p})_{\a}{}^{\b} \, , \\[2mm]
		(\g_{m})_{\a}{}^{\r} (\g_{n})_{\r}{}^{\s} (\g_{p})_{\s}{}^{\b} &= \eta_{mn} (\g_{p})_{\a}{}^{\b} - \eta_{mp} (\g_{n})_{\a}{}^{\b} + \eta_{np} (\g_{m})_{\a}{}^{\b} + \e_{mnp} \d_{\a}{}^{\b} \, ,
	\end{align}
\end{subequations}
where we have introduced the 3D Levi-Civita tensor $\e$, with $\e^{012} = - \e_{012} = 1$. We also have the orthogonality and completeness relations for the $\g$-matrices
\begin{equation}
	(\g^{m})_{\a \b} (\g_{m})^{\r \s} = - \d_{\a}{}^{\r} \d_{\b}{}^{\s}  - \d_{\a}{}^{\s}  \d_{\b}{}^{\r} \, , \hspace{8mm} (\g_{m})_{\a \b} (\g_{n})^{\a \b} = -2 \eta_{mn} \, .
\end{equation}
The $\g$-matrices are used to swap from vector indices to spinor indices. For example, given some three-vector $x_{m}$, it may equivalently be expressed in terms of a symmetric second-rank spinor $x_{\a \b}$ as follows:
\begin{subequations}
	\begin{align}
		x_{\a \b} = (\g^{m})_{\a \b} x_{m}  \, , \hspace{5mm} x_{m} = - \frac{1}{2} (\g_{m})^{\a \b} x_{\a \b} \, , \\[2mm]
		\det (x_{\a \b}) = \frac{1}{2} x^{\a \b} x_{\a \b} = - x^{m} x_{m} = -x^{2} \, .
	\end{align}
\end{subequations}
The same conventions are also adopted for the spacetime partial derivatives $\partial_{m}$
\begin{subequations}
	\begin{align}
		\partial_{\a \b} = (\g^{m})_{\a \b} \partial_{m}  \, , \hspace{5mm} \partial_{m} = - \frac{1}{2} (\g_{m})^{\a \b} \partial_{\a \b} \, , \\[2mm]
		\partial_{m} x^{n} = \d_{m}^{n} \, , \hspace{5mm} \partial_{\a \b} x^{\r \s} = - \d_{\a}{}^{\r} \d_{\b}{}^{\s}  - \d_{\a}{}^{\s}  \d_{\b}{}^{\r} \, ,
	\end{align}
\end{subequations}
\begin{equation}
	\x^{m} \partial_{m} = - \frac{1}{2} \x^{\a \b} \partial_{\a \b} \, .
\end{equation}
We also define the supersymmetry generators $Q_{\a}$
\begin{equation}
	Q_{\a} = \text{i} \frac{\partial}{\partial \q^{\a}} + (\g^{m})_{\a \b} \q^{\b} \frac{\partial}{\partial x^{m}} \, , \label{Supercharges}
\end{equation}
and the covariant spinor derivatives
\begin{equation}
	D_{\a} = \frac{\partial}{\partial \q^{\a}} + \text{i} (\g^{m})_{\a \b} \q^{\b} \frac{\partial}{\partial x^{m}} \, , \label{Covariant spinor derivatives}
\end{equation}
which anti-commute with the supersymmetry generators, $\{ Q_{\a} , D_{\b}\} = 0$, and obey the standard anti-commutation relations
\begin{equation}
	\{ D_{\a} , D_{\b} \} = 2 \text{i} \, (\g^{m})_{\a \b} \partial_{m} \, .
\end{equation}

\section{Conservation identities}\label{AppB}

For imposing superfield conservation equations on three-point correlation functions, the following identities are essential:
\begin{subequations} \label{Derivative identities}
	\begin{align}
		\cD^{\a} Q_{1} &= \frac{\text{i}}{X^{1/2}} \Big\{ v^{\a} R_{3} + w^{\a} R_{2} - Q_{1} \, 	(\hat{X}\cdot\hat{\Q})^{\a} \Big\} \, , \\
		\cD^{\a} Q_{2} &= \frac{\text{i}}{X^{1/2}} \Big\{ u^{\a} R_{3} + w^{\a} R_{1} - Q_{2} \, 	(\hat{X}\cdot\hat{\Q})^{\a} \Big\} \, , \\
		\cD^{\a} Q_{3} &= \frac{\text{i}}{X^{1/2}} \Big\{ u^{\a} R_{2} + v^{\a} R_{1} - Q_{3} \, 	(\hat{X}\cdot\hat{\Q})^{\a} \Big\} \, ,
	\end{align}
	\vspace{-5mm}
	\begin{align}
		\cD^{\a} Z_{1} &= \frac{\text{i}}{X^{1/2}} \Big\{ 2 u^{\a} R_{1} - Z_{1} \, (\hat{X}\cdot\hat{\Q})^{\a} \Big\} \, , \\
		\cD^{\a} Z_{2} &= \frac{\text{i}}{X^{1/2}} \Big\{ 2 v^{\a} R_{2} - Z_{2} \, (\hat{X}\cdot\hat{\Q})^{\a} \Big\} \, , \\
		\cD^{\a} Z_{3} &= \frac{\text{i}}{X^{1/2}} \Big\{ 2 w^{\a} R_{3} - Z_{3} \, (\hat{X}\cdot\hat{\Q})^{\a} \Big\} \, ,
	\end{align}
	\vspace{-5mm}
	\begin{align}
		\cD^{\a} R_{1} &= \frac{1}{X^{1/2}} \Big\{ - u^{\a} - \frac{\text{i}}{4} (\hat{X} \cdot u)^{\a} \boldsymbol{J} \Big\} \, , \\
		\cD^{\a} R_{2} &= \frac{1}{X^{1/2}} \Big\{ - v^{\a} - \frac{\text{i}}{4} (\hat{X} \cdot v)^{\a} \boldsymbol{J} \Big\} \, , \\
		\cD^{\a} R_{3} &= \frac{1}{X^{1/2}} \Big\{ - w^{\a} - \frac{\text{i}}{4} (\hat{X} \cdot w)^{\a} \boldsymbol{J} \Big\} \, ,
	\end{align}
	\vspace{-5mm}
	\begin{align}
		\cD^{\a} S_{1} &= \frac{1}{X^{1/2}} \Big\{  (\hat{X} \cdot 	u)^{\a} - \frac{3\text{i}}{4} \, u^{\a} \boldsymbol{J} \Big\} \, , \\
		\cD^{\a} S_{2} &= \frac{1}{X^{1/2}} \Big\{  (\hat{X} \cdot 	v)^{\a} - \frac{3\text{i}}{4} \, v^{\a} \boldsymbol{J} \Big\} \, , \\
		\cD^{\a} S_{3} &= \frac{1}{X^{1/2}} \Big\{ (\hat{X} \cdot 	w)^{\a} - \frac{3\text{i}}{4} \, w^{\a} \boldsymbol{J} \Big\} \, .
	\end{align}
\end{subequations} \\
Similar relations hold for the action of $\cQ^{\a}$ on the basis structures.



\printbibliography[heading=bibintoc,title={References}]



\end{document}